\documentclass[prd,preprint,tightenlines,floatfix,showpacs,preprintnumbers,nofootinbib,eqsecnum]{revtex4-1}

\pdfoutput=1

\usepackage[T1]{fontenc} 
\usepackage[utf8]{inputenc}

\usepackage{hyperref}
\usepackage{graphicx}
\usepackage{amssymb} 
\usepackage{xspace}
\usepackage{xcolor}
\usepackage{bm}         
\usepackage{array}
\usepackage{booktabs}   
\usepackage{mleftright}

\usepackage{bbold}
\usepackage{amsmath,amscd}
\usepackage{placeins}
\usepackage{soul}
\usepackage{braket}
\usepackage{tikz-cd}
\usepackage{empheq}
\usepackage[most]{tcolorbox}
\usetikzlibrary{arrows, shapes,shadows}
\usepackage{mathtools,slashed}
\usepackage{dsfont}
\usepackage[makeroom]{cancel}
\usepackage{multirow}

\usepackage[capitalise]{cleveref}

\crefname{section}{Sec.}{Secs.}
\crefname{table}{Tab.}{Tabs.}
\crefname{figure}{Fig.}{Figs.}
\crefname{equation}{Eq.}{Eqs.}
\crefname{appendix}{Appendix\ }{Appendix\ }

\DeclareMathAlphabet{\mathpzc}{OT1}{pzc}{m}{it}

\newcommand{\vev}[0]{VEV\xspace}
\newcommand{\vevs}[0]{VEVs\xspace}

\newcommand{\del}{\partial}

\newcommand*{\Scale}[2][4]{\scalebox{#1}{$#2$}}%
\renewcommand{\(}{\left(}
\renewcommand{\)}{\right)}
\renewcommand{\[}{\left[}
\renewcommand{\]}{\right]}

\newcommand{\alf}[1]{\alpha^{-1}_{\mathrm{#1}}}
\newcommand{\al}[2]{\alpha^{-1}_{\mathrm{#1}}\left(#2\right)}

\newcommand{\U}[1]{\mathrm{U}(1)_{\mathrm{#1}}}			
\newcommand{\SU}[2]{\mathrm{SU}(#1)_{\mathrm{#2}}}		
\newcommand{\SO}[2]{\mathrm{SO}(#1)_{\mathrm{#2}}}		
\newcommand{\E}[1]{\mathrm{E}_{#1}}		

\newcommand{\lam}[2]{\lambda_{\Scale[0.50]{#1 \-- #2}}}
\newcommand{\la}[1]{\lambda_{\Scale[0.5]{#1}}}
\newcommand{\lap}[1]{\overline{\lambda}_{\Scale[0.5]{#1}}}
\newcommand{\y}[1]{\mathrm{y}_{\Scale[0.5]{#1}}}
\newcommand{\yp}[1]{\overline{\mathrm{y}}_{\Scale[0.5]{#1}}}

\newcommand{\LLR}[3]{\left(\bm{L}^{ #1} \right)^{ #2 }{}_{ #3 }}
\newcommand{\QL}[3]{\left(\bm{Q}_{\mathrm{L}}^{ #1} \right)^{ #2 }{}_{ #3 }}
\newcommand{\QR}[3]{\left(\bm{Q}_{\mathrm{R}}^{ #1} \right)^{ #2 }{}_{ #3 }}

\definecolor{bostonuniversityred}{rgb}{0.8, 0.0, 0.0}

\newcommand{\abs}[1]{\left| #1 \right| }
\newcommand{\mean}[1]{\left \langle #1 \right \rangle }

\newcommand{\ro}{\textrm}

\begin{document}

\title{Prospects for New Physics from gauge Left-Right-Colour-Family Grand Unification hypothesis}

\author{Ant{\'o}nio~P.~Morais}
\email{aapmorais@ua.pt}
\affiliation{Departamento de F\'isica, Universidade de Aveiro and CIDMA, Campus de Santiago, 
3810-183 Aveiro, Portugal}
\affiliation{Department of Astronomy and Theoretical Physics, Lund University, 221 00 Lund, Sweden}

\author{Roman~Pasechnik}
\email{Roman.Pasechnik@thep.lu.se}
\affiliation{Department of Astronomy and Theoretical Physics, Lund University, 221 00 Lund, Sweden}

\author{Werner~Porod}
\email{porod@physik.uni-wuerzburg.de}
\affiliation{Institut für Theoretische Physik und Astrophysik, Uni Würzburg, Germany\vspace{1cm}}

\begin{abstract}
\vspace{0.5cm}
Given the tremendous phenomenological success of the Standard Model (SM) 
framework, it becomes increasingly important to understand to what extent its specific 
structure dynamically emerges from unification principles. In this study, we present 
a novel anomaly-free supersymmetric (SUSY) Grand Unification model based upon gauge trinification 
$[\SU{3}{}]^3$ symmetry and a local $\SU{2}{F}\times \U{F}$ family symmetry, with particle
spectra and gauge symmetries inspired by a possible reduction pattern $\mathrm{E}_8 \to 
\mathrm{E}_6\times \SU{2}{F}\times \U{F}$, with subsequent $\mathrm{E}_6\to [\SU{3}{}]^3$ 
symmetry breaking step. In this framework, higher-dimensional operators of $\mathrm{E}_6$ 
induce the threshold corrections in the gauge and Yukawa interactions leading, 
in particular, to only two distinct Yukawa couplings in the fundamental 
sector of the resulting $[\SU{3}{}]^3\times \SU{2}{F}\times \U{F}$ Lagrangian. 
Among the appealing features emergent in this framework are the Higgs-matter unification 
and a unique minimal three Higgs doublet scalar sector 
at the electroweak scale as well as tree-level hierarchies in the light fermion 
spectra consistent with those observed in nature. In addition, 
our framework reveals a variety of prospects for New Physics searches at the LHC and 
future colliders such as vector-like fermions, as well as rich scalar, gauge and 
neutrino sectors.
\end{abstract}

\maketitle


\section{Introduction}
\label{Sect:Intro}

After the discovery of the Higgs boson at the LHC by the ATLAS and CMS collaborations \citep{Aad:2012tfa,Chatrchyan:2012xdj} our
current understanding for the origin of the mass of the fundamental particles, as described by
the Standard Model (SM), has finally met an experimental confirmation. Despite the great success achieved, a consensual explanation for the observed features of the particle spectra and interactions observed in nature 
is still lacking. Along these lines, while over the past forty years the strong and electroweak (EW) interactions have been extensively probed and confirmed in various experiments, their origin at a more fundamental level is still unknown. Besides, the existing SM framework is not capable of explaining some of the observed phenomena such as the specific patterns and hierarchies in its fermion spectra nor contains a suitable candidate for Dark Matter. At last, but not least, it cannot explain the observed matter-antimatter asymmetry in the Universe.

Typically, these problems are addressed separately in different contexts. In order to describe the origin of the SM gauge interactions one typically refers to Grand Unified Theories (GUTs) where larger continuous symmetries contain the SM gauge group, e.g.~$\SU{5}{}$, $\SO{10}{}$, or $\mathrm{E}_6$  \cite{Georgi:1974sy,Fritzsch:1974nn,Chanowitz:1977ye,Georgi:1978fu,
Georgi:1979dq,Georgi:1979ga,Georgi:1982jb,Gursey:1975ki,Gursey:1981kf,Achiman:1978vg,Pati:1974yy}. A common procedure to resolve the flavour problem in minimal extensions of the SM or in GUT theories is by introducing new discrete or continuous family symmetries at high-energy scales. For a few most recent and representative implementations, see e.g.~Refs.~\cite{Ordell:2019zws,CarcamoHernandez:2019cbd,Vien:2019zhs,CarcamoHernandez:2019vih,CarcamoHernandez:2018iel,Bjorkeroth:2019csz,Gui-JunDing:2019wap} and references therein. Most of the studies focus on the neutrino sector combined with a variation of the seesaw mechanisms~\cite{Schechter:1980gr,Schechter:1981cv} (see also Refs.~\cite{Boucenna:2014zba,Ma:2014qra,Chulia:2016ngi,CentellesChulia:2018gwr,CentellesChulia:2018bkz,Mohapatra:1979ia,Foot:1988aq,Babu:1988ki}).

In this work, we propose a new look into such fundamental questions as 1) the origin of the gauge interactions in the SM, and 2) the origin of the quark, lepton and neutrino families' replication experimentally observed in nature. These questions are addressed by tying together in a common framework both flavour physics and Grand Unification, which are typically treated on a different footing. Furthermore, we explore which new physics scenarios are expected to emerge at phenomenologically relevant energy scales as sub-products of our framework and investigate theoretical possibilities for both the gauge couplings unification as well as Yukawa couplings unification.

In previous work by some of the authors \cite{Camargo-Molina:2016bwm,Camargo-Molina:2016yqm,Camargo-Molina:2017kxd}, a philosophy of family-gauge unification has been introduced based upon a trinification-GUT $\[\SU{3}{}\]^3$ model, or T-GUT for short, where the gauge sector is extended by a global $\SU{3}{F}$ family symmetry. A supersymmetric (SUSY) version of this theory is called as the SUSY Higgs-Unified Theory (SHUT) due to an emergent SM Higgs-matter unification property inspired by an embedding of $\[\SU{3}{}\]^3 \times \SU{3}{F}$ symmetry into $\mathrm{E}_8$. The SHUT framework reveals several interesting features such as, e.g.~the radiative nature of the Yukawa sector of SM leptons and lightest quarks as well as the absence of the $\mu$-problem. However, its first particular realisation in Ref.~\cite{Camargo-Molina:2017kxd} relies on a few simplifying assumptions such as the presence of a $\mathbb{Z}_3$ cyclic permutation symmetry acting upon the $\[\SU{3}{}\]^3$ subgroup of $\mathrm{E}_6$, as has been proposed initially by Glashow \cite{original}, and an approximately global family symmetry. These assumptions are not necessary and will be consistently avoided in the framework presented in this work leading to several relevant features to be discussed in what follows.

The $\mathrm{E}_6$-based GUTs, also accompanied with family symmetries, have received a lot of attention in the literature due to a number of attractive features 
(see e.g.~Refs.~\cite{Gursey:1975ki,Achiman:1978vg,Barbieri:1980vc,Bando:1999km,Bando:2000gs,Bando:2001bj,Maekawa:2002eh,Maekawa:2002bk,Maekawa:2004qj,Ishiduki:2009vr,Kawase:2010na}).
In variance to the previous implementation of the flavored T-GUT realised by some of us in Ref.~\cite{Camargo-Molina:2017kxd}, in this work we abandon the $\mathbb{Z}_3$ 
symmetry at the T-GUT scale and consider a minimal anomaly-free realisation of the $\mathrm{E}_6 \times \SU{2}{F} \times \U{F}$ SUSY theory followed by $\mathrm{E}_6\to \[\SU{3}{}\]^3$ 
symmetry breaking \cite{Morais:2020odg}\footnote{For alternative realisations of $\mathrm{E}_6$ GUT with $\SU{2}{}$ or $\SU{3}{}$ family symmetry, 
see Refs.~\cite{Maekawa:2002eh,Maekawa:2004qj,Ishiduki:2009vr,Kawase:2010na}}. We also 
consider the family $\SU{2}{F} \times \U{F}$ symmetry as a gauge group on the same footing of the Left, Right and Colour symmetries. Under the key hypothesis of a full 
Left-Right-Colour-Family unification, the gauge and matter sectors of our model are inspired by the reduction pattern $\mathrm{E}_8 \to \mathrm{E}_6 \times \SU{2}{F} \times \U{F}$
that may be realised in extra-dimensional scenarios via e.g. the Wilson-line breaking and orbifolding techniques \cite{Dixon:1985jw,Dixon:1986jc}\footnote{For example, an orbifolding mechanism
has also been used to unify the gauge and family symmetries into $SO(18)$ in Refs.~\cite{Reig:2017nrz,Reig:2018ocz} and, combined with a Wilson-line breaking technique, 
into $\mathrm{E}_8$ in Refs.~\cite{Aranda:2020noz,Aranda:2020zms}.}. Starting with vector-like $\mathrm{E}_8$ representations, the extra-dimensional symmetry breaking mechanisms enable one to remove mirror antichiral components yielding a chiral anomaly-free theory in four dimensions \cite{Adler:2002yg,Adler:2004uj}. In the framework of orbifolding scenarios, such an approach typically yields many light unobservable states that are difficult to make consistent with phenomenology in a conventional field-theoretical way, which is an open problem. In this work, we do not rely on any particular extra-dimensional scenario of $\mathrm{E}_8$ breaking. Instead, we adopt a phenomenologically motivated approach and postulate a minimal anomaly-free superfield content in the effective four-dimensional $\mathrm{E}_6 \times \SU{2}{F} \times \U{F}$ SUSY theory, where the sector of light chiral superfields (containing the SM matter) is inspired by its possible embedding into the lowest $248$ representation of $\mathrm{E}_8$, and then explore its overall phenomenological consistency by studying its symmetry breaking, particle spectra and gauge coupling evolution down to the SM energy scale. If the $\mathrm{E}_8$ and $\mathrm{E}_6$ 
breaking scales are not too far apart, one expects that high-dimensional operators of the $\mathrm{E}_6$ theory are sizeable. Indeed, we show that such operators are important for 
both gauge coupling unification as well as for explaining the observed hierarchy of the fermion mass spectrum. We also show that taking the measured gauge couplings as input one obtains 
an $\mathrm{E}_8$ scale of a few times $10^{17}$ GeV as expected for the string scale.

The first notable consequence of dimension-5 operators in the $\mathrm{E}_6$ gauge-kinetic function 
is the existence of sizeable threshold corrections to the gauge couplings, see e.g.~\cite{Chakrabortty:2008zk}. Thus, the universality among Left-Right-Colour 
$\SU{3}{}$ gauge interactions previously imposed by a $\mathbb{Z}_3$ permutation 
group \cite{Camargo-Molina:2017kxd} does not hold any longer. Therefore, 
the mass scale for the soft SUSY breaking terms gets considerably lowered compared
to the previous attempt. This is intrinsically connected to a second notable effect, where dimension-4 operators in the superpotential of the $\E{6}$ theory, in combination with a $\SU{2}{F}\times \U{F}$ flavour symmetry, only allow for two distinct Yukawa couplings in the $\[\SU{3}{}\]^3 \times \SU{2}{F}\times \U{F}$ theory.
This, in turn, together with a slight hierarchy in the vacuum expectation values of
the low-energy scale Higgs bosons allows for an explanation of the top-charm and bottom-strange mass hierarchies
at tree-level. Besides second and third generation quark Yukawa couplings, Majorana neutrino mass terms are also tree-level generated. All other Yukawa couplings in the SHUT model are loop induced as a consequence of soft-SUSY breaking, potentially offering a first-principles explanation for the fermion mass hierarchies and mixing angles observed in nature. As a by-product, the SHUT model also provides specific new physics scenarios involving additional Higgs doublets, new vector-like fermions and even flavour non-universal gauge bosons, possibly, at the reach of the LHC or future colliders.
The size of the soft-SUSY breaking terms and the freedom that they add to the model with a total of 35 mass-dimensional parameters provides enough freedom to make the SM Higgs and Yukawa sectors 
consistent with phenomenology and potentially realisable with not too strong fine-tuning.

The article is organised as follow. In \cref{Sect:model-defs}, we give a detailed 
discussion of the high-scale SUSY model structure focussing on specific conditions on the $\mathrm{E}_8$ reduction pattern
that need to be satisfied in order to obtain a minimal working model based upon $\mathrm{E}_6 \times \SU{2}{F} \times \U{F}$ symmetry.
Besides, we show how the $\left[\SU{3}{}\right]^3$ gauge group emerges along an $\SU{2}{F} \times \U{F}$ family symmetry and 
also how the corresponding representations emerge from those of $\mathrm{E}_8$. We also demonstrate the crucial 
role of high-dimensional operators in generating the T-GUT superpotential which
only contains two unified Yukawa couplings. In \cref{Sect:soft-sector}, we introduce 
the most generic soft SUSY breaking sector in the left-right (LR) symmetric phase 
of the model emerging from the considered high-scale T-GUT. We also discuss 
the spontaneous gauge symmetry breaking (SSB) scheme induced by these soft 
interactions. We find that a new parity emerges, an analogue to $R$-parity in MSSM, 
which forbids the Yukawa-driven proton decay channels. This model is a multi-Higgs 
model and in \cref{Sect:fermion-sector} we give a first analysis of the fermion 
spectra and mixings for both, the `light' SM-like chiral fermions as well as the
vector-like states. A particular focus will be on the interplay of tree-level 
and loop-induced contributions. In \cref{Sect:scales}, we demonstrate how 
the measured gauge couplings lead to an $\mathrm{E}_8$ scale of $O(10^{17})$~GeV 
taking into account the tree-level threshold corrections due to $\mathrm{E}_6$ 
high-dimensional operators. In \cref{Sect:Conclusions}, a brief summary and 
an outlook for future studies is given. In a series of appendices we collect 
the most important details on group-theoretical aspects and 
$\mathrm{E}_6$ representations, the evolution of the gauge couplings at 
different stages of symmetry breaking including the tree-level matching conditions.
Moreover, we present a generic structure of the effective Lagrangian below 
the trinification breaking scale.

\section{Defining the model from unification principles}
\label{Sect:model-defs}  
  
In order to consistently unify the SM gauge and non-SM family interactions one 
needs a simple group  with high enough rank whose reduction down to the SM 
gauge symmetry should occur in several symmetry breaking steps. 
An ambitious goal here is to construct a GUT theory where both types of 
couplings' unification, in gauge and Yukawa sectors, are a dynamically 
emergent phenomena.

A promising candidate to play such a role is the exceptional $\E{8}$ symmetry that has long been motivated as the one describing the dynamics of massless sectors in superstring theories~\cite{Gursey:1975ki,Achiman:1978vg,Gursey:1981kf,Candelas:1985en} (see also Ref.~\cite{Green:2012oqa}). However, it is known that $\E{8}$ is a vector-like symmetry due to the presence of chiral and anti-chiral $\E{6}$ $\bm{27}$-plets in its fundamental representation. Therefore, to obtain the chiral nature of known matter one typically relies on the geometrical extra-dimensional symmetry breaking mechanisms
such as the orbifold compatification and Wilson-line breaking \cite{Dixon:1985jw,Dixon:1986jc}. In this framework, the breaking of a single $\E{8}$ or superstring-inspired $\E{8}\times \E{8}'$ symmetries
via $\mathbb{Z}_N$ orbifold compactification can occur in several distinct ways containing, for example, an $\E{6}$ symmetry or other $\E{8}$ subgroups and leading to different possibilities for massless chiral matter in four-dimensions depending on the orbifold order $N$ \cite{Katsuki:1989kd}. In a class of scenarios with $\E{6}$ remnant of such a compactification, the usual gauge interactions of the SM belong to the $\E{6}$ gauge group, while the remaining group factors can be regarded as candidates for the family symmetry i.e.\ candidates for a new ``horizontal'' gauge symmetry that acts in the space of SM fermion generations and is present below the $\E{8}$ energy scale.

In one of the possible realisations considered earlier in Refs.~\cite{Camargo-Molina:2016yqm,Camargo-Molina:2017kxd},
 \begin{equation}
     \E{8} \supset \E{6} \times \SU{3}{F} \supset \[\SU{3}{}\]^3\times \SU{3}{F} \to \dots \,.
     \label{E8-red-typeI}
 \end{equation}
the family symmetry $\SU{3}{F}$ has been treated as a global symmetry, for simplicity, while the gauge couplings unification has been imposed already at the level of trinification $\[\SU{3}{}\]^3$.
The particle content has been inspired by an embedding of the chiral superfields into a single vector-like $\E{8}$ representation, while no particular extra-dimensional mechanism leading to such
a content has been discussed. The model offers a number of emergent distinct features such as the absence of $\mu$-problem, accidental baryon symmetry, tree-level Cabbibo mixing in the quark sector etc. On the other hand, the universality of the Yukawa interactions in the superpotential imposed by the $\[\SU{3}{}\]^3\times \SU{3}{F}$ symmetry and the gauge couplings unification at the $\[\SU{3}{}\]^3$ breaking scale, both require a significant fine-tuning in the scalar sector in order to push the soft SUSY breaking scale to much larger values than the electroweak scale and to enhance the one-loop radiative correction to the third-generation quark Yukawa couplings and, hence, to split the second- and third-generation quark masses for consistency with experimental data.
\begin{table}[htb!]
	\begin{center}
		\begin{tabular}{c|c|c}
			\toprule                     
		      sector       &   I   & II  \\    
			\midrule
		      untwisted & $\left(\bm{27},\bm{2} \right)^{(+)} + \left(\bm{1},\bm{2} \right)^{(-)} $  & $\left(\bm{27},\bm{1} \right)^{(+)} + \left(\bm{\overline{27}},\bm{1} \right)^{(-)} $ \\    
			\midrule
		      twisted     & $16\left(\bm{27},\bm{1} \right)^{(-)} + 32\left(\bm{1},\bm{2} \right)^{(-)} $ & $10\left(\bm{27},\bm{1} \right)^{(-)} + 6 \left(\bm{\overline{27}},\bm{1} \right)^{(-)}$  \\
		                      & $+ 80\left(\bm{1},\bm{1} \right)^{(-)} $  & $ + 32\left(\bm{1},\bm{2} \right)^{(-)} + 16\left(\bm{1},\bm{1} \right)^{(-)} $ \\    
			\bottomrule
		\end{tabular} 
		\caption{Fundamental representations of the $\E{6} \times \SU{2}{F} \times \U{F}$ SUSY EFT 
		inspired by the $\mathbb{Z}_4$ orbifold compactification scenario \cite{Katsuki:1989kd}.		
	   For simplicity, antichiral states are not listed and notation of $\U{F}$ charges is suppressed.
		To comply with phenomenological motivation, an extra $\mathbb{Z}_2$ symmetry is imposed forbiding a mixing between $\mathbb{Z}_2$-even 
		(``visible'') and $\mathbb{Z}_2$-odd (``hidden'') sectors. We assume that $\{\dots\}^{(+)}$ remain at low energies and contain the SM sectors,
		while all $\{\dots\}^{(-)}$ acquire large masses and are integrated out below $\E{8}$ energy scale.}
		\label{tab:Z4}  
	\end{center}
\end{table}

In this work, we discuss another promising scenario that generalised the one in Eq.~\ref{E8-red-typeI}
\begin{equation}
     \E{8} \supset \E{6} \times \SU{2}{F} \times \U{F} \to \[\SU{3}{}\]^3 \times \SU{2}{F} \times \U{F} \to \dots \,,
     \label{E8-red-typeII} 
\end{equation}
with the fully gauged family $\SU{2}{F} \times \U{F}$ symmetry, and the corresponding scale hierarchies
 \begin{equation}
 M_8 \gtrsim M_6 > M_3 \,,
     \label{E8-red-typeII-scales}
 \end{equation}
where $M_8$, $M_6$ and $M_{3}$ are the $\E{8}$, $\E{6}$ and $\[\SU{3}{}\]^3$ breaking scales, respectively. Here, we do not impose the $\[\SU{3}{}\]^3$ gauge couplings unification readily
at $M_3$ scale as was done in Refs.~\cite{Camargo-Molina:2016yqm,Camargo-Molina:2017kxd}, instead considering their natural unification at $M_6$. While we do not
consider any particular extra-dimensional $\E{8}$ reduction scenario, we fully rely on the conventional effective field theory (EFT) approach to $\E{6} \times \SU{2}{F} \times \U{F}$ theory 
in four dimensions and introduce the minimal anomaly-free chiral superfield content inspired by the $\mathbb{Z}_4$ orbifold compactification scheme \cite{Katsuki:1989kd}
following the setup proposed in Ref.~\cite{Morais:2020odg}, see Table~\ref{tab:Z4}. The gauge and Witten anomalies are cancelled in the considered scheme. Specifically, below the GUT scale $M_8$, the resulting four-dimensional SUSY EFT would contain
only two massless fundamental $(\bm{27},\bm{2})_{(1)}$ and $ (\bm{27},\bm{1})_{(-2)}$ superfields (``visible'' sector, denoted as $\{\dots\}^{(+)}$), 
while all the other representations that play a critical role in 
anomalies' cancellation are assumed to acquire mass through yet unknown mechanism (``hidden'' sector, denoted as $\{\dots\}^{(-)}$). The imposed exact $\mathbb{Z}_2$ parity 
forbids any mixing between the ``visible'' ($\mathbb{Z}_2$-even) and ``hidden'' ($\mathbb{Z}_2$-odd) sectors. We disregard the latter at lower 
energy scales in this work (so that the parity signs $\{\dots\}^{(+)}$ can then be suppressed in the considered EFT). The massless ``visible'' superfields 
from the untwisted sector are then considered to emerge from a single vector-like $\E{8}$ representation,
\begin{eqnarray}
\bm{248}^{(+)} = \left(\bm{1}, \bm{8} \right)^{(+)} \oplus \left( \bm{78},\bm{1} \right)^{(+)} \oplus \left( \bm{27},\bm{3} \right)^{(+)} \oplus \left(\bm{\overline{27}} \,,
\bm{\overline{3}} \right)^{(+)} \,,
\label{248}
\end{eqnarray}
more specifically, from $\left( \bm{27},\bm{3} \right)^{(+)}$ part of it, while the antichiral representation $\left(\bm{\overline{27}} \,,\bm{\overline{3}} \right)^{(+)} $
is assumed to be projected away by the orbifold compactification procedure. Such a unification of fermion generations implies that the SUSY flavor problem 
\cite{Altmannshofer:2009ne,Gabbiani:1996hi} can be resolved even in the case of large neutrino mixings, while realistic Yukawa matrices can be straightforwardly 
obtained by the spontaneous breaking of the horizontal family symmetry \cite{Berezhiani:1985in,Maekawa:2002eh,Maekawa:2004qj}.
Besides the massless fundamental superfields $(\bm{27},\bm{2})_{(1)}$ and $ (\bm{27},\bm{1})_{(-2)}$, the same $248$ representation contains 
also a massive $\E{6}$-adjoint $\(\bm{78},\bm{1}\)_{(0)}$ superfield that is kept in the spectrum until the trinification symmetry gets broken
by \vevs in its components (for details of the symmetry breaking chain, see below).

Below the $M_8$ scale, quadratic and cubic terms of heavy superfields from large $\E{6}$
representations such as bi-fundamental $\(\bm{650},\bm{1}\)_{(0)}$ etc are generated
in the superpotential. The latter fields may develop \vevs effectively triggering further 
breaking of the $\E{6}$ symmetry down to the trinification group \cite{Morais:2020odg}.
In the geometrical approach, \vevs in large $\E{6}$ representaions may mimic the effect 
of Wilson line \vevs in extra dimensions. Eventually, an analogical process also induces 
a further breaking down to a SUSY LR-symmetric theory. 
In particular, \vevs in heavy modes generate effective $\mu$-terms for adjoint 
$\(\bm{78},\bm{1}\)_{(0)}$ (and hence for $\Delta_{\rm L,R,C}$) superfield 
(called $\mu_{78}$ in Ref.~\cite{Camargo-Molina:2017kxd}), setting up $M_3$ scale. 
The mechanism of generation of large massive representations such as 
$\(\bm{650},\bm{1}\)_{(0)}$ or $\(\bm{78},\bm{1}\)_{(0)}$ from $\E{8}$ is beyond 
the scope of this article and left for future work. However, we assume that their 
size is given by the GUT scale. This way, the orbifolding mechanism may, 
in principle, be responsible for dynamical generation of all the scales in the 
high-scale SUSY theory given in \cref{E8-red-typeII-scales}, with a mild hierarchy 
between those.

Following a close analogy with the previous work \cite{Camargo-Molina:2017kxd}, in this scenario every 
$\SU{3}{}$ gets broken by a rank- and SUSY-preserving \vev in the corresponding adjoint superfield.
All $\SU{3}{L,R,C}$-adjoint superfields $\Delta_{\rm L,R,C}$ that emerge from the $\(\bm{78},\bm{1}\)_{(0)}$-rep 
upon $\E{6}$ breaking will gain a mass of the order of trinification $\[\SU{3}{}\]^3$ (T-GUT) breaking scale $M_3$, 
and thus do not play any role below that scale.

We recall that the unification condition of the $\SU{3}{C}$, $\SU{3}{L}$ and $\SU{3}{R}$ gauge couplings at $M_3$ scale ($g_\mathrm{C}$, $g_\mathrm{L}$ and $g_\mathrm{R}$ respectively), typically discussed in trinification-based scenarios \cite{original}, emerges due to an imposed cyclic $\mathbb{Z}_3$-permutation symmetry acting on the trinification gauge fields. However, as was demonstrated in Ref.~\cite{Camargo-Molina:2017kxd}, such a restriction comes with a price, namely, a too large soft-SUSY breaking scale, approximately $10^{11}~\mathrm{GeV}$, is unavoidable. This makes it rather challenging to generate a consistent Higgs sector at the EW scale without a significant fine-tuning. Alternatively, noting that the trinification gauge group $\SU{3}{C} \times \SU{3}{L} \times \SU{3}{R}$ is a maximal symmetry of $\E{6}$, the corresponding gauge couplings can instead become universal at (and beyond) the $\E{6}$ breaking scale, $M_6$. In this work, we thoroughly explore this new possibility without the simplifying assumption of a $\mathbb{Z}_3$ symmetry but incorporating the effect of high-dimensional operators that introduce a splitting between the $g_\mathrm{C}$, $g_\mathrm{L}$ and $g_\mathrm{R}$ gauge couplings at $M_6$.

\subsection{$\E{6}$ breaking effects}

\subsubsection{Gauge coupling unification}
\label{sec:unification}

In this article, we consider that both the trinification and flavour symmetries are remnants of a fundamental $\E{8}$ unifying force emerging via the following symmetry breaking chain
\begin{equation}
\begin{aligned}
\E{8} \overset{M_{8}}{\longrightarrow}~\E{6} \times \SU{2}{F} \times \U{F} \overset{M_{6}}{\longrightarrow}~ \[\SU{3}{}\]^3 \times \SU{2}{F} \times \U{F}.
\label{eq:brkE6}
\end{aligned}
\end{equation}
If $M_{8}$ and $M_{6}$ scales are not too far off, then the effects coming from $\E{8}$ breaking via orbifold compactifications can play a relevant role and should be taken into account. The dominant dimension-5 corrections to the gauge-kinetic terms $-\tfrac{1}{4 C}\, {\rm Tr}\(\bm{F}^{\mu \nu}\cdot \bm{F}_{\mu \nu}\)$ 
are of the form \cite{Chakrabortty:2008zk} 
\begin{equation}
	\mathcal{L}_{5D} = -\frac{\xi}{M_{\mathrm{3F}}} \[\frac{1}{4C}\, {\rm Tr}\(\bm{F}_{\mu \nu}\cdot \tilde{\Phi}_{\E{6}} \cdot \bm{F}^{\mu \nu}\)\] 
	\label{eq:5D}
\end{equation}
where $\bm{F}_{\mu \nu}$ is the $\E{6}$ field strength tensor, $C$ is a charge normalization factor, $\xi$ is a coupling constant and $\tilde{\Phi}_{\E{6}}$ is a linear combination of Higgs multiplets transforming under the symmetric product of two $\E{6}$ adjoint representations
\begin{equation}
	\tilde{\Phi}_{\E{6}} \in \( \bm{78} \otimes \bm{78} \)_\mathrm{sym} = \bm{1} \oplus \bm{650} \oplus \bm{2430}\,.
	\label{eq:PhiE6}
\end{equation}
Here we refer to the relevant formalism developed in Ref.~\cite{Chakrabortty:2008zk} 
for more details on the dimension-5 corrections such as those in \cref{eq:5D}.

For our purposes, we need two $\bm{650}$ multiplets as the minimal content required for generation of sufficient hierarchies in the SM fermion spectra already at tree level. The emergence of these two representations from $\E{8}$ is described in \cref{app:E6-operator}. A generic breaking of $\E{6}$ down to trinification can follow a linear combination of the following orthogonal directions
\begin{equation}
   \bm{\sigma} \equiv \bm{1}\,, \qquad \bm{\Sigma} \equiv \bm{650}\,, \qquad \bm{\Sigma^\prime} \equiv \bm{650^\prime}\,, \qquad \bm{\Psi} \equiv \bm{2430}\,.
\end{equation}
While a \vev in $\E{6}$-singlet $\bm{\sigma}$ superfield would not break $\E{6}$ alone by itself, it mixes with $\[\SU{3}{}\]^3$-singlets contained in the other representations, and hence affects the breaking in a generic case, so it must be taken into consideration. The corresponding generic \vev setting obeys the relation
\begin{equation}
    v^2_\mathrm{\E{6}} = v_\sigma^2 + v^2_\Sigma + v^2_{\Sigma^\prime} + v^2_\Psi \equiv \(k_\sigma^2 + k_{\Sigma}^2 + k_{\Sigma^\prime}^2 + k_{\Psi}^2\) v^2_{\E{6}}
    \label{eq:vevE6}
\end{equation}
with
\begin{equation}
	k_\sigma^2 + k_{\Sigma}^2 + k_{\Sigma^\prime}^2 + k_{\Psi}^2 = 1\,.
	\label{eq:sumk}
\end{equation}
 The modified gauge coupling unification conditions after the breaking in
 \cref{eq:brkE6} induced by the dimension-5 operators \eqref{eq:5D} 
 read as \cite{Chakrabortty:2008zk}
 \begin{equation}
 \alpha_{\mathrm{C}}^{-1} \(1+ \zeta \delta_\mathrm{C}\)^{-1} = \alpha_{\mathrm{L}}^{-1} \(1+ \zeta \delta_\mathrm{L}\)^{-1} = \alpha_{\mathrm{R}}^{-1} \(1+ \zeta \delta_\mathrm{R}\)^{-1}
 \label{eq:mod}
 \end{equation}
where
\begin{eqnarray}
\alpha^{-1}_i = \frac{4 \pi}{g_i^2} \,, \qquad \zeta \sim \frac{M_6}{M_8} \,,
\label{eq:zeta}
\end{eqnarray} 
and $\delta_{\mathrm{C},\mathrm{L},\mathrm{R}}$ are the group theoretical factors for each \vev given in table 4 of Ref.~\cite{Chakrabortty:2008zk}. 

Note that for a large hierarchy $M_6 \ll M_8$ the gauge coupling unification conditions, \cref{eq:mod}, reduce to the standard unification relations $\alpha_{\mathrm{C}}^{-1} \simeq \alpha_{\mathrm{L}}^{-1} \simeq \alpha_{\mathrm{R}}^{-1}$, thus, recovering an approximate $\mathbb{Z}_3$-permutation symmetry in the gauge sector of the T-GUT, previously imposed in Ref.~\cite{Camargo-Molina:2017kxd}. 
However, if $M_6 \sim M_8$ then sizeable threshold corrections on the gauge couplings emerge with a significant impact on the subsequent RG evolution. Here we will further consider that $\E{6}$ breaking towards the trinification symmetry can proceed through the generic vacuum direction obeying \cref{eq:vevE6} such that the $\delta_{\mathrm{C,L,R}}$ factors are given by the following relations:
 \begin{equation}
 \begin{aligned}
 \delta_\mathrm{C} &= -\frac{1}{\sqrt{2}}k_{\Sigma} - \frac{1}{\sqrt{26}} k_{\Psi}\,,
 \\
 \delta_\mathrm{L,R} &= \frac{1}{2\sqrt{2}}k_{\Sigma} \pm \frac{3}{2\sqrt{2}}k_{\Sigma^\prime} - \frac{1}{\sqrt{26}} k_{\Psi}\,.		
 \end{aligned}
 \label{eq:deltas}
 \end{equation}

Note that the singlet direction $v_\sigma$ does not participate in deviations from non-universality at one-loop level. As we will see below in \cref{Sect:scales}, the relations in \cref{eq:deltas} above modify the boundary values of the $g^{}_{\mathrm{L,R,C}}$ couplings at the $M_6$ scale in such a way that their one-loop running allows for low-scale soft-SUSY breaking interactions in overall consistency with the SM phenomenology.

\subsubsection{Origin of Yukawa interactions}

Denoting the fundamental chiral representations in the $\E{6} \times \SU{2}{F} \times \U{F}$ phase as
\begin{equation}
   \(\bm{27},\bm{2}\)_{(1)} \equiv \bm{\psi}^{\mu\,i}\,, \qquad \(\bm{27},\bm{1}\)_{(-2)} \equiv \bm{\psi}^{\mu\,3} \,,
   \label{eq:27s}
\end{equation}
where $\mu = 1\,, \ldots\,,27$ is a fundamental $\E{6}$ index, $i = 1,2$ is a $\SU{2}{F}$ doublet index and the subscripts are $\U{F}$ charges, the superpotential for the massless sector vanishes due to the anti-symmetry of family contractions, i.e.
\begin{equation}
    W_{27} = \frac{1}{2} \mathcal{Y}_{27} d_{\mu \nu \lambda} \varepsilon_{ij} \bm{\psi}^{\mu\,i} \bm{\psi}^{\nu\,j} \bm{\psi}^{\lambda\,3} = 0
    \label{eq:Super0}
\end{equation}
where $d_{\mu \nu \lambda}$ is a completely symmetric $\E{6}$ tensor, the only invariant tensor corresponding to $\bm{27}\times \bm{27}\times \bm{27}$ product, see Ref.~\cite{Kephart:1981gf,Deppisch:2016xlp}, and $\varepsilon_{ij}$ is the totally anti-symmetric $\SU{2}{}$ Levi-Civita tensor. Note that the vanishing superpotential in Eq.~\eqref{eq:Super0}, on its own, cannot generate a non-trivial Yukawa structure in the considered $\E{6} \times \SU{2}{F} \times \U{F}$ theory. This means that renormalisable $\E{6}$ interactions in this theory are not capable of generating the Yukawa sector in a form similar to $\bm{L}\cdot \bm{Q}_\mathrm{L}\cdot \bm{Q}_\mathrm{R}$ in the SHUT theory emerging after $\E{6}$ breaking, i.e.~in the trinification theory supplemented with $\SU{2}{F} \times \U{F}$, see \cref{E8-red-typeII}, or in the trinification theory supplemented with $\SU{3}{F}$ introduced in Ref.~\cite{Camargo-Molina:2017kxd}.
However, such vanishing terms imply that effects from high-dimensional operators become relevant and should be considered in detail. In particular, the product of three $\bm{27}$-plets forms invariant contractions with the bi-fundamental $\bm{650}$-plets $\bm{\Sigma}^\mu_\nu$ and ${\bm{\Sigma^\prime}}^\mu_\nu$ generated below the $\E{8}$ breaking scale $M_8$ as follows
\begin{equation} \label{eq:super4D}
    \begin{aligned}
   W_{4D} &= \frac{1}{2} \frac{1}{M_8} \varepsilon_{ij} \bm{\psi}^{\mu\,i} \bm{\psi}^{\nu\,j} \bm{\psi}^{\lambda\,3} \[ \tilde{\lambda}_1 \bm{\Sigma}^\alpha_\mu d_{\alpha \nu \lambda} + 
   \tilde{\lambda}_2 \bm{\Sigma}^\alpha_\nu d_{\alpha \mu \lambda} + \tilde{\lambda}_3 \bm{\Sigma}^\alpha_\lambda d_{\alpha \mu \nu} +
   \right.
   \\
   &+ \left.
   \tilde{\lambda}_4 {\bm{\Sigma^\prime}}^\alpha_\mu d_{\alpha \nu \lambda} + 
   \tilde{\lambda}_5 {\bm{\Sigma^\prime}}^\alpha_\nu d_{\alpha \mu \lambda} + \tilde{\lambda}_6 {\bm{\Sigma^\prime}}^\alpha_\lambda d_{\alpha \mu \nu}  \]
    \end{aligned}
\end{equation}
where the $\tilde{\lambda}_{1,2,4,5}$ terms are no longer completely symmetric under $\E{6}$ contractions, thus no longer vanishing. Once the $\bm{650}$-plets develop the \vevs (see Ref.~\cite{Chakrabortty:2008zk} for more details),
\begin{equation}
    \begin{aligned}
    \mean{\bm{\Sigma}} &= \frac{k_\Sigma v_{\E{6}}}{\sqrt{18}} \mathrm{diag}\(\underset{\textrm{9 entries}}{-2,...,-2},\,\underset{\textrm{9 entries}}{1,\ldots,1},\,\underset{\textrm{9 entries}}{1,\ldots,1}\) \\  
    \mean{\bm{\Sigma^\prime}} &= \frac{k_{\Sigma^\prime} v_{\E{6}}}{\sqrt{6}} \mathrm{diag}\(\underset{\textrm{9 entries}}{0,...,0},\,\underset{\textrm{9 entries}}{1,\ldots,1},\,\underset{\textrm{9 entries}}{-1,\ldots,-1}\)    
    \end{aligned}
\end{equation}
breaking $\E{6}$ to its trinification subgroup, an effective superpotential
  	\begin{equation}
  	\begin{aligned}
  	W_{3} =&\Scale[0.96]{\varepsilon_{ij} \(
  	\mathcal{Y}_{1} \bm{L}^i \cdot \bm{Q}^3_\mathrm{L} \cdot \bm{Q}^j_\mathrm{R}
  	- \mathcal{Y}_{2} \bm{L}^i \cdot \bm{Q}^j_\mathrm{L} \cdot \bm{Q}^3_\mathrm{R} 
  	+ \mathcal{Y}_{2}\bm{L}^3 \cdot \bm{Q}^i_\mathrm{L}\cdot \bm{Q}^j_\mathrm{R} \)}
  	\end{aligned}\label{super2}
  	\end{equation}
is generated\footnote{We have used \texttt{E6-Tensors} package \cite{Deppisch:2016xlp} to verify the form of the superpotential \cref{super2}.} reproducing a new version of the SHUT model 
with local family symmetry $\SU{2}{F} \times \U{F}$ and with Yukawa couplings
\begin{equation}
    \begin{aligned}
    \mathcal{Y}_1 &= \zeta \frac{k_{\Sigma^\prime}}{\sqrt{6}} \( \tilde{\lambda}_4 - \tilde{\lambda}_5 \)\,, \\
    \mathcal{Y}_2 &= \zeta \frac{k_\Sigma}{2 \sqrt{2}} \( \tilde{\lambda}_2 - \tilde{\lambda}_1 \) - \frac{\sqrt{3} k_\Sigma}{2 k_{\Sigma^\prime}} \mathcal{Y}_1\,,
    \end{aligned}
\end{equation}
where $k_{\Sigma},$ $k_{\Sigma^\prime}$ and $\zeta$ are defined in 
Eqs.~(\ref{eq:vevE6}) and (\ref{eq:zeta}), respectively. Note that the superpotential in Eq.~\eqref{super2} contains an accidental Abelian $\U{W} \times \U{B}$ symmetry whose charges can be chosen as in \cref{tab:AccSym}.  	
\begin{table}[htb!]
	\begin{center}
		\begin{tabular}{ccc}
			\toprule                     
			& $\U{W}$ & $\U{B}$  	\\    
			\midrule
			$\bm{L}$     			    							& $+1$		& $0$			\\
			$\bm{Q}_\mathrm{L}$  						& $-1/2$	& $+1/3$		\\
			$\bm{Q}_\mathrm{R}$ 	 					& $-1/2$	& $-1/3$ 		\\
			\bottomrule
		\end{tabular} 
		\caption{Charge assignment of the light bi-triplets in the trinification theory under the accidental symmetries of the superpotential \cref{super2}. 
		The family index is implicit.}
		\label{tab:AccSym}  
	\end{center}
\end{table} 

It is instructive to notice that the SUSY theory exhibits a new accidental $\mathbb{Z}_2$ parity which can be equivalently associated with either $\U{W}$ or $\U{B}$ symmetries of the superpotential
\begin{equation}
    \mathbb{P}_{\rm B} = \(-1\)^{2 W + 2 S} = \(-1\)^{3 B + 2 S} \,,
\end{equation}
where $S$ is the spin, while $W$ and $B$ are the $\U{W}$ and $\U{B}$ charges, respectively, given in \cref{tab:AccSym}. The corresponding $\mathbb{P}_{\rm B}$-parity of the underlying fields is provided in \cref{tab:PB}.
\begin{table}[htb!]
	\begin{center}
		\begin{tabular}{ccccccccc}
			\toprule                     
			& $L$ & $\widetilde{L}$  & $Q_\ro{L}$ & $\widetilde{Q}_\ro{L}$ & $Q_\ro{R}$ & $\widetilde{Q}_\ro{R}$ & $V_\mu$ & $g$	\\
			\midrule
		$\mathbb{P}_\ro{B}$ & $-$ & $+$  & $+$ & $-$ & $+$ & $-$ & $+$ & $-$	\\
			\bottomrule
		\end{tabular} 
		\caption{$\mathbb{P}_\ro{B}$ parity charges of the SHUT fields. Scalar fields are denoted with tildes, $V_\mu$ corresponds to vector bosons while $g$ are the gaugino fermions.}
		\label{tab:PB}  
	\end{center}
\end{table} 

In analogy to conventional R-parity, we may denote $\mathbb{P}_\ro{B}$ as $B$-parity and its relevance will become evident below. In particular, Higgs bosons, which are embedded in $\widetilde{L}$, are even while squarks are odd under $B$-parity. This is quite relevant since triple-Higgs and Higgs-fermion Yukawa interactions are allowed whereas triple-squark or quark-quark-squark terms are forbidden. This means that the only fundamental interactions that could destabilise the proton in the considered SHUT framework would come from $B$-parity violating $\E{6}$ gauge interactions at the $M_6$ scale.

Note that the origins of (non-universal) gauge and Yukawa interactions in the SHUT model are interconnected and emerge due to the $\E{6}$-breaking effects by means of the high-dimensional operators. We also see from Eqs.~\eqref{eq:super4D} and \eqref{super2} that the $M_8$ and $M_6$ scales cannot be too far off. If so, the SM quark and lepton masses would be strongly suppressed by a small ratio $v_{\E{6}}/M_8$ which, in turn, would make it challenging to reproduce the observed fermion spectrum. Interestingly, as we will notice in \cref{Sect:scales}, the measured values of the gauge couplings at the EW-scale imply that the $M_6$ and $M_8$ scales are indeed almost degenerate making both the Yukawa and gauge sectors self-consistent without any artificial tuning.

The massless superfields resulting from the $\(\bm{27},\bm{2}\)_{(1)}$ and $\(\bm{27},\bm{1}\)_{(-2)}$ supermultiplets form bi-triplets of the trinification group and transform according to the quantum numbers specified in \cref{tab:ChiralSuper}, where we cast the components of the lepton and quark superfields as 	
	\begin{equation}
	\begin{aligned}
	&	 \qquad \qquad \quad\Scale[0.95]{\LLR{i,3}{l}{r} =\begin{pmatrix}
  		\bm{\mathcal{N}}_{\mathrm{R}} & \bm{\mathcal{E}}_{\mathrm{L}} & \bm{e}_{\mathrm{L}}\\
  		\bm{\mathcal{E}}_{\mathrm{R}} & \bm{\mathcal{N}}_{\mathrm{L}} & \bm{\nu}_{\mathrm{L}}\\
  		\bm{e}_{\mathrm{R}} & \bm{\nu}_{\mathrm{R}} & \bm{\phi}
  		\end{pmatrix}^{i,3}\,,}
  		\\
  		&\Scale[0.90]{\QL{i,3}{x}{l}=\begin{pmatrix}\bm{u}_{\mathrm{L}}^x & \bm{d}_{\mathrm{L}}^x & \bm{D}_{\mathrm{L}}^x
  		\end{pmatrix}^{i,3}},
  		\quad  \Scale[0.90]{\QR{i,3}{r}{x}=\begin{pmatrix}\bm{u}_{\mathrm{R}x} & \bm{d}_{\mathrm{R}x} & \bm{D}_{\mathrm{R}x}
  		\end{pmatrix}^{\top\;\;i,3}}\,,
	\end{aligned}
  	\label{eq:L-tri-triplet}
  	\end{equation}
with $l$, $r$ and $x$ denoting $\SU{3}{L}$, $\SU{3}{R}$ and $\SU{3}{C}$ triplet indices, respectively. Note that the $\ro{L}$ and $\ro{R}$ subscripts do not denote left and right chiralities and the fermionic components of the superfields are defined as left-handed Weyl spinors.
\begin{table}[htb!]
	\begin{center}
		\begin{tabular}{cccccc}
			\toprule                     
			& $\SU{3}{L}$ & $\SU{3}{R}$  & $\SU{3}{C}$  & $\SU{2}{F}$ & $\U{F}$ \\    
			\midrule
			$\bm{L}^{i}$ & $\bm{3}$ & $\bm{\overline{3}}$ & $\bm{1}$  						& $\bm{2}$	& $1$	\\
			$\bm{L}^{3}$ & $\bm{3}$ & $\bm{\overline{3}}$ & $\bm{1}$  						& $\bm{1}$	& $-2$	\\
			$\bm{Q}^{i}_\mathrm{L}$ & $\bm{\overline{3}}$ & $\bm{1}$ & $\bm{3}$  						& $\bm{2}$	& $1$
			\\
			$\bm{Q}^{3}_\mathrm{L}$ & $\bm{\overline{3}}$ & $\bm{1}$ & $\bm{3}$  						& $\bm{1}$	& $-2$\\
			$\bm{Q}^{i}_\mathrm{R}$ & $\bm{1}$ & $\bm{3}$ & $\bm{\overline{3}}$  						& $\bm{2}$	& $1$
			\\
			$\bm{Q}^{3}_\mathrm{R}$ & $\bm{1}$ & $\bm{3}$ & $\bm{\overline{3}}$  						& $\bm{1}$	& $-2$\\
			\bottomrule
		\end{tabular} 
		\caption{Fundamental chiral superfields in the SHUT model.}
		\label{tab:ChiralSuper}  
	\end{center}
\end{table}
As was thoroughly investigated in an earlier work \cite{Camargo-Molina:2017kxd}, the breaking of the trinification symmetry takes place once the scalar components of the heavy 
adjoint octet superfields $\bm{\Delta}_\mathrm{L}$ and $\bm{\Delta}_\mathrm{R}$ acquire \vevs, i.e. $v_\mathrm{L} = v_\mathrm{R} \equiv M_3$, respectively,
\begin{equation}
\SU{3}{L} \times \SU{3}{R} \overset{v_\mathrm{L,R}}{\to} \SU{2}{L} \times \SU{2}{R} \times \U{L} \times \U{R} \,.
\label{3222111}
\end{equation}
Such superfields are embedded in a heavy $\E{6}$ adjoint $\bm{78}$-plet which also contains two trinification tri-triplets that we denote as $\bm{\Xi}$ and $\bm{\Xi}^\prime$. Since both the octets and the tri-triplets have a common origin in $\E{6}$ they can share a universal mass and hence be kept in the trinification spectrum. Furthermore, since they are not gauge singlets, their effect to the one-loop running of the gauge couplings must be considered. The quantum numbers of the $\(\bm{78}, \bm{1}_{(0)}\)$ components are shown in \cref{tab:ChiralSuperAdj} and
\begin{table}[htb!]
	\begin{center}
		\begin{tabular}{cccccc}
			\toprule                     
			 & $\SU{3}{L}$ & $\SU{3}{R}$  & $\SU{3}{C}$  & $\SU{2}{F}$ & $\U{F}$ \\    
			\midrule
			$\bm{\Delta}_\mathrm{L}$ & $\bm{8}$ & $\bm{1}$ & $\bm{1}$ & $\bm{1}$	& $0$	\\
			$\bm{\Delta}_\mathrm{R}$ & $\bm{1}$ & $\bm{8}$ & $\bm{1}$ & $\bm{1}$	& $0$ 
			\\
			$\bm{\Delta}_\mathrm{C}$ & $\bm{1}$ & $\bm{1}$ & $\bm{8}$ & $\bm{1}$	& $0$ 
			\\
			$\bm{\Xi}$ & $\bm{3}$ & $\bm{\overline{3}}$ & $\bm{3}$ & $\bm{1}$	& $0$ 
			\\
			$\bm{\Xi^\prime}$ & $\bm{\overline{3}}$ & $\bm{3}$ & $\bm{\overline{3}}$ & $\bm{1}$	& $0$ 
			\\
			\bottomrule
		\end{tabular} 
		\caption{Components of the adjoint chiral superfield $\(\bm{78},\bm{1}_{(0)}\)$.}
		\label{tab:ChiralSuperAdj}  
	\end{center}
\end{table}
the part of the superpotential containing massive trinification representations reads
\begin{equation}
    \begin{aligned}
    W_{78} = \sum_{A = \mathrm{L,R,C}} \[ \frac{1}{2} \mu_{78} \mathrm{Tr} \bm{\Delta}^2_A + \frac{1}{3!} \mathcal{Y}_{78} \mathrm{Tr} \bm{\Delta}^3_A \] 
    + \mu_{78} \mathrm{Tr} \(\bm{\Xi} \bm{\Xi}^\prime\) + \sum_{A = \mathrm{L,R,C}} \mathcal{Y}_{78} \mathrm{Tr}\(\bm{\Xi} \bm{\Xi}^\prime \bm{\Delta}_A\)\,.
    \end{aligned}
\end{equation}
  	\tikzstyle{blockU} = [rectangle, draw, fill=red!8, 
  	text width=3em, text centered, rounded corners, minimum height=3em, drop shadow]
  	\tikzstyle{block0} = [rectangle, draw, fill=red!8, 
  	text width=10em, text centered, rounded corners, minimum height=3em, drop shadow]
  	\tikzstyle{block1} = [rectangle, draw, fill=red!8, 
  	text width=16em, text centered, rounded corners, minimum height=3em, drop shadow]
  	\tikzstyle{block2} = [rectangle, draw, fill=red!8, 
  	text width=24em, text centered, rounded corners, minimum height=3em, drop shadow]
  	\tikzstyle{block3} = [rectangle, draw, fill=green!8, 
  	text width=24em, text centered, rounded corners, minimum height=3em, drop shadow]
  	\tikzstyle{block4} = [rectangle, draw, fill=green!8, 
  	text width=20em, text centered, rounded corners, minimum height=3em, drop shadow]
  	\tikzstyle{block5} = [rectangle, draw, fill=green!8, 
  	text width=16em, text centered, rounded corners, minimum height=3em, drop shadow]
	\tikzstyle{block6} = [rectangle, draw, fill=blue!10, 
  	text width=16em, text centered, rounded corners, minimum height=3em, drop shadow]
	\tikzstyle{blockLOW} = [rectangle, draw, fill=blue!10, 
  	text width=8em, text centered, rounded corners, minimum height=3em, drop shadow]
  	\tikzstyle{line} = [draw, -latex']
  	\tikzstyle{cloud} = [draw, ellipse,fill=red!20, node distance=3cm,
  	minimum height=2em]
  	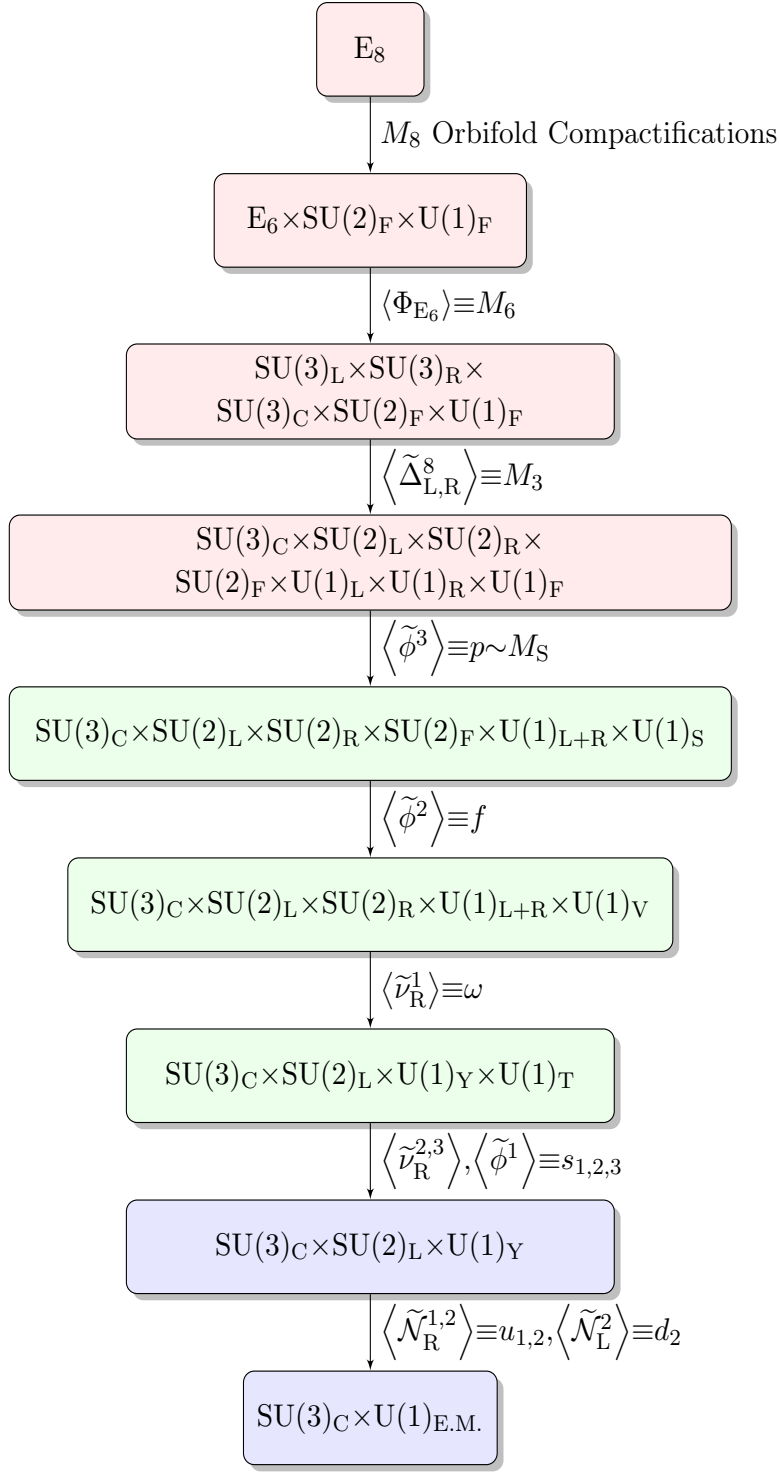
\begin{figure}[ht!]
  		\centering
  		\resizebox{10.5cm}{19.4cm}{
  			\begin{tikzpicture}[node distance = 2.0cm, auto]
  			\node [blockU] (E8) {$\E{8}$};
  			\node [block0, below of=E8, node distance=2.1cm] (E6) {$\E{6} \times \SU{2}{F} \times \U{F}$};
  			\node [block1, below of=E6, node distance=2.1cm](identify) {$\mathrm{SU}(3)_{\mathrm{L}} \times \mathrm{SU}(3)_{\mathrm{R}} \times \mathrm{SU}(3)_{\mathrm{C}} \times  
  				\SU{2}{F}\times \U{F} $};
  			\node [block2, below of=identify, node distance=2.1cm] (stop) {$\mathrm{SU}(3)_{\mathrm{C}} \times \mathrm{SU}(2)_{\mathrm{L}} \times 
  				\mathrm{SU}(2)_{\mathrm{R}}\times
  				\mathrm{SU}(2)_{\mathrm{F}} \times 
  				\mathrm{U}(1)_{\mathrm{L}} \times \mathrm{U}(1)_{\mathrm{R}} \times \mathrm{U}(1)_{\mathrm{F}} $};
  			\node [block3, below of=stop, node distance=2.1cm] (stop2) {$\mathrm{SU}(3)_{\mathrm{C}} \times \mathrm{SU}(2)_{\mathrm{L}} \times 
  				\mathrm{SU}(2)_{\mathrm{R}}\times
  				\mathrm{SU}(2)_{\mathrm{F}} \times 
  				\mathrm{U}(1)_{\mathrm{L+R}} \times \mathrm{U}(1)_{\mathrm{S}}$};
  			\node [block4, below of=stop2, node distance=2.1cm] (stop3) {$\mathrm{SU}(3)_{\mathrm{C}} \times \mathrm{SU}(2)_{\mathrm{L}} \times 
  				\mathrm{SU}(2)_{\mathrm{R}}\times
  				\mathrm{U}(1)_{\mathrm{L+R}} \times \mathrm{U}(1)_{\mathrm{V}}$};  			
  			\node [block5, below of=stop3, node distance=2.1cm] (stop4) {$\mathrm{SU}(3)_{\mathrm{C}} \times \mathrm{SU}(2)_{\mathrm{L}}\times 
  				\mathrm{U}(1)_{\mathrm{Y}} \times  \mathrm{U}(1)_{\mathrm{T}}$};
  			\node [block6, below of=stop4, node distance=2.1cm] (stop5) {$\mathrm{SU}(3)_{\mathrm{C}} \times \mathrm{SU}(2)_{\mathrm{L}}\times 
  				\mathrm{U}(1)_{\mathrm{Y}}$};
  			\node [blockLOW, below of=stop5, node distance=2.1cm] (final) {$\mathrm{SU}(3)_{\mathrm{C}} \times 
  				\mathrm{U}(1)_{\ro{E.M.}}$};	
  			\path [line] (E8) -- node {$M_8$~Orbifold Compactifications}(E6);
  			\path [line] (E6) -- node {$\Braket{\Phi_{\E{6}}} \equiv M_6$}(identify);  	
  			\path [line] (identify) -- node {$\Braket{\widetilde{\Delta}_{\mathrm{L,R}}^8} \equiv M_3$}(stop);
  			\path [line] (stop) -- node {$\Braket{\widetilde{\phi}^3} \equiv p \sim  M_{\rm S}$}(stop2);
  			\path [line] (stop2) -- node {$\Braket{\widetilde{\phi}^2} \equiv f$}(stop3);  			
  			\path [line] (stop3) -- node {$\Braket{\widetilde{\nu}_{\rm R}^1} \equiv \omega$}(stop4);
  			\path [line] (stop4) -- node {$\Braket{\widetilde{\nu}_{\rm R}^{2,3}},\Braket{\widetilde{\phi}^{1}} \equiv s_{1,2,3}$}(stop5);
  			\path [line] (stop5) -- node {$\Braket{\widetilde{\mathcal{N}}_{\rm R}^{1,2}} \equiv u_{1,2},\,\Braket{\widetilde{\mathcal{N}}_{\rm L}^{2}} \equiv d_2$}(final);
  			\end{tikzpicture}
  		}
  		\caption{\label{fig:1112abc} The gauge symmetry breaking scheme considered in this work. In the red blocks SUSY is approximately unbroken while in the green ones it is softly broken. 
  		The blue blocks represent the SM gauge symmetry and below with only the lightest states included. The $\tilde{\nu}_\mathrm{R}^{2,3}$ and $\tilde{\phi}^2$ scalars are allowed 
  		to mix forming two physical scalars and a Goldstone boson.}
  	\end{figure} 

After trinification symmetry breaking, \cref{3222111}, we are left with the left-right symmetric theory whose tree-level superpotential can be written as
\begin{equation}
\begin{aligned}
\hspace*{-2mm}W =& \mathcal{Y}_{1} \varepsilon_{ij}\[ \bm{\chi}^i \cdot \bm{q}^3_\mathrm{L}\cdot \bm{q}^{j}_\mathrm{R} 
+ \bm{\ell}^{i}_\mathrm{R} \cdot \bm{D}^3_\mathrm{L} \cdot \bm{q}^{j}_\mathrm{R} 
+ \bm{\ell}^i_\mathrm{L} \cdot \bm{q}^3_\mathrm{L} \cdot \bm{D}^{j}_\mathrm{R} 
+ \bm{\phi}^i \cdot \bm{D}^3_\mathrm{L} \cdot \bm{D}^{j}_\mathrm{R} \] \\ 
-& \mathcal{Y}_{2} \varepsilon_{ij}\[ \bm{\chi}^i \cdot \bm{q}^j_\mathrm{L}\cdot \bm{q}^{3}_\mathrm{R} 
+ \bm{\ell}^{i}_\mathrm{R} \cdot \bm{D}^j_\mathrm{L} \cdot \bm{q}^{3}_\mathrm{R} 
+ \bm{\ell}^i_\mathrm{L} \cdot \bm{q}^j_\mathrm{L} \cdot \bm{D}^{3}_\mathrm{R} 
+ \bm{\phi}^i \cdot \bm{D}^j_\mathrm{L} \cdot \bm{D}^{3}_\mathrm{R} \] \\
+& \mathcal{Y}_{2} \varepsilon_{ij}\[ \bm{\chi}^3 \cdot \bm{q}^i_\mathrm{L}\cdot \bm{q}^{j}_\mathrm{R} 
+ \bm{\ell}^{3}_\mathrm{R} \cdot \bm{D}^i_\mathrm{L} \cdot \bm{q}^{j}_\mathrm{R} 
+ \bm{\ell}^3_\mathrm{L} \cdot \bm{q}^i_\mathrm{L} \cdot \bm{D}^{j}_\mathrm{R} 
+ \bm{\phi}^3 \cdot \bm{D}^i_\mathrm{L} \cdot \bm{D}^{j}_\mathrm{R} \] \,.
\end{aligned}\label{super3}
\end{equation}
where we recast the chiral superfields in \cref{eq:L-tri-triplet} as
\begin{eqnarray}
\begin{aligned}
\Scale[0.95]{\LLR{i}{l}{r} =\begin{pmatrix}
	\bm{\chi}^{\bar{l}}{}_{\bar{r}} & \bm{\ell}^{\bar{l}}_{\mathrm{L}}\\
	\bm{\ell}_{\mathrm{R}\bar{r}} & \bm{\phi}
	\end{pmatrix}^{i}\,,} 
\quad
\Scale[0.95]{\QL{i}{x}{l}=\begin{pmatrix}\bm{q}_{\mathrm{L}\bar{l}}^x & \bm{D}_{\mathrm{L}}^x
	\end{pmatrix}^{i}}, \label{eq:tri-triplets}
\quad
\Scale[0.95]{\QR{i}{r}{x}=\begin{pmatrix}\bm{q}_{\mathrm{R}x}^{\bar{r}} & \bm{D}_{\mathrm{R}x}
	\end{pmatrix}^{\top\;\;i}}\,,
\end{aligned}
\label{eq:tri-triplet2}
\end{eqnarray}  	 	
and where $\bar{l}$ and $\bar{r}$ are the $\SU{2}{L}$ and $\SU{2}{R}$ doublet indices, respectively.

\section{Soft-SUSY breaking interactions}
\label{Sect:soft-sector}

The choice of the $\E{8}$ symmetry breaking pattern down to a LR-symmetric SUSY theory, with three distinct but relatively compressed breaking scales, $M_8$, $M_6$ and $M_3$, introduced above, leaves the $\bm{27}$-plet components $\bm{L}$, $\bm{Q}_\mathrm{L}$ and $\bm{Q}_\mathrm{R}$ massless. The latter, hence, contain the light SM matter sectors naturally decoupled from the trinification breaking scale $M_3$. Indeed, the subsequent breaking steps towards the SM gauge group should be induced by a new energy scale which originates from another sector, in particular, the sector of soft-SUSY breaking interactions.

The existence of the soft SUSY breaking sector triggers the breaking of the remaining gauge symmetries down to the SM gauge group. The most generic \vev setting that leaves the SM gauge symmetry unbroken at low energies reads
  	\begin{equation} 
  	\label{eq:LVEVs}
  	\begin{aligned}
   \Scale[0.95]{\mean{\tilde{L}^1} =\begin{pmatrix}
  		0 & 0 & 0\\
  		0 & 0 & 0\\
  		0 & \omega & s_1
  		\end{pmatrix}\,,} \quad
  	\Scale[0.95]{\mean{\tilde{L}^2} =\begin{pmatrix}
  		0 & 0 & 0\\
  		0 & 0 & 0\\
  		0 & s_2 & f
  		\end{pmatrix}\,,} \quad
  	 \Scale[0.95]{\mean{\tilde{L}^3} =\begin{pmatrix}
  	 	0 & 0 & 0\\
  	 	0 & 0 & 0\\
  	 	0 & s_3 & p
  	 	\end{pmatrix}}
  	  \quad \, , 
  	\end{aligned} 
  	\end{equation}
where we adopt the following hierarchy
\begin{equation}
	M_\mathrm{EW} ~< ~ s_{1,2,3} ~ \leq ~\omega ~ \leq ~ f ~ \leq ~ p ~ \ll ~ M_3 \leq M_6 \lesssim M_8\,,
	\label{hierarchies}
\end{equation}
with the lowest EW symmetry breaking (EWSB) scale, $M_\mathrm{EW}$. The corresponding full symmetry breaking chain down to the SM gauge group is represented in \cref{fig:1112abc}. In the red blocks, the role of the soft SUSY breaking parameters on the heavy spectrum is negligible while in the green ones it becomes relevant for the remaining light states. The blue blocks further indicate the SM gauge symmetry and below with the lightest states kept in the spectrum. We have considered the minimal realistic realisation of the Higgs sector to trigger the breaking of the EW symmetry in the last step. For further details see \cref{sec:CKM}.

The allowed soft-SUSY trilinear interactions preserving $\U{W}$ and $\U{B}$ read
\begingroup
\small
\begin{equation}
\begin{aligned}
\mathcal{L}_\mathrm{WB}^\mathrm{soft} &= -\varepsilon_{ij} \left(
a_{10} \tilde{q}_{{\rm L}\,l}^{i}\tilde{q}_{{\rm R}}^{j\,r} \tilde{\chi}^{3\,l}_{r}
+
a_{11} \tilde{q}_{{\rm L}\,l}^{i}\tilde{q}_{\rm R}^{3\,r} \tilde{\chi}^{j\,l}_{r}
+
a_{12} \tilde{q}^3_{{\rm L}\,l}\tilde{q}_{{\rm R}}^{i\,r} \tilde{\chi}^{l\,j}_{r}
+
a_{13}\tilde{D}_{\rm L}^{i} \tilde{D}_{\rm R}^{j} \tilde{\phi}^3\right)
\\
&
-\varepsilon_{ij} \left(
a_{14}\tilde{D}_{\rm L}^{i} \tilde{D}^3_{\rm R}  \tilde{\phi}^{j}
+
a_{15}\tilde{D}^3_{\rm L} \tilde{D}_{\rm R}^{i} \tilde{\phi}^{j}
+
a_{16}\tilde{q}_{{\rm L}\,l}^{i} \tilde{D}_{\rm R}^{j}\tilde{\ell}_{\rm L}^{3\,l}
+
a_{17}\tilde{q}_{{\rm L}\,l}^{i} \tilde{D}^3_{\rm R}\tilde{\ell}_{\rm L}^{j\,l}
\right)
\\
& 
-\varepsilon_{ij}\left(
a_{18}\tilde{q}^3_{{\rm L}\,l}\tilde{D}_{\rm R}^{i}\tilde{\ell}_{\rm L}^{j\,l}
+
a_{19} \tilde{D}_{\rm L}^{i}\tilde{q}_{{\rm R}}^{j\,r}\tilde{\ell}^3_{{\rm R}\,r} 
+
a_{20} \tilde{D}_{\rm L}^{i}\tilde{q}_{{\rm R}}^{3\,r}\tilde{\ell}_{{\rm R}\,r}^{j} 
+
a_{21} \tilde{D}^3_{\rm L} \tilde{q}_{{\rm R}}^{j\,r}\tilde{\ell}_{{\rm R}\,r}^{i} 
\right) + \mathrm{c.c.}\,,
\label{softbrkg2}
\end{aligned}
\end{equation} 
\endgroup
whereas the mass terms are of the form
\begin{equation}
	m^2_{\varphi} \varphi^\ast \varphi
	\label{softmass}
\end{equation}
where $\varphi$ represents any of the scalars contained in the superfields~\eqref{eq:tri-triplets} with the appropriate group contractions left implicit.

The $\U{W}$-violating soft interactions are given by the following trilinear terms allowed by the gauge symmetry and $B$-parity of the SUSY LR-symmetric theory
\begin{equation}
	\begin{aligned}
	\mathcal{L}_\mathrm{\slashed{W}}^\mathrm{soft} =& -\varepsilon_{ij}\varepsilon^{rr^\prime}\varepsilon_{ll^\prime}
	\left(a_1 \tilde{\chi}^{il}{}_{r}\tilde{\ell}_{\rm L}^{j\,l^\prime}\tilde{\ell}_{{\rm R}\,r^\prime}^3
	+
	a_2 \tilde{\chi}^{i\,l}_{r}\tilde{\ell}_{\rm L}^{3\,l^\prime} \tilde{\ell}_{{\rm R}\,r^\prime}^{j}
	\right. \\
	&+\left.
	a_3 \tilde{\chi}^{3\,l}{}_{r}\tilde{\ell}_{\rm L}^{i\,l^\prime}\tilde{\ell}_{{\rm R}\,r^\prime}^j 
	+
	a_4 \tilde{\chi}^{i\,l}_{r}\tilde{\chi}^{3\,l^\prime}_{r^\prime} \tilde{\phi}^{j} 
	+
	a_5 \tilde{\chi}^{i\,l}_{r}\tilde{\chi}^{j\,l^\prime}_{r^\prime} \tilde{\phi}^{3} 
	\right)+ \mathrm{c.c.}
	\label{softbrkgW}
	\end{aligned}
\end{equation}
Note that soft trilinear $\U{B}$-violating interactions are not allowed by 
the $B$-parity in the considered theory.

With the superpotential \eqref{super3} and the soft-SUSY breaking interactions \eqref{softbrkg2}, \eqref{softmass} and \eqref{softbrkgW} we have all relevant
ingredients necessary to consistently generate a SM-like low-energy EFT through the breaking chain shown in \cref{fig:1112abc}.

Indeed, in the considered LR symmetric SUSY theory the main ballpark of free parameters comes from the soft-SUSY breaking sector, in particular, 
17 trilinear terms (5 with sleptons and 12 with squarks), 16 soft mass terms of $\tilde{L}\tilde{L}$- and $\tilde{Q}\tilde{Q}$-type as well as 
2 high-scale gaugino mass parameters corresponding to those in the $E_6$ and gauge-family sectors. On top of that, there are also four gauge 
couplings in the gauge sector of the SUSY theory while all the low-scale Yukawa couplings are matched to only two universal high-scale Yukawa 
terms in the superpotential that govern the strongest hierarchies between the SM quarks already at tree level (see also below). Note, as will be clear 
from the forthcoming sections the radiative corrections to the Yukawa couplings are determined by the soft-SUSY breaking parameters and gauge couplings,
whose number is sufficient to accommodate the measured SM fermion masses and mixing angles.

\section{Implications for the fermion sector}
\label{Sect:fermion-sector}

\subsection{Quark masses and CKM mixing}
\label{sec:CKM}

In what follows, at the first stage we would like to discuss the properties of the SM quark spectrum neglecting the effect of vector-like quarks (VLQs) $D_{\mathrm L,R}$.
With the $p$, $f$, $\omega$ and $s_{1,2,3}$ \vev setting, one generates the gauge group of the SM at low energy scales according to the breaking scheme 
schematically illustrated in \cref{fig:1112abc}. In this scheme, the subsequent EW symmetry can only be broken by $\SU{2}{L}$ doublet \vevs, 
in the spirit of  N-Higgs doublet models. 
Thus, the most generic \vev setting that one can have in the SHUT model consistent with 
the considered symmetry breaking scheme reads
\begin{equation} 
\begin{aligned}
\mean{\tilde{L}^1} ~=~ \frac{1}{\sqrt{2}}\begin{pmatrix}
	{u_1} & 0 & 0\\
	0 & {d_1} & {e_1}\\
	0 & \omega & s_1
	\end{pmatrix}\,, \quad
\mean{\tilde{L}^2} ~=~ \frac{1}{\sqrt{2}}\begin{pmatrix}
	{u_2} & 0 & 0\\
	0 & {d_2} & {e_2}\\
	0 & s_2 & f
	\end{pmatrix}\,, 
\quad
\mean{\tilde{L}^3} ~=~ \frac{1}{\sqrt{2}}\begin{pmatrix}
	{u_3} & 0 & 0\\
	0 & {d_3} & {e_3}\\
	0 & s_3 & p
	\end{pmatrix}
\quad \, , 
\end{aligned} 
\end{equation}
where $u_i$, $d_i$ and $e_i$ denote up-type, down-type and sneutrino-type EWSB \vevs, respectively. It is instructive to consider only those minimal \vevs settings 
that roughly reproduce the viable quark mass and mixing parameters in the SM already at tree level. Note, if one considers both $d_i$ and $e_i$ \vevs, they contribute to a non-trivial mixing between the down-type $D_\mathrm{R}$ and $d_\mathrm{R}$ quarks. In what follows and unless noted otherwise, we would like to align our EW-breaking \vevs in such a way that $e_i = 0$ corresponding to a small mixing between $D_\mathrm{R}$ and $d_\mathrm{R}$ quarks suppressed by a strong hierarchy between the EW scale and the higher intermediate scales associated with $\omega,f,p$ \vevs.

The quark mass sector in the SHUT model reveals a number of interesting features. The up-quark mass matrix takes the following form
\begin{equation}
\label{eq:Mu_generic}
M_\mathrm{u} ~=~ 
\frac{1}{\sqrt{2}}\begin{pmatrix}
0 & u_3 \mathcal{Y}_2 & u_2 \mathcal{Y}_2\\
-u_3 \mathcal{Y}_2 & 0 & -u_1 \mathcal{Y}_2\\
-u_2 \mathcal{Y}_1 & u_1 \mathcal{Y}_1 & 0
\end{pmatrix}\,,
\end{equation}
yielding the generic mass spectrum with one massless quark, the would-be $u$-quark in the SM,
	\begin{equation}\label{eq:mu}
		m_\ro{u} = 0 \qquad m_\ro{c}^2 = \tfrac{1}{2} \mathcal{Y}_2^2 \(u_1^2 + u_2^2 + u_3^2\)  \qquad m_\ro{t}^2 = \tfrac{1}{2} \[\mathcal{Y}_1^2 \(u_1^2 + u_2^2\) + \mathcal{Y}_2^2 u_3^2 \]\,.
	\end{equation} 	
Here we notice that the proper charm-top mass hierarchy is realised if and only if $\mathcal{Y}_2 \ll \mathcal{Y}_1$. This condition will further be employed in the analysis of the down-type quark spectrum and mixing.

In fact, as it is explicit in the field decomposition \eqref{eq:L-tri-triplet} there are six down-type quarks, three $\SU{2}{L}$-doublet (chiral) components $d_\mathrm{L,R}^{1,2,3}$ and three $\SU{2}{L}$-singlet (vector-like) fields $D_\mathrm{L,R}^{1,2,3}$ which acquire large masses above the EW scale. The generic down-type quark mass form thus takes the following structure
\begin{equation}
	 M^{\rm 6 \times 6}_\mathrm{d} =\frac{1}{\sqrt{2}}\begin{pmatrix}
	 0 &  d_3 \mathcal{Y}_2 & d_2 \mathcal{Y}_2 & 0 & 0 & 0\\
	  -d_3 \mathcal{Y}_2 & 0 &  - d_1 \mathcal{Y}_2 & 0 & 0 & 0\\
	 -d_2 \mathcal{Y}_1 &  d_1 \mathcal{Y}_2 & 0 & 0 & 0 & 0\\
	 0 & s_3 \mathcal{Y}_2 & s_2 \mathcal{Y}_2 & 0 & p \mathcal{Y}_2 & f \mathcal{Y}_2\\
	 -s_3 \mathcal{Y}_2 & 0 & -\omega \mathcal{Y}_2 & -p \mathcal{Y}_2 & 0 & - s_1 \mathcal{Y}_2\\
	 -s_2 \mathcal{Y}_1 & w \mathcal{Y}_1 & 0 & -f \mathcal{Y}_1 & s_1 \mathcal{Y}_1 & 0\\
	 \end{pmatrix} \,,
	 \label{eq:Md_generic}
\end{equation}
written in the basis $\(d_\mathrm{L}^i~~D_\mathrm{L}^i\)^\top M_d~ \(d_\mathrm{R}^i~~D_\mathrm{R}^i\)$ with $i=1,2,3$. 

The VLQs acquire their masses as soon as the $p$, $f$ and $\omega$ \vevs are generated corresponding 
to the fifth, sixth and seventh boxes in \cref{fig:1112abc}. Before the EWSB (Higgs doublet) and $s_i$
\vevs are developed the total down-type quark mass matrix reads
\begin{equation} \label{eq:Md-heavy}
	 M^{\rm 6 \times 6}_\mathrm{d} \simeq \frac{1}{\sqrt{2}}\begin{pmatrix}
	 0 & 0 & 0 & 0 & 0 & 0\\
	 0 & 0 & 0 & 0 & 0 & 0\\
	 0 & 0 & 0 & 0 & 0 & 0\\
	 0 & 0 & 0 & 0 & p \mathcal{Y}_2 & f \mathcal{Y}_2\\
	 0 & 0 & -\omega \mathcal{Y}_2 & -p \mathcal{Y}_2 & 0 & 0\\
	 0 & w \mathcal{Y}_1 & 0 & -f \mathcal{Y}_1 & 0 & 0\\
	 \end{pmatrix} \,,
\end{equation}
yielding the following VLQ mass spectrum where we kept the first-order terms 
in $\mathcal{Y}_2 \ll \mathcal{Y}_1$ as needed for a realistic $u$-quark mass spectrum,
\begin{eqnarray} 
&& m_\ro{D/S}^2\simeq \frac{1}{2} (f^2+p^2)\mathcal{Y}_2^2 \,, \quad
m_\ro{S/D}^2\simeq \frac{\omega^2(f^2+p^2+\omega^2)}{2(f^2+\omega^2)}\mathcal{Y}_2^2 \,, \\
&& m_\ro{B}^2\simeq \frac{1}{2} (f^2+\omega^2)\mathcal{Y}_1^2 + \frac{f^2p^2}{2(f^2+\omega^2)}\mathcal{Y}_2^2 \,.
\end{eqnarray}
Here, we adopt that the lightest VLQ is $D$-quark, such that $m_\ro{D} < m_\ro{S}$, so which of the first two states is $D$-quark and which is $S$-quark depends on relative magnitudes of $f$, $p$ and $\omega$ (see below). 

As can clearly be seen from \cref{eq:Md-heavy}, the massless (before the EWSB) states will consist of $d_L$ and an admixture of $d_R$ with $D_R$ states. After diagonalising this mass form we can use the resulting matrix to bring \cref{eq:Md_generic} in a block-diagonal structure where the three light states can be properly identified. This way we obtain the mass matrix of the light down-type quark states in the following approximate form
\begin{equation} \label{eq:Md3x3}
M_\mathrm{d} ~\approx~ 
\frac{1}{\sqrt{2}}\begin{pmatrix}
0 & 0 & \mathcal{Y}_2  \frac{d_3 f - d_2 p}{\sqrt{f^2+p^2+\omega^2}}\\
-d_3 \mathcal{Y}_2 & 0 & d_1 \mathcal{Y}_2 \frac{p}{\sqrt{f^2+p^2+\omega^2}}\\
-d_2 \mathcal{Y}_1 & 0 & d_1 \mathcal{Y}_1 \frac{f}{\sqrt{f^2+p^2+\omega^2}}
\end{pmatrix}\,,
\end{equation}
It is obvious that there is one massless state, the would-be SM down-quark $d$, in analogy to the zeroth up-quark mass found above in \cref{eq:mu}.  While the mass spectrum and mixing can be, in principle, calculated analytically for 
the most generic case with six nonzero Higgs doublet \vevs $u_i$ and $d_i$, 
the resulting formulas are rather lengthy and not very enlightening. Instead,
we have analysed three distinct scenarios with five nonzero Higgs doublet \vevs 
by setting one of the down-type \vevs $d_i$ to zero. We have analysed the down-type 
mass spectra and CKM in each of such scenarios and found that only one of them 
(with $d_1=0$) provides the physical CKM matrix and spectrum compatible with 
those in the SM. Other two scenarios corresponding to $d_2=0$ or $d_3=0$ render
unphysical CKM mixing, and hence are no longer discussed here.

Thus, setting $d_1=0$ in \cref{eq:Md3x3} one arrives at the following physical down-type quark spectrum
	\begin{equation}\label{eq:md}
		m_\ro{d} = 0 \,, \qquad   
		m_\ro{s}^2 = \frac{(d_3 f - d_2 p)^2}{2(f^2+p^2+\omega^2)} \mathcal{Y}_2^2 \,, \qquad
		m_\ro{b}^2 = \frac{1}{2} (d_2^2\mathcal{Y}_1^2 + d_3^2\mathcal{Y}_2^2) \,,
	\end{equation}
which is exact i.e.~no hierarchies between the \vevs and Yukawa couplings are imposed at this step. Note, the SM-like down-type quark masses \cref{eq:md} represent the leading contributions as emerge from the full $6 \times 6$ down-type mass matrix in \cref{eq:Md_generic}. Remarkably, even for the maximal number of possible Higgs \vevs 
the first generation $u$ and $d$ quarks appear as massless states at tree level. Therefore, the origin of their mass is purely radiative, in consistency with their observed decoupling in the quark mass spectrum. As will be shown numerically below, $s_i$ produce only a minor effect on the tree-level down-type masses and mixing, thus, justifying the approximate procedure employed here.

It is clear from \cref{eq:md} that in the realistic \vev hierarchy $p,f,\omega \gg d_{2,3}$, one recovers a strong hierarchy $m_s\ll m_b$ in consistency with the charm-top mass hierarchy in the up-quark sector. In fact, taking $u_3 = d_3 = 0$ for simplicity we observe that the ratio of both Yukawa couplings reads
\begin{equation}\label{eq:ratio}
	\frac{\mathcal{Y}_1}{\mathcal{Y}_2} ~=~ \frac{m_\ro{t}}{m_\ro{c}} ~\approx~ \frac{m_\ro{b}}{m_\ro{s}} ~\sim~ \mathcal{O}\(100\) \,,
\end{equation}
Indeed, the second and third quark generations acquiring their masses already at tree-level such that their hierarchy is controlled by the only two Yukawa couplings in the SHUT superpotential, $\mathcal{Y}_1$ and $\mathcal{Y}_2$. This demonstrates that leading order terms in our model can potentially explain the quark masses and their hierarchies without significant fine tuning of the underlying model parameters.

Let us now consider the realistic quark mixing starting from the light-quark mass forms \cref{eq:Mu_generic} and \cref{eq:Md3x3} by setting $d_1=0$. The corresponding left quark mixing matrices $L_\ro{u}$ and $L_\ro{d}$ defined as $m_\ro{u,d}^2 = L_\ro{u,d}^\dagger (M_\ro{u,d}M_\ro{u,d}^\dagger) L_\ro{u,d}$ provide the CKM mixing matrix in analytic form 
\begin{equation}
\begin{aligned}
V_{\rm CKM} \equiv L_\ro{u} L_\ro{d}^\dagger =
\begin{pmatrix}
	 \frac{d_2 u_2 \mathcal{Y}_1^2 + d_3 u_3 \mathcal{Y}_2^2}{\sqrt{\mathcal{A}\mathcal{B}}} & -\frac{u_1 \mathcal{Y}_1 }{\sqrt{\mathcal{A}} } & 
	 \frac{(d_2 u_3 - d_3 u_2)\mathcal{Y}_1 \mathcal{Y}_2 }{\sqrt{\mathcal{A}\mathcal{B}}} \\
	 -\frac{d_2 u_1 \mathcal{Y}_1}{\sqrt{\mathcal{B}\mathcal{C}}} & -\frac{u_2}{\sqrt{\mathcal{C}}} & 
	 \frac{d_3 u_1 \mathcal{Y}_2}{\sqrt{\mathcal{B}\mathcal{C}}}\\
	 \frac{(\mathcal{C} d_3 - d_2 u_2 u_3)\mathcal{Y}_1 \mathcal{Y}_2 }{\sqrt{\mathcal{A}\mathcal{B}\mathcal{C}}} & 
	 \frac{u_1 u_3 \mathcal{Y}_2}{\sqrt{\mathcal{A}\mathcal{C}}} & 
	 \frac{\mathcal{C} d_2 \mathcal{Y}_1^2 + d_3 u_2 u_3 \mathcal{Y}_2^2}{\sqrt{\mathcal{A}\mathcal{B}\mathcal{C}}}
\end{pmatrix}\label{eq:exact-CKM}
\end{aligned}
\end{equation}
where the ordering of rows and columns is consistent with the ordering of the mass states in Eqs.~(\ref{eq:mu}) and (\ref{eq:md}), and
\begin{eqnarray}
\mathcal{A} = \mathcal{C} \mathcal{Y}_1^2 + u_3^2 \mathcal{Y}_2^2 \,, \qquad 
\mathcal{B} = d_2^2 \mathcal{Y}_1^2 + d_3^2 \mathcal{Y}_2^2 \,, \qquad
\mathcal{C} = u_1^2 + u_2^2 \,.
\end{eqnarray}
We have explicitly imposed the positivity of all the \vevs and Yukawa couplings, $u_i>0$, $d_{2,3}>0$, $\mathcal{Y}_{1,2}>0$, for simplicity. Note, the CKM matrix in \cref{eq:exact-CKM} is exact for the $3\times 3$ quark 
mass forms (\ref{eq:Mu_generic}) and (\ref{eq:Md3x3}) in a sense that no hierarchies between the \vevs and Yukawa couplings are imposed here. Remarkably, while the down-type quark masses in \cref{eq:md} contain an explicit dependence 
on the high-scale \vevs $p,f,\omega$, the corresponding CKM mixing does not contain any information about $p,f,\omega$ \vevs at all as long as the approximate down-type matrix \cref{eq:Md3x3} is concerned.

Accounting for the first subleading term only, the top-bottom mixing element $[V_{\rm CKM}]_{33} \equiv V_\ro{tb}$ in the limit of small $\mathcal{Y}_2 \ll \mathcal{Y}_1$ reads
\begin{eqnarray}
V_{tb} \simeq 1 - \left(\frac{\mathcal{Y}_2}{\mathcal{Y}_1}\right)^2 
\frac{d_3^2 \mathcal{C} + d_2 u_3(d_2 u_3 - 2 d_3 u_2)}{2d_2^2\mathcal{C}} \,,
\label{eq:Vtb}
\end{eqnarray}
whose deviation from unity is well under control due to a very small ratio 
$\mathcal{Y}_2/\mathcal{Y}_1\ll 1$. Apparently, the same ratio is responsible 
for a strong suppression of $V_\ro{td}$, $V_\ro{ts}$, $V_\ro{bu}$ and $V_\ro{bc}$ CKM elements.

In the limit $u_3 \to 0$ and $d_3 \to 0$, the top-bottom mixing approaches unity from below, i.e. $V_{tb} \to 1^-$. Furthermore, in this case the CKM matrix takes a particularly simple Cabibbo form
\begin{equation}\label{eq:VCKM}
	 |V_{\rm CKM}| =
	 \begin{pmatrix}
	 \cos \theta_C & \sin \theta_C & 0\\
	 \sin \theta_C & \cos \theta_C & 0\\
	 0 & 0 & 1
	 \end{pmatrix} \,,
\end{equation}
where the Cabibbo angle is directly related to the ratio of the up-type 
Higgs doublet \vevs as follows
\begin{eqnarray}
\theta_C = \arctan \left(\frac{u_1}{u_2}\right) \,.
\end{eqnarray}
Thus, while the small ratio $\mathcal{Y}_2/\mathcal{Y}_1\ll 1$ imposes a strong suppression on mixing between
the third generation with the other two already at the classical level of 5HDM, one acquires an additional suppression also in the effective 3HDM limit corresponding to very small (or zero) third-generation Higgs \vevs, $u_3$ and $d_3$. Due to the very specific structure of the CKM matrix and the masses, one cannot impose a limit of small first- and/or second-generation Higgs \vevs $u_{1,2}$ and $d_2$ without destroying the realistic quark mixing. This fact renders an interesting possibility for a unique minimal effective 3HDM scenario of the SHUT theory with dominant $u_{1,2}$ and $d_2$ \vevs only. This also gives rise to a nearly Cabibbo quark mixing, realistic hierarchies between the second- and third-generation quark masses and a new physics decoupled sector of heavy VLQs already at the classical level.
\begin{table}[htb!]
	\begin{center}
		\scalebox{.9}{ \begin{tabular}{c|ccc|ccc}
			\toprule                     
			Scenarios & $\omega~\ro{[TeV]}$ & $f~\ro{[TeV]}$ & $p~\ro{[TeV]}$ & $m_\ro{D}~\ro{[TeV]}$ & $m_\ro{S}~\ro{[TeV]}$& $m_\ro{B}~\ro{[TeV]}$\\    
			\midrule
			$\omega \sim f \sim p$ & $100 - 1000$ & $100 - 1000$ & $100 - 1000$ & $1 - 10$ & $1 - 10$& $100 - 1000$ \\
			$\omega \sim f \ll p$ & $10 - 100$ & $10 - 100$ & $100 -1000
			$& $1 - 10$ & $1 - 10$& $10 - 100$\\
			$\omega \ll f \sim p$ & $100$ & $1000$ & $1000$& $1$ & $10$& $1000$ \\		
			\bottomrule
		\end{tabular} }
		\caption{An example for an order of magnitude estimation of VLQ mass scales relevant for numerical considerations in this work.}
		\label{tab:VLQmass}
	\end{center}
\end{table} 

Thus, a realistic low-scale EFT of the SHUT model may only contain either five (with $u_i$, $d_{2,3}$), four (with $u_{1,2}$, $d_{2,3}$) or the minimum of three (with $u_{1,2}$, $d_2$) Higgs doublets yielding the realistic tree-level quark spectra and mixing, and each such scenario is unique. Any other scenario is incompatible with the SM at tree level. Recall that our calculations so far did not include sub-dominant radiative effects. As will be discussed below, such effects will be necessary for a full description of the quark sector.

\subsection{Numerical analysis}
\label{sec:numerics}

\subsubsection{VLQ hierarchies}
\label{sec:VLQnum}

The three distinct realistic examples of possible hierarchies among the $\omega$, $f$ and $p$ scales with their effects on the VLQ masses are shown in \cref{tab:VLQmass}. We have chosen for these examples that the $\omega$, $f$ and $p$ \vevs are such that the lightest VLQ mass scale is at or above $1~\ro{TeV}$. In fact, the soft scales $\omega,f,p$ cannot be too close to the EWSB scale since otherwise the lightest VLQs would become unacceptably light. In essence, the benchmark scenarios in \cref{tab:VLQmass} show that the low-scale EFT limit of the SHUT model may contain either one light VLQ generation at the $\ro{TeV}$ scale (last row) or, alternatively, two light generations (second and third row). This illustrates that a hypothetical discovery of VLQs at the LHC or at a future collider would become a smoking gun of the SHUT model and a way to indirectly probe its symmetry breaking scales above the EWSB one.

\subsubsection{Tree-level deviations from unitarity}
\label{sec:UNInum}

The Cabibbo-like CKM mixing discussed above in \cref{sec:CKM} can be considered as a good approximation in the case of vanishing third-generation Higgs \vevs, $u_3$ and $d_3$ and in the VLQs decoupling limit. Retaining the latter limit, for a particular parameter space point in the realistic 3HDM $\(u_1,u_2,d_2\)$ scenario,
\begin{equation}
\label{eq:benchTree}
    \mathcal{Y}_1 ~=~ 0.98\,,~~\mathcal{Y}_2 ~=~ 0.0068\,,~~u_1 ~=~ 59.65~\ro{GeV}\,,~~u_2 ~=~ 238.6~\ro{GeV}\,,~~d_2 ~=~ 6~\ro{GeV}\,,
\end{equation}
chosen such that $u_1^2 + u_2^2 + d_2^2 = \(246~\ro{GeV}\)^2$ and $u_1/u_2 \approx 0.25$,
one obtains the following quark mass spectrum and mixing at tree level,
\begin{equation}
    \begin{aligned}
     m_\ro{t}& ~=~ 170.4~\ro{GeV}\,,~~m_\ro{c} ~=~ 1.18~\ro{GeV}\,,~~m_\ro{b} ~=~ 4.15~\ro{GeV}\,,~~m_\ro{s} ~=~ 0.017~\ro{GeV}\,,
     \\
     &~\qquad \qquad \qquad
     |V_\ro{CKM}| ~=~ 
     \begin{pmatrix}
	 0.97 & 0.24 & 0\\
	 0.24 & 0.97 & 0\\
	 0 & 0 & 1
	 \end{pmatrix} \,,
    \end{aligned}
\end{equation}
which appear in a reasonably close vicinity of the experimentally measured values.

It is instructive to study the impact of VLQs on the light quark masses and mixing in the case of exact $6\times 6$ down-type quark mass matrix \cref{eq:Md_generic}. The generalized $3\times 6$ CKM mixing matrix is defined as
\begin{equation}
    V_\ro{CKM}~=~ L_\ro{u}\cdot P \cdot L_\ro{d}^\dagger ~=~ \(V_\ro{CKM}^\ro{SM}~|~V_\ro{CKM}^\ro{VLQs}\)~~\textrm{with}~~P ~=~ 
    \( \bm{\mathbb{1}}_{3\times3}~~\bm{0}_{3 \times 3}\) \,.
\end{equation}
It generally depends on the Yukawa couplings $\mathcal{Y}_{1,2}$ and on the symmetry breaking scales $p$, $f$, $\omega$ and $s_i$. In the full down-quark mass form in \cref{eq:Md_generic} we will now fix $s = s_i = 10~\ro{TeV}$, $i=1,2,3$, and consider the benchmark points in the 3HDM EFT $\(u_1,u_2,d_2\)$ for each of the three soft-scale \vev hierarchies summarised in \cref{tab:VLQmass}.

\subsubsection*{Fully compressed $\omega \sim f \sim p$ scenario}

In this first example, let us consider that the $p$, $f$ and $\omega$ scales are not too far off and are set to, e.g.
\begin{equation}
    p ~=~ 220~\ro{TeV}\,,~~ f ~=~ 210~\ro{TeV}\,,~~\omega ~=~ 200~\ro{TeV}\,,
\end{equation}
from where the down-type quark mass spectrum becomes
\begin{equation}
    m_\ro{s} ~=~ 0.017~\ro{GeV}\,,~~ m_\ro{b} ~=~ 4.15~\ro{GeV}\,, ~~ m_\ro{D} ~=~ 1.3~\ro{TeV}\,, ~~ m_\ro{S} ~=~ 1.5~\ro{TeV}\,, ~~ m_\ro{B} ~=~ 211.0~\ro{TeV}\,.
\end{equation}
Note that this scenario contains two light VLQs at the $\ro{TeV}$ scale and a heavy one well beyond the reach of the LHC. The total quark mixing matrix reads
\begin{equation}
    \begin{aligned}
   \abs{V_\ro{CKM}} &\simeq
    \(
\begin{array}{ccc|ccc}
0.97 & 0.24 & 2.31 \times 10^{-5} & 4.36 \times 10^{-6} & 7.29 \times 10^{-7} & \sim 0\\
0.24 & 0.97 & 9.23 \times 10^{-5} & 1.74 \times 10^{-5} & 2.92 \times 10^{-6} & \sim 0\\
0 & 9.51 \times 10^{-5} & 1 &  5.55 \times 10^{-5} & 1.15 \times 10^{-5} & 6.47 \times 10^{-7}
\end{array}\label{eq:CKM-full}
\)\,.
    \end{aligned}
\end{equation}
With this example we observe that the SM-like $3\times 3$ CKM quark mixing is no longer unitary with small deviations induced via a small tree-level mixing with VLQs. It also generates small elements in $V_\ro{CKM}^\ro{VLQs}$, with the largest entry being $V_{\rm tD} = 5.55 \times 10^{-5}$. The correct values of the light quark masses as well as the mixing between the third with the first two generations is expected to be generated at one-loop level as will be discussed below. On the other hand, such effects are sub-leading contributions to VLQ masses.

\subsubsection*{$\omega$-$f$ compressed $\omega \sim f \ll p$ scenario}

For the second example, we fix
\begin{equation}\label{eq:wf}
    p ~=~ 600~\ro{TeV}\,,~~ f ~=~ 110~\ro{TeV}\,,~~\omega ~=~ 100~\ro{TeV}\,,
\end{equation}
which results in the following quark mass spectrum 
\begin{equation}
    m_\ro{s} ~=~ 0.028~\ro{GeV}\,,~~ m_\ro{b} ~=~ 4.14~\ro{GeV}\,, ~~ m_\ro{D} ~=~ 1.99~\ro{TeV}\,, ~~ m_\ro{S} ~=~ 2.94~\ro{TeV}\,, ~~ m_\ro{B} ~=~ 103.4~\ro{TeV}\,.
\end{equation}
and the corresponding total quark mixing matrix reads
\begin{equation}
    \begin{aligned}
   \abs{V_\ro{CKM}} &\simeq
    \(
\begin{array}{ccc|ccc}
0.97 & 0.24 & 1.92 \times 10^{-5} & 8.33 \times 10^{-7} & 8.34 \times 10^{-8} & \sim 0\\
0.24 & 0.97 & 7.65 \times 10^{-5} & 3.33 \times 10^{-6} & 3.34 \times 10^{-7} & \sim 0\\
0 & 7.91 \times 10^{-5} & 1 &  1.0 \times 10^{-4} & 2.03 \times 10^{-6} & 2.69 \times 10^{-6}
\end{array}\label{eq:CKM-full-2}
\)\,.
    \end{aligned}
\end{equation}
In consistency with the estimations of \cref{tab:VLQmass}, such a scenario with compressed $\omega$ and $f$ scales yields two light VLQs and a heavy one. A larger $p$-scale also induces a further suppression in the CKM mixing elements in comparison with the fully compressed scenario discussed above. However, the $V_\ro{tD}$ element is still the largest one and is of the order $10^{-4}$.

\subsubsection*{$f$-$p$ compressed $\omega \ll f \sim p$ scenario}

For the final benchmark scenario, let us keep the $p$ value fixed and shift $f$ 
towards the $p$-scale, e.g.
\begin{equation}\label{eq:fp}
    p ~=~ 600~\ro{TeV}\,,~~ f ~=~ 590~\ro{TeV}\,,~~\omega ~=~ 180~\ro{TeV}\,.
\end{equation}
The quark mass spectrum becomes
\begin{equation}
    m_\ro{s} ~=~ 0.020~\ro{GeV}\,,~~ m_\ro{b} ~=~ 4.15~\ro{GeV}\,, ~~ m_\ro{D} ~=~ 1.2~\ro{TeV}\,, ~~ m_\ro{S} ~=~ 4.05~\ro{TeV}\,, ~~ m_\ro{B} ~=~ 427.6~\ro{TeV}\,,
\end{equation}
and the mixing matrix reads
\begin{equation}
    \begin{aligned}
   \abs{V_\ro{CKM}} &\simeq
    \(
\begin{array}{ccc|ccc}
0.97 & 0.24 & 3.34 \times 10^{-6} & 4.16 \times 10^{-6} & 2.91 \times 10^{-8} & \sim 0\\
0.24 & 0.97 & 1.34 \times 10^{-5} & 1.66 \times 10^{-5} & 1.16 \times 10^{-7} & \sim 0\\
0 & 1.38 \times 10^{-5} & 1 &  9.54 \times 10^{-6} & 2.70 \times 10^{-7} & 1.58 \times 10^{-7}
\end{array}\label{eq:CKM-full-3}
\)\,.
    \end{aligned}
\end{equation}

This case shows a larger relative suppression in $V_\ro{CKM}^\ro{VLQs}$ elements due to larger $p$ and $f$ scales. However, contrary to the previous two scenarios, here we have $\abs{V_\ro{cD}} > \abs{V_\ro{tD}}$ which follows from non-trivial details of the mixing.

\begin{figure}[!htb]
		\centering
		\includegraphics[width=.50\textwidth]{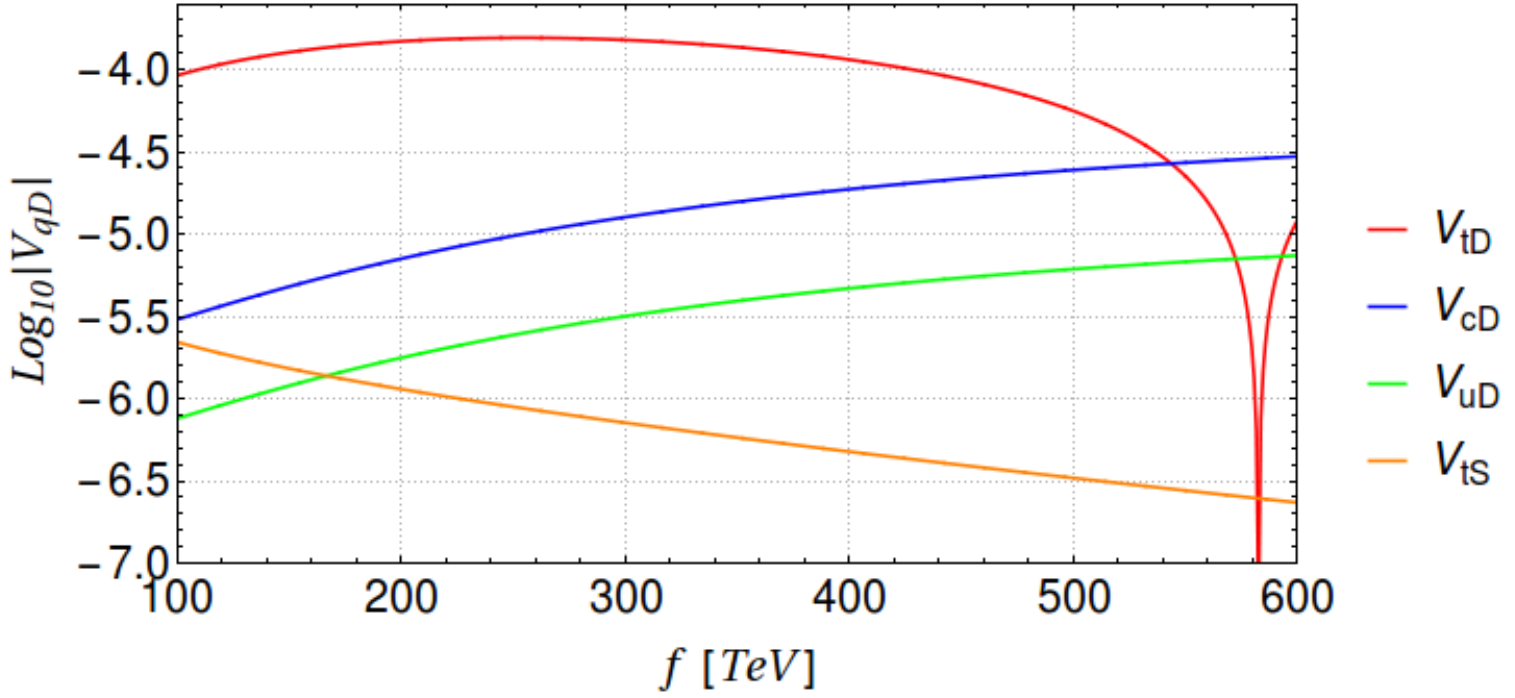}~
		\includegraphics[width=.50\textwidth]{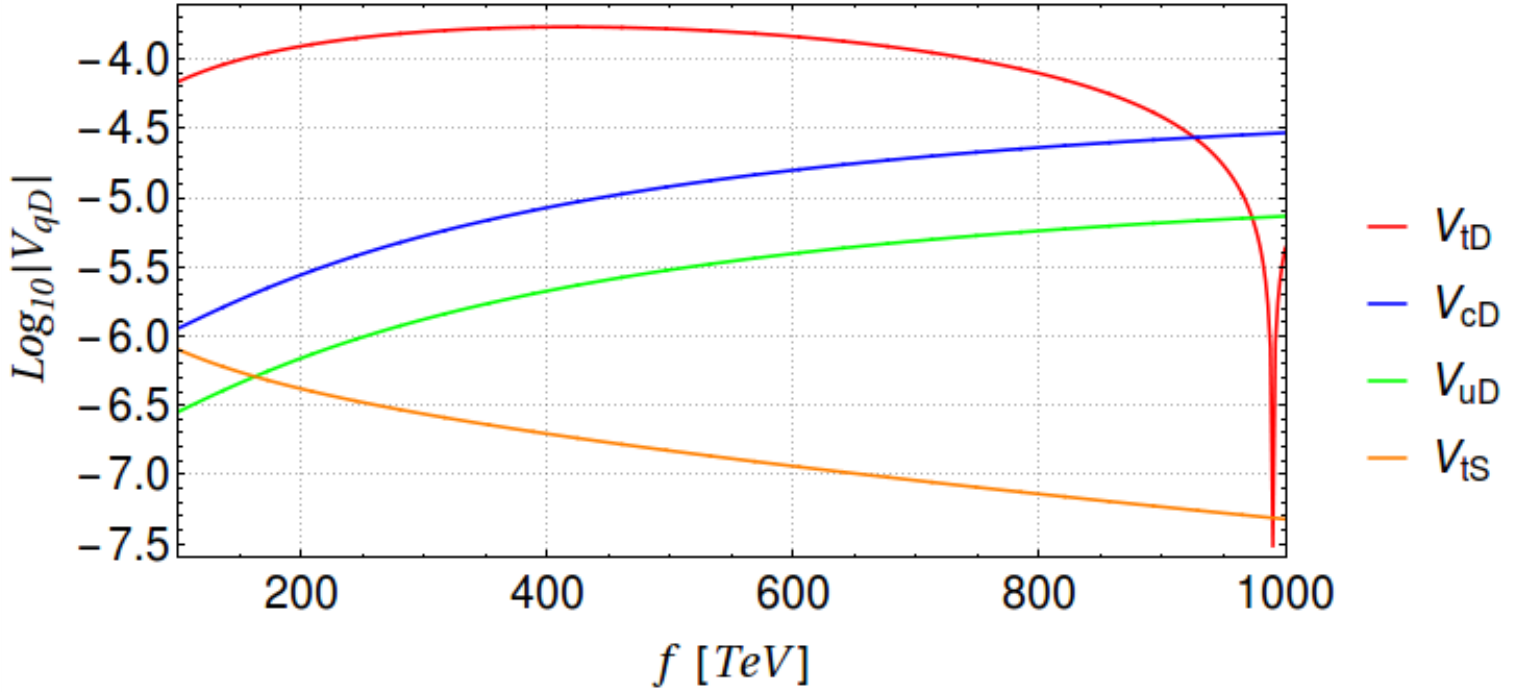}
		\caption{The four largest quark mixing matrix elements between the VLQs and up-type SM-like quarks. In these plots $s~=~10~\ro{TeV}$ and $\omega ~=~ 100~\ro{TeV}$ while on the left $p ~=~ 600~\ro{TeV}$ and on the right $p ~=~ 1000~\ro{TeV}$. The $f$-scale varies between the limiting $\omega$-$f$ and the $p$-$f$ compressed scenarios.}
		\label{fig:fp_to_wp}
	\end{figure}

\begin{figure}[!htb]
		\centering
		\includegraphics[width=.50\textwidth]{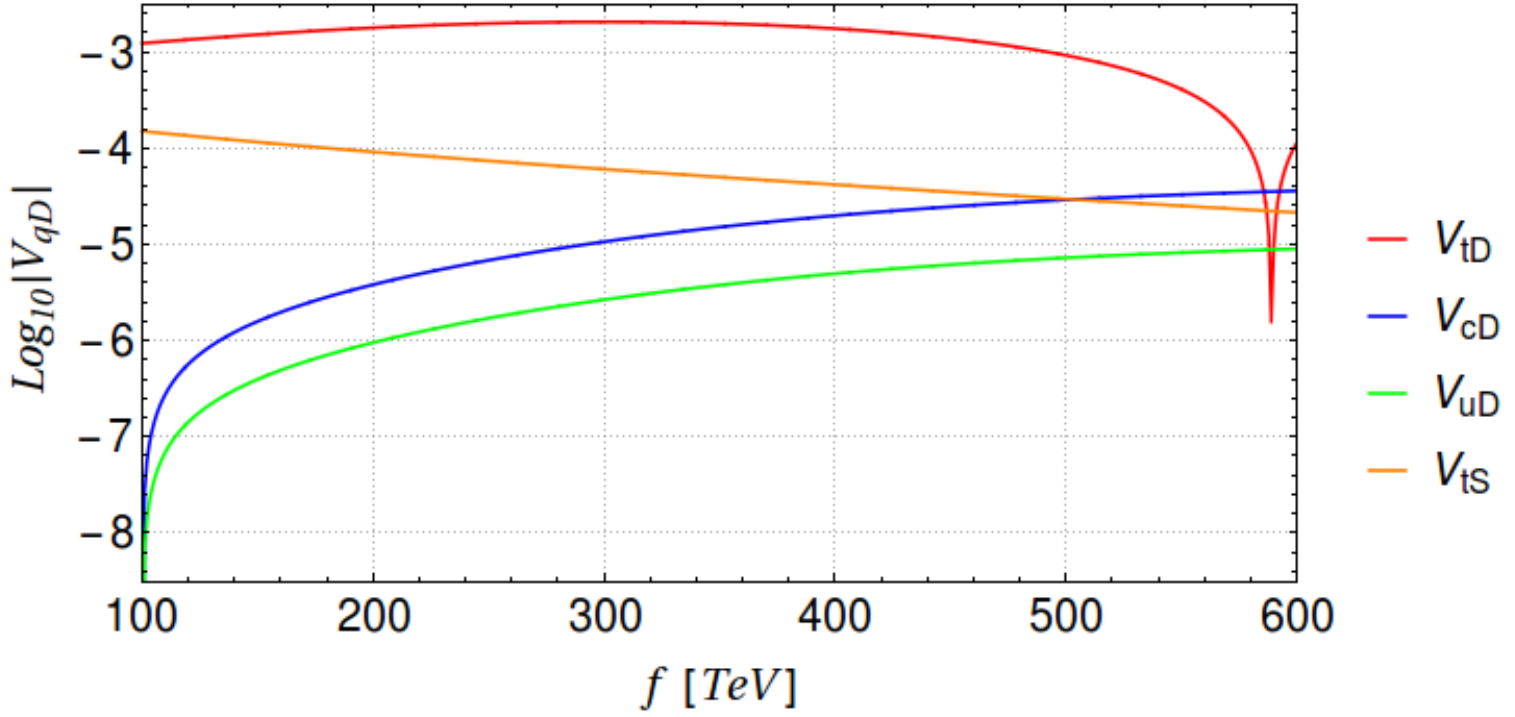}~
		\includegraphics[width=.50\textwidth]{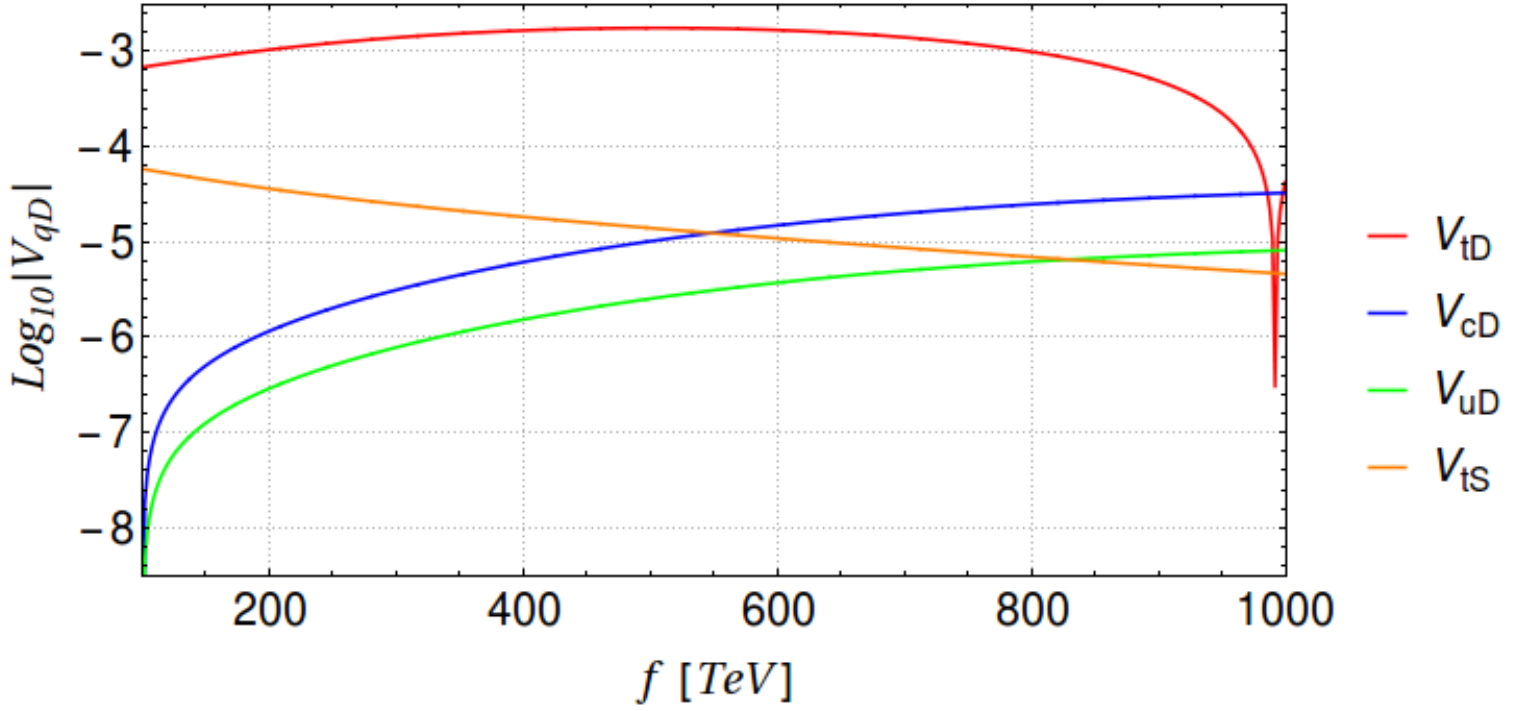}
		\caption{The four largest quark mixing elements between VLQs and up-type SM-like quarks. In these plots $s~=~\omega~=~100~\ro{TeV}$. On the left panel $p ~=~ 600~\ro{TeV}$ and on the right panel $p ~=~ 1000~\ro{TeV}$. The $f$-scale varies between the limiting $\omega$-$f$ and the $p$-$f$ compressed scenarios.}
		\label{fig:fp_to_swp}
	\end{figure}

In order to visualise the behaviour of $V_\ro{CKM}^\ro{VLQs}$ elements between different regimes we show in \cref{fig:fp_to_wp} the absolute values of $V_\ro{tD}$, $V_\ro{cD}$, $V_\ro{uD}$ and $V_\ro{tS}$ elements.
We have fixed the $\omega$ scale to $100~\ro{TeV}$ and considered two possibilities for the $p$-\vev, $600~\ro{TeV}$ (left panel) and $1000~\ro{TeV}$ (right panel). In both cases we keep $s~=~10~\ro{TeV}$ as mentioned above. By inspecting \cref{fig:fp_to_wp} we observe the following:
\begin{itemize}
    \item In the $\omega$-$f$ compressed regime, in absolute value, the $V_\ro{tD}$ element is the largest $V_\ro{CKM}^\ro{VLQs}$ element of order $\mathcal{O}\(10^{-4}\)$ followed by 
    $V_\ro{cD} \gtrsim V_\ro{tS} > V_\ro{uD}$;
    \item Approximately half way between the limiting $\omega$-$f$ and the $f$-$p$ compressed scenarios $V_\ro{tD}$ reaches a maximum of approximately $10^{3.8}$;
    \item While approaching the $p$-$f$ compressed regime, the $V_\ro{tD}$ element crosses zero leading to a spiky structure in the log-plot, while $V_\ro{cD}$ and $V_\ro{uD}$ continuously grow and $V_{tS}$ continuously decreases;
    \item In the $p$-$f$ compressed regime $V_{cD}$ becomes the largest $V_\ro{CKM}^\ro{VLQs}$ element of order $\mathcal{O}\(10^{-4.5}\)$ followed by $V_\ro{uD} ~\approx~ V_\ro{tD} ~\sim~ \mathcal{O}\(10^{-5}\)$, all these are well above $V_\ro{tS}$;
    \item A growing $p$-scale generically imposes a stronger suppression on the $V_\ro{CKM}^\ro{VLQs}$ elements as expected.
\end{itemize}
Note the $\omega$-$s$ degeneracy enhances the $V_\ro{CKM}^\ro{VLQs}$ elements as shown in \cref{fig:fp_to_swp}.

In the case of $p~=~600~\ro{TeV}$, the $\omega$-$f$ compressed regime yields $V_\ro{tD}\sim \mathcal{O}\(10^{-2.8}\)$ and $V_\ro{tS} \sim \mathcal{O}\(10^{-3.8}\)$ while the remaining elements become very small. Furthermore, the down-type quark mass spectrum in this limit reads
\begin{eqnarray} \nonumber 
    && m_\ro{s} ~=~ 0.028~\ro{GeV}\,,~~ m_\ro{b} ~=~ 2.9~\ro{GeV}\,, \\ 
    && m_\ro{D} ~=~ 1.7~\ro{TeV}\,, ~~ m_\ro{S} ~=~ 3.08~\ro{TeV}\,, ~~ m_\ro{B} ~=~ 138.6~\ro{TeV}\,,
\end{eqnarray}
with $m_\ro{s}$ and $m_\ro{b}$ being tantalisingly close to their running values at the $Z$-boson mass scale according to Ref.~\cite{Bora:2012tx}. Alternatively, the $p$-$f$ compressed scenario predicts $V_\ro{tD} > V_\ro{cD} > V_\ro{tS} > V_\ro{uD}$, with all of them being between $10^{-4}$ and $10^{-5}$. The same behaviour is seen for $p~=~ 1000~\ro{TeV}$ with a slight suppression in the mixing between the VLQs and SM-like quarks.

\subsection{Radiative effects}

\subsubsection{Light quark and lepton sectors}

\begin{figure}[t]
	\centering
	\includegraphics[width=.50\textwidth]{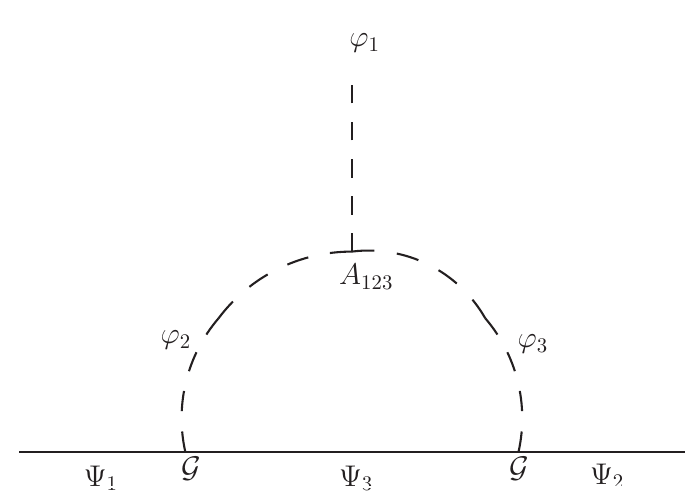}~
	\caption{One-loop topology contributing to the radiatively generated Yukawa interactions. $\mathcal{G}$ denotes a Yukawa coupling with D-term origin.}
	\label{fig:loop-YukA}
\end{figure}

The dominant one-loop contribution to the Yukawa couplings for both, leptons and quarks, in our model is given in \cref{fig:loop-YukA}. 
In the zero external momentum limit, the one-loop amplitude reads as
\begin{align}
\label{eq:loop-Yuk}
\kappa = 2  i \,\mathcal{G}^2 \, C_A A_{123} m_{\Psi_3} f(m^2_{\Psi_3},m^2_{\varphi_2},m^2_{\varphi_3})
 \,.
\end{align}
where $C_A=(N^2-1)/(2 N)$ in case of an $SU(N)$ gaugino, $\mathcal{G}$ denotes a Yukawa coupling with D-term origin and $m_{\Psi_3}$ is the gaugino mass. $m_{\varphi_2}$ and $m_{\varphi_3}$ are the scalar masses. 
The effective trilinear coupling 
$A_{123} $ gets contribution from two sources: (i) After
the breaking of a gauge symmetry as an effective
coupling $\lambda_{1236} \mean{\varphi_6}$.
Here $\lambda_{1236}$ is a quartic coupling
originating either from an $F$- or $D$-term
as discussed in detail in \cref{app:FullEffectiveLs}.
(ii) A trilinear coupling $a_{123}$ from the soft
SUSY breaking sector. Hence, the SUSY breaking and the fermion sectors are interconnected via radiative corrections to the corresponding Yukawa interactions. The radiative threshold contributions to the Yukawa couplings have to be calculated at the scale where the fermions and scalars in the loop propagators acquire their masses. Such a scale can be of the order of any of the $p$, $f$, $\omega$ or $s$ \vevs introduced above.

The loop integral in \cref{eq:loop-Yuk} is  given by
\begin{align}
f(x,y,z) &= \frac{1}{16 \pi^2} \frac{1}{x-y} 
 \left( \frac{y\, \log\left( \frac{z}{y}\right)}{y-z}
  - \frac{x\, \log\left( \frac{z}{x}\right)}{x-z} \right) \,.
\end{align}
To get a better understanding for the dependence
on the involved masses, it is useful to consider
certain limit, in particular the case where
all masses are equal or the case where there
is a sizable hierarchy between scalars and fermions
in the loop. The different limits read as
\begin{align}
f(x,x,x) & =  \frac{1}{32 \pi^2 x}  \\
 f(x,y,y) &= \frac{1}{16 \pi^2} \frac{-x+y + x \log(x/y)}{(x-y)^2} \\
f(x,y,y) & \simeq 
 \left\{
\begin{array}{ccc}
\frac{1}{16 \pi^2} \left[ \frac{1}{y} + \frac{x}{y^2} \left(1 +  \log\left(\frac{x}{y}\right)\right) \right]  & \text{for} & x \ll y \\
 \frac{1}{16 \pi^2} \left[ \frac{-1}{x}  \left(1 +  \log\left(\frac{x}{y}\right)\right)
 -  \frac{y}{x^2} \left(1 +  2\log\left(\frac{y}{2}\right)\right)\right]
 & \text{for} & y \ll x
\end{array}
\right.
\end{align}
Independent of the precise hierarchy,
we see  that $\kappa$ scales roughly like 
\begin{eqnarray}
\kappa \sim \frac{A_{123} m_{\Psi_3}}{\max(m^2_{\Psi_3},m^2_{\varphi_2})} \,.
\end{eqnarray}
This implies, that in scenarios where
the scalars are heavier than the gauginos, we can
get an additional suppression of the corresponding
Yukawa coupling beside the loop suppression.

In the following we estimate the size of $A_{123}$
due to the $\lambda_{1236} \mean{\varphi_6}$
contributions for the scenario discussed in the
previous section. There, we have focussed on a particular EWSB-\vev setting $\(u_1,\,u_2,\,d_2\)$, which means that $\varphi_1$ in \cref{fig:loop-YukA} should be identified with the corresponding first and second generation Higgs doublets. The corresponding quartic couplings
are given in \cref{app:FullEffectiveLs}.
These Higgs doublets originate from 
$\SU{2}{L} \times \SU{2}{R} \times \SU{2}{F}$ tri-doublets. For the generation of the quark
Yukawa couplings the internal scalars have to
be squarks. 
By inspection of the scalar potential in \cref{app:FullEffectiveLs} we 
notice that the only possibilities for the $\lambda_{1236}$ vertex are the
couplings $\lambda_{69-70}$, $\lambda_{170}$ and $\lambda_{177}$. The scalar $\varphi_6$ which
obtains the \vev is one of the $\widetilde{\ell}_\ro{R}^{i,\,3}$ implying that
this \vev is  either $\omega$ or $s$. 
\begin{figure}[!htb]
	\centering
	\includegraphics[width=.99\textwidth]{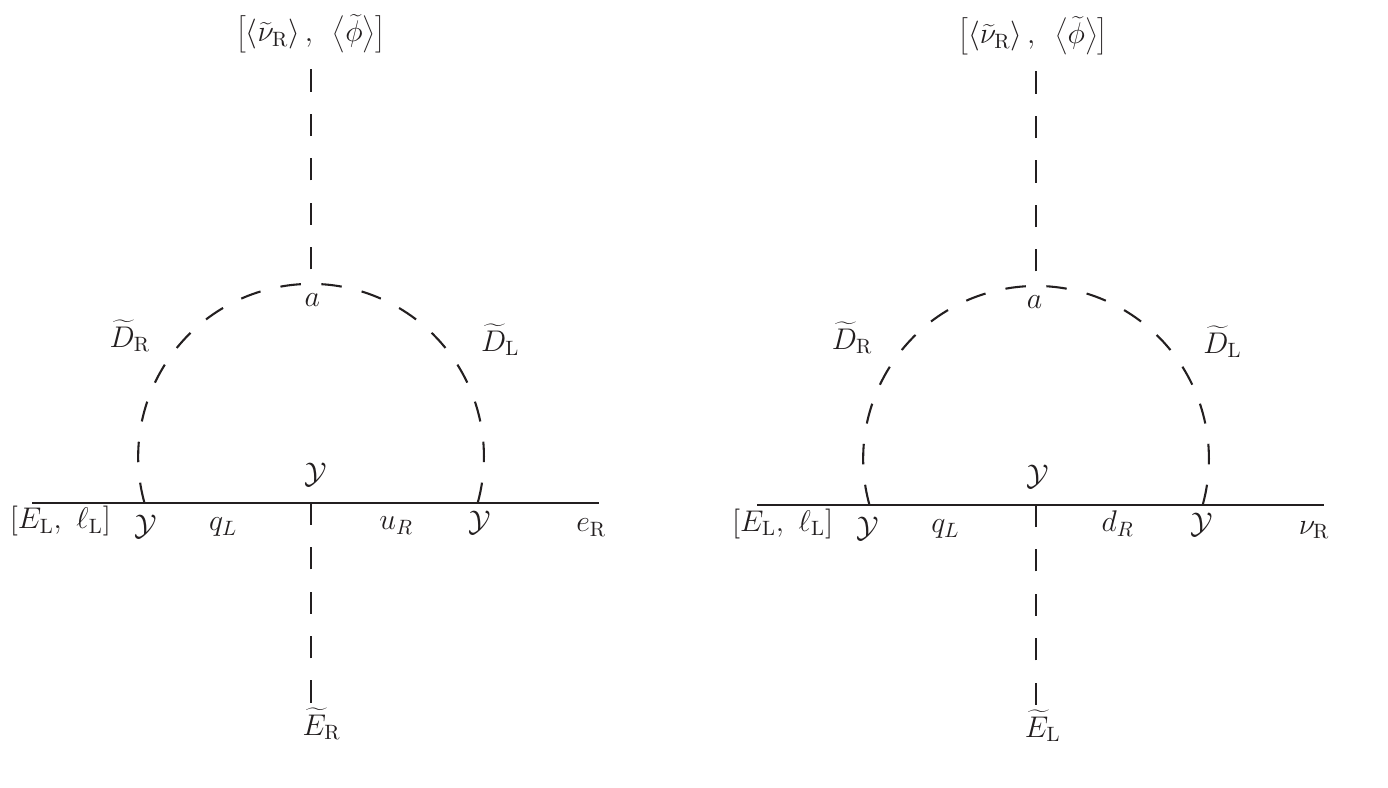}~
	\caption{Examples of one-loop diagrams contributing to the radiatively generated lepton Yukawa couplings, in addition to the topologies in \cref{fig:loop-YukA}. The indices $\ro{L,R}$ refer to the original tri-triplets given in \cref{eq:tri-triplet2}. $\mathcal{Y}$ denotes a Yukawa couplings with F-term origin.
	}
	\label{fig:loop-Yuk-leptons}
\end{figure}

We emphasise at this stage, that in case of second- and third-generation quarks we have contributions to the Yukawa couplings at tree-level and at the one-loop
from the strong and electroweak gauginos. Furthermore, we see from \cref{tab:match-4} in  \cref{app:FullEffectiveLs} that only $F$-terms
yield relevant couplings upon tree-level matching,
where typical orders of magnitude are for exmaple
\begin{eqnarray}
\lambda_{69} = \abs{\mathcal{Y}_2}^2 \sim 10^{-4}\,,
\quad
\lambda_{170} = \abs{\mathcal{Y}_1}^2 \sim 1 \,, \quad
\lambda_{177} = \mathcal{Y}_1 \mathcal{Y}_2^\ast \sim 10^{-2} \,.
\end{eqnarray}
Setting at this stage for simplicity $\omega = s$
we see, that the various possibilities
for $A_{123}$ can easily vary over
four orders of magnitude, e.g.\ 
$A_{123} = O[(10^{-4} ~\textrm{to}~ 1) \cdot \omega]$. Using again $\omega = 100~\ro{TeV}$ we obtain $A_{123} \sim \mathcal{O}\(0.01 - 100~\ro{TeV} \)$. 
Note that, if the ratio $m_{\Psi_3}/m_{\varphi_2}^2 \sim \mathcal{O}\(10^{-4}~\ro{TeV}^{-1}\)$, then the radiative corrections to SM-like quark Yukawa couplings coming from the diagram in \cref{fig:loop-YukA} can be as small as $10^{-8}$ (or even smaller). Conversely, if $m_{\varphi_{2,3}} \sim m_{\Psi_3}$ then $m_{\Psi_3}/m_{\varphi_{2,3}}^2 \sim \mathcal{O}\(\ro{TeV}^{-1}\)$, thus, for $A_{123} \sim 100~\ro{TeV}$, such radiative corrections can be as large as ${\cal O}(1)$. This result is rather relevant as it offers the possibility for large hierarchies in the fermion sectors, potentially reproducing the observed fermion masses and mixing angles without the need for a significant fine tuning.

In variance to quarks, in the charged lepton and Dirac neutrino sectors, all the Yukawa couplings are purely radiative. They receive several distinct contributions at one-loop. One type of contributions is generated by the same one-loop topologies as for the quark Yukawa couplings illustrated in \cref{fig:loop-YukA} but with the electroweak gauginos in the fermionic propagators only. Besides, there are also additional one-loop contributions, via new topologies with quark and squark propagators shown in \cref{fig:loop-Yuk-leptons}, and two more obtained from the latter by simultaneous replacements of the fields in propagators and legs:
\begin{eqnarray}
&& \tilde{D}_{\rm L} \to \tilde{u}_{\rm R},\,\tilde{d}_{\rm R}\,, \qquad
u_{\rm R},\,d_{\rm R} \to D_{\rm L}\,, \qquad
\tilde{D}_{\rm R} \to \tilde{q}_{\rm L}\,, \qquad
q_{\rm L} \to D_{\rm R} \,, \\
&& \widetilde{E}_\ro{R},\,\widetilde{E}_\ro{L} \to [\langle \tilde\nu_{\rm R} \rangle,\, \langle \tilde\phi \rangle] \,, \qquad
[\langle \tilde\nu_{\rm R} \rangle, \langle \tilde\phi \rangle] \to \widetilde{E}_\ro{R},\,\widetilde{E}_\ro{L}\,,
\end{eqnarray}
where the quark fields are the gauge eigenstates defined before the quark mixing. Note, here we do not specify the generation indices and hence the type of the trilinear coupling $a$ which should be extracted from the soft SUSY-breaking Lagrangian. Similarly, the Yukawa couplings commonly denoted as ${\cal Y}$ due to its F-term origin are, in general, different in each vertex. Finally, replacing the $\nu_{\rm R}$ fermion leg by $\phi$ and, simultaneously, $d_{\rm R}\to D_{\rm R}$ in the propagator of the right diagram in \cref{fig:loop-Yuk-leptons} we obtain two additional one-loop induced bilinear operators, $E_{\rm L}\phi$ and $l_{\rm L}\phi$.

Thus, due to different origins of the Yukawa interactions, we have an understanding why the second- and third-generation quark Yukawa couplings are larger than the first-generation quark and leptonic ones (including both the charged leptons and neutrinos). Before considering the SM-like leptons in more detail we have to investigate their mixing with the heavy vector-like leptons.

\subsubsection{Vector-like lepton sector}
\label{sec:vlls}

Another interesting feature of our model is the presence of nine copies of fermion 
$\SU{2}{L}$ doublets as one notices in \cref{eq:L-tri-triplet}. In the following, we 
denote them as
\begin{align}
\begin{aligned}
E^{i}_{\rm R} &= \( 
\begin{array}{c}
\mathcal{N}_\mathrm{R}^{i}    \\
\mathcal{E}_\mathrm{R}^{i}   
\end{array}
\)
\quad
E^{i}_{\rm L} = \( 
\begin{array}{c}
\mathcal{E}_\mathrm{L}^{i}    \\
\mathcal{N}_\mathrm{L}^{i}   
\end{array}
\)
\quad
\ell^{i}_{\rm L} = \( 
\begin{array}{c}
e_\mathrm{L}^i    \\
\nu_\mathrm{L}^i   
\end{array}
\) \,,
\end{aligned}
\end{align}
where $i = 1,2,3$ is the generation index, consitently with the notation introduced in \cref{eq:L-tri-triplet}. As will be discussed below, as soon as 
the $p$, $f$ and $\omega$ \vevs are generated, three doublets remain massless 
while the other six acquire a large mass and hence become vector-like with respect 
to $\SU{2}{L}$. Recalling that all lepton masses are purely radiative, such vector-like 
leptons (VLLs) are expected to be lighter than VLQs. However, they cannot be arbitrarily 
light in order to comply with the direct searches at collider experiments \cite{Tanabashi:2018oca}.

\begin{figure}[t]
	\centering
	\includegraphics[width=1.\textwidth]{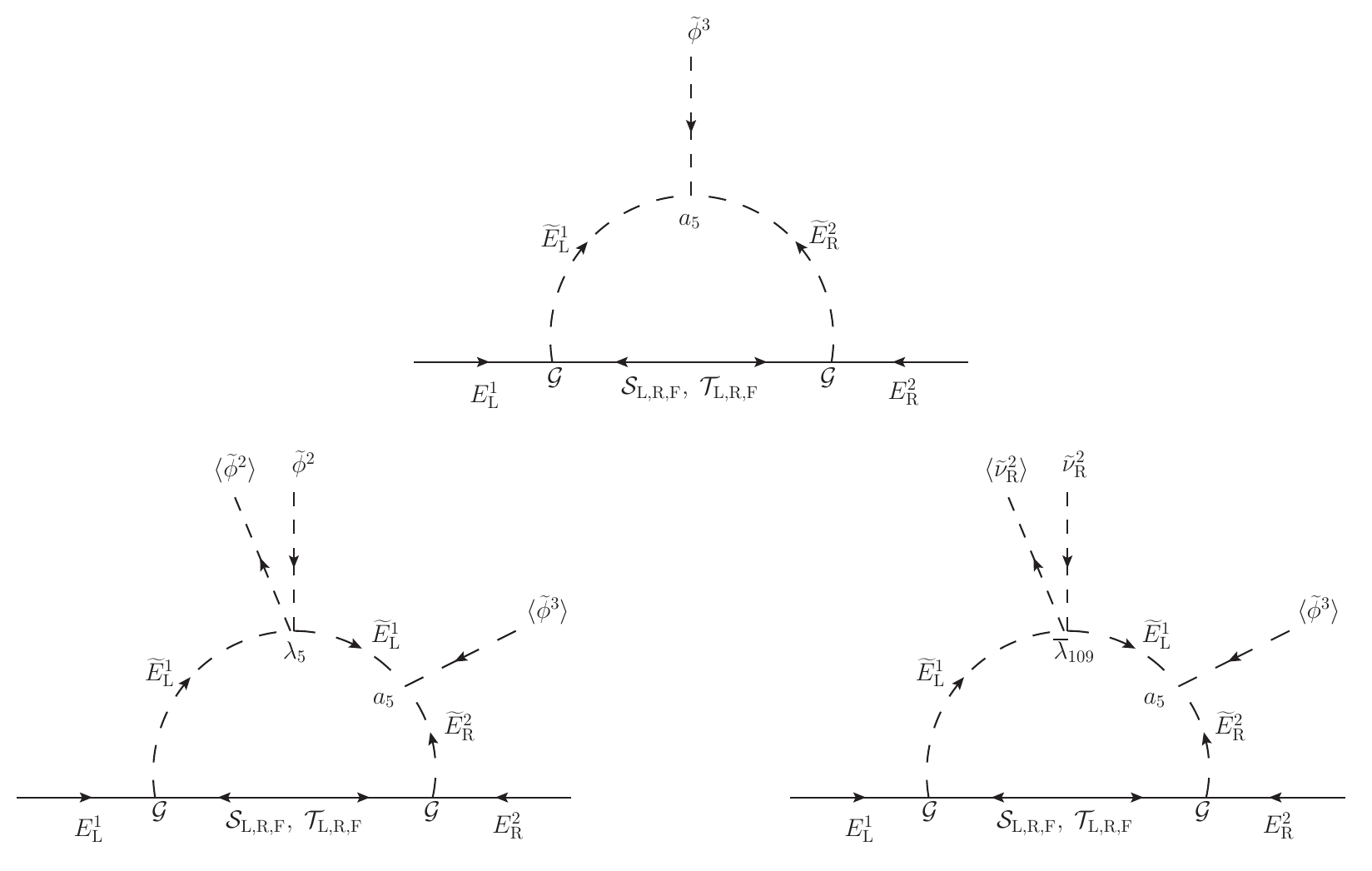}~
	\caption{One-loop diagrams contributing to the radiative generation of $\kappa_1$ (top diagram), $\kappa_1^\prime$ (bottom-left diagram) and $\kappa_1^{\prime \prime}$ (bottom-right diagram) VLL Yukawa interactions. The Yukawa coupling $\mathcal{G}$ is equal to $\y{25}^{\Scale[0.5]{\ro{L,R,F}}}$ for the $\mathcal{S}_\ro{L,R,F}$ propagators, or to $\y{31,37,43}$ -- for the $\mathcal{T}_\ro{L,R,F}$ propagators, respectively. For more details on the corresponding Yukawa operators, 
	see \cref{App:Eff-fermion}.
	}
	\label{fig:loop-VLL}
\end{figure}

The allowed Yukawa interactions involving lepton-doublets can be separated in two main 
groups. While the first will be responsible for mass terms proportional to the $\omega$, 
$f$ and $p$ \vevs and read
  	\begin{equation}
  	\begin{aligned}
  	\mathcal{L}^{1-\mathrm{loop}}_{\mathrm{VLL},1} =& E_\mathrm{L}^1 E_\mathrm{R}^2 
  	\(\kappa_1 \tilde{\phi}^3 + \kappa_1^\prime \tilde{\phi}^2 + \kappa_1^{\prime \prime} \tilde{\nu}^1_\ro{R}\) 
  	+ E_\mathrm{L}^2 E_\mathrm{R}^1 \(\kappa_2 \tilde{\phi}^3 + \kappa_2^\prime 
  	\tilde{\phi}^2 + \kappa_2^{\prime \prime} \tilde{\nu}^1_\ro{R}\)
  	\\
  	&
  	+ E_\mathrm{L}^1 E_\mathrm{R}^3 \(\kappa_3 \tilde{\phi}^2 + \kappa_3^\prime 
  	\tilde{\phi}^3 + \kappa_3^{\prime \prime} \tilde{\nu}^1_\ro{R}\)
  	+ E_\mathrm{L}^3 E_\mathrm{R}^1 \(\kappa_4 \tilde{\phi}^2 + \kappa_4^\prime 
  	\tilde{\phi}^3 + \kappa_4^{\prime \prime} \tilde{\nu}^1_\ro{R}\)
  	\\
  	&
  	+ \ell_\mathrm{L}^2 E_\mathrm{R}^3 \(\kappa_5 \tilde{\nu}^1_\ro{R} + 
  	\kappa_5^\prime \tilde{\phi}^2 + \kappa_5^{\prime \prime} \tilde{\phi}^3 \)
  	+ \ell_\mathrm{L}^3 E_\mathrm{R}^2 \(\kappa_6 \tilde{\nu}^1_\ro{R} + 
  	\kappa_6^\prime \tilde{\phi}^2 + \kappa_6^{\prime \prime} \tilde{\phi}^3 \)
  	\\
  	&
  	+ \ell_\mathrm{L}^2 E_\mathrm{R}^2 \(\kappa_7 \tilde{\nu}^1_\ro{R} + 
  	\kappa_7^\prime \tilde{\phi}^2 + \kappa_7^{\prime \prime} \tilde{\phi}^3 \)
  	+ \ell_\mathrm{L}^3 E_\mathrm{R}^3 \(\kappa_8 \tilde{\nu}^1_\ro{R} + 
  	\kappa_8^\prime \tilde{\phi}^2 + \kappa_8^{\prime \prime} \tilde{\phi}^3 \)\,,
  	\end{aligned}\label{Lfermi-p}
  	\end{equation}
the second group contains Yukawa interactions responsible for generating VLL mass terms proportional to the $s_i$ \vevs,
\begin{equation}
  	\begin{aligned}
  	\mathcal{L}^{1-\mathrm{loop}}_{\mathrm{VLL},2} =&  E_\mathrm{L}^1 E_\mathrm{R}^1 \(\kappa_9 \tilde{\nu}^2_\ro{R} + \kappa_9^\prime \tilde{\nu}^3_\ro{R} + \kappa_9^{\prime \prime} \tilde{\phi}^{1\,\ast} \)
  	+ \ell_\mathrm{L}^2 E_\mathrm{R}^1 \(\kappa_{10} \tilde{\nu}^3_\ro{R} + \kappa_{10}^\prime \tilde{\nu}^2_\ro{R} + \kappa_{10}^{\prime \prime} \tilde{\phi}^{1\,\ast} \)
  	\\
  	&
  	+ \ell_\mathrm{L}^3 E_\mathrm{R}^1 \(\kappa_{11} \tilde{\nu}^2_\ro{R} + \kappa_{11}^\prime \tilde{\nu}^3_\ro{R} + \kappa_{11}^{\prime \prime} \tilde{\phi}^{1\,\ast} \)
  	+ E_\mathrm{L}^2 E_\mathrm{R}^2 \(\kappa_{12} \tilde{\nu}^2_\ro{R} + \kappa_{12}^\prime \tilde{\nu}^3_\ro{R} + \kappa_{12}^{\prime \prime} \tilde{\phi}^{1\,\ast} \)
  	\\
  	&
  	+ E_\mathrm{L}^2 E_\mathrm{R}^3 \(\kappa_{13} \tilde{\phi}^{1\,\ast} + \kappa_{13}^\prime \tilde{\nu}^3_\ro{R} + \kappa_{13}^{\prime \prime} \tilde{\nu}^2_\ro{R} \)
  	+ E_\mathrm{L}^3 E_\mathrm{R}^2 \(\kappa_{14} \tilde{\phi}^{1\,\ast} + \kappa_{14}^\prime \tilde{\nu}^3_\ro{R} + \kappa_{14}^{\prime \prime} \tilde{\nu}^2_\ro{R} \)
  	\\
  	&
  	+ E_\mathrm{L}^3 E_\mathrm{R}^3 \(\kappa_{15} \tilde{\nu}^2_\ro{R} + \kappa_{15}^\prime \tilde{\nu}^3_\ro{R} + \kappa_{15}^{\prime \prime} \tilde{\phi}^{1\,\ast} \)
  	+ \ell_\mathrm{L}^1 E_\mathrm{R}^2 \(\kappa_{16} \tilde{\nu}^3_\ro{R} + \kappa_{16}^\prime \tilde{\nu}^2_\ro{R} + \kappa_{16}^{\prime \prime} \tilde{\phi}^{1\,\ast} \)
  	\\
  	&
  	+ \ell_\mathrm{L}^1 E_\mathrm{R}^3 \(\kappa_{17} \tilde{\nu}^2_\ro{R} + \kappa_{17}^\prime \tilde{\nu}^3_\ro{R} + \kappa_{17}^{\prime \prime} \tilde{\phi}^{1\,\ast} \)\,.
  	\end{aligned}\label{Lfermi-s}
  	\end{equation}
The different diagrams contributing to the generation of the Yukawa couplings $\kappa_\alpha$, $\kappa_\alpha^\prime$ and $\kappa_\alpha^{\prime \prime}$ are displayed
in \cref{fig:loop-VLL}. We stress that all of them involve $\U{W}$-breaking soft SUSY terms,
given in \cref{softbrkgW}, which is essential as otherwise all charged leptons would remain
massless even after electroweak symmetry breaking.

After the corresponding symmetry breaking the charged lepton mass matrix written in the basis
\begin{equation}
\begin{aligned}
\label{eq:DiracM}
\mathcal{L}_{C} ~=~
\begin{pmatrix} e_\mathrm{L}^1 & e_\mathrm{L}^2 & e_\mathrm{L}^3 & \mathcal{E}_\mathrm{L}^1 & 
\mathcal{E}_\mathrm{L}^2 & \mathcal{E}_\mathrm{L}^3 \end{pmatrix} 
M_{\ell} \begin{pmatrix} e_\mathrm{R}^1 & e_\mathrm{R}^2 & e_\mathrm{R}^3 & \mathcal{E}_\mathrm{R}^1 & 
\mathcal{E}_\mathrm{R}^2 & \mathcal{E}_\mathrm{R}^3 \end{pmatrix}^{\mathrm{T}}
+~{\rm c.c.}\,. 
\end{aligned}
\end{equation}
takes the following  form
\begin{align}
\label{eq:MLTp}
M_{\ell} & ~\approx~  \(
\begin{array}{cccccc}
0  \; & \; 0 \; & \; 0 \; & \; 0\; & \; \kappa^{}_{16} s^{}_3  + \kappa_{16}^\prime s^{}_2  + \kappa_{16}^{\prime \prime} s^{}_1\; &\; \kappa^{}_{17} s^{}_2  + \kappa_{17}^\prime s^{}_3  + \kappa_{17}^{\prime \prime} s^{}_1 \\
0  \;      &\; 0\; &\; 0\; &\; \kappa^{}_{10} s^{}_3  + \kappa_{10}^\prime s^{}_2  + \kappa_{10}^{\prime \prime} s^{}_1\; &\; \kappa^{}_7 \omega + \kappa_{7}^\prime f  + \kappa_{7}^{\prime \prime} p\; &\; \kappa^{}_5 \omega + \kappa_{5}^\prime f  + \kappa_{5}^{\prime \prime} p \\
0 \;      &\; 0\; &\; 0\; &\; \kappa^{}_{11} s^{}_2  + \kappa_{11}^\prime s^{}_3  + \kappa_{11}^{\prime \prime} s^{}_1 \; &\; \kappa^{}_6 \omega + \kappa_{6}^\prime f  + \kappa_{6}^{\prime \prime} p\; &\; \kappa^{}_8 \omega + \kappa_{8}^\prime f  + \kappa_{8}^{\prime \prime} p \\
0 \;      &\; 0\; &\; 0\; &\; \kappa^{}_{9} s^{}_2  + \kappa_{9}^\prime s^{}_3  + \kappa_{9}^{\prime \prime} s^{}_1 \; &\; \kappa^{}_1 p + \kappa_{1}^\prime f  + \kappa_{1}^{\prime \prime} \omega\; &\; \kappa^{}_3 f + \kappa_{3}^\prime p  + \kappa_{3}^{\prime \prime} \omega \\
0 \;   &\; 0\; &\; 0\; &\; \kappa^{}_2 p + \kappa_{2}^\prime f  + \kappa_{2}^{\prime \prime} \omega\; &\; \kappa^{}_{12} s^{}_2  + \kappa_{12}^\prime s^{}_3  + \kappa_{12}^{\prime \prime} s^{}_1 \; &\; \kappa^{}_{13} s^{}_1  + \kappa_{13}^\prime s^{}_3  + \kappa_{13}^{\prime \prime} s^{}_2  \\
0       & \; 0\;  & \; 0\;  & \; \kappa^{}_4 f + \kappa_{4}^\prime p  + \kappa_{4}^{\prime \prime} \omega\;  & \; \kappa^{}_{14} s^{}_1  + \kappa_{14}^\prime s^{}_3  + \kappa_{14}^{\prime \prime} s^{}_2 \;  & \; \kappa^{}_{15} s^{}_2  + \kappa_{15}^\prime s^{}_3  + \kappa_{15}^{\prime \prime} s^{}_1 
\end{array}
\)\,.
\end{align}
Similar to what has been done in the extended quark sector, we investigate the case where the
$s_i$ can be neglected with respect to $\omega$, $f$ and $p$ that are assumed to be of similar 
size in what follows. Note, in this first consideration we assume a sufficiently heavy gaugino-mass 
scale enabling us to a first approximation to neglect a small effect of the gaugino-lepton mixing, and 
hence the lepton flavor violation induced by such a mixing, in the decoupling limit. 
In a future study, such an effect could be added and any modifications 
to the current results should be analysed setting bounds on parameters of the model 
from the LEP constraints and to explore further potential for phenomenological 
explorations (for such a discussion in other SUSY models with the Higgs as a slepton, 
see e.g.~Refs.~\cite{Biggio:2016sdu,Okada:2016bqx}).

It is possible to further simplify $M_\ell$ by noting that $\kappa_i$ 
dominates over $\kappa_i^\prime$ and $\kappa_i^{\prime \prime}$ in $\mathcal{L}^{1-\mathrm{loop}}_{\mathrm{VLL},1}$. 
To see this, let us consider the first term in \cref{Lfermi-p} where $\kappa_1$ is generated from 
the top diagram in \cref{fig:loop-VLL} while $\kappa_1^\prime$ and $\kappa_1^{\prime \prime}$ 
from the bottom-left and bottom-right diagrams respectively. While the top diagram is linear in $a_5$, defined in \cref{softbrkgW}, the bottom ones contain suppression factors of the order of the scalar quartic couplings $\lambda_5$ and $\overline{\lambda}_{109}$, which are defined in \cref{eq:Vsc1,eq:Vsc2}. From the tree-level matching conditions in \cref{tab:match-1} we see that both $\lambda_5$ and $\overline{\lambda}_{109}$ are of D-term origin and can be written as
\begin{equation}
    \begin{aligned}
    \lambda_5 &= -\dfrac{\pi}{6} \( 3 \alpha_\ro{F} - 2 {\alpha^\prime_\ro{F}} + 2 {\alpha^\prime_\ro{L}} + 2 {\alpha^\prime_\ro{R}}\)\,, \\
    \overline{\lambda}_{109} &= -\dfrac{\pi}{6} \( 3 \alpha_\ro{R} + 3 \alpha_\ro{F} - {\alpha^\prime_\ro{R}} - {\alpha^\prime_\ro{F}} \)\,.
    \end{aligned}
    \label{eq:quartics}
\end{equation}
We can now estimate the size of both quartic couplings using \cref{eq:quartics} and typical values for the inverse structure constants at the $p$-scale. 
Taking the second example point in \cref{tab:bench}, shown in \cref{fig:uni-2}, that we will discuss in \cref{Sect:scales}, we have
\begin{equation}
    \alpha_\ro{F} = \dfrac{1}{50.9}\;, \quad \alpha^\prime_\ro{F} = \dfrac{1}{76.3}\;, \quad \alpha_\ro{R} = \dfrac{1}{44.5}\;, \quad \alpha_\ro{R}^\prime = \dfrac{1}{69.6}\;, \quad \alpha_\ro{L}^\prime = \dfrac{1}{57.6}\;,
\end{equation}
from where we get
\begin{equation}
    \lambda_5 = -0.05 \qquad \textrm{and} \qquad \overline{\lambda}_{109} = -0.04\;.
\end{equation}
Note, that there are no $F$-term contributions
to the quartic interactions as these would involve
squarks. The additional $D$-term suppression leads
to the estimate:
\begin{equation}
    \dfrac{\kappa_{i}^\prime}{\kappa_{i}} \sim \dfrac{\kappa_{i}^{\prime \prime}}{\kappa_{i}} \sim \mathcal{O}\(0.01\)\qquad \textrm{with} \qquad i=1,\ldots,8\,.
\end{equation}
If the hierarchy between the $\omega$ and $p$ \vevs does not go beyond an order of magnitude, in the limit of $s_i \to 0$ we can approximate $M_\ell$ as follows
\begin{align}
\label{eq:MLTp-2}
M_{\ell} & ~\approx~  \(
\begin{array}{cccccc}
0  \; & \; 0 \; & \; 0 \; & \; 0\; & \; 0\; &\; 0 \\
0  \;      &\; 0\; &\; 0\; &\; 0\; &\; \kappa_7 \omega\; &\; \kappa_5 \omega \\
0 \;      &\; 0\; &\; 0\; &\; 0\; &\; \kappa_6 \omega\; &\; \kappa_8 \omega \\
0 \;      &\; 0\; &\; 0\; &\; 0\; &\; \kappa_1 p\; &\; \kappa_3 f \\
0 \;   &\; 0\; &\; 0\; &\; \kappa_2 p\; &\; 0\; &\; 0 \\
0       & \; 0\;  & \; 0\;  & \; \kappa_4 f\;  & \; 0\;  & \; 0
\end{array}
\)\,.
\end{align}

The VLL masses then become
  	\begin{equation}
  	\begin{aligned}
  	m^2_{T} &~=~  p^2 \kappa_2^2 + f^2 \kappa_4^2 \,, \\ 
  	m^2_{M,E} &~=~ \frac{1}{2}\Big( \omega^2 \Lambda_1 + p^2 \kappa_1^2 + f^2 \kappa_3^2 \pm 
  	\Big[\big(\omega^2 \Lambda_1 + p^2 \kappa_1 + f^2 \kappa_3^2\big)^2 \\ 
  	& - 
  	4 \omega^2 \big( \omega^2 \Lambda_2 -2 f p \Lambda_3 + p^2 \Lambda_4 + f^2 \Lambda_5 \big) \Big]^{1/2} \Big)\,,
  	\end{aligned}\label{vll}
  	\end{equation} 
  	where
  	\begin{equation}
  	\begin{aligned}
  	\Lambda_1 &~=~  \kappa_7^2 + \kappa_5^2 +\kappa_6^2 + \kappa_8^2 \,, \\ 
  	\Lambda_2 &~=~  \(\kappa_5\kappa_6 - \kappa_7\kappa_8\)^2 \,, \\
  	\Lambda_3 &~=~  \(\kappa_7\kappa_5 + \kappa_6\kappa_8\) \kappa_1 \kappa_3\,, \\
  	\Lambda_4 &~=~  \(\kappa_5^2 + \kappa_8^2\) \kappa_1^2\,, \\
  	\Lambda_5 &~=~ \(\kappa_7^2 + \kappa_6^2\) \kappa_3^2\,.  
  	\end{aligned}\label{Lambdas}
  	\end{equation} 
If once again we choose $\omega$ to be the smallest of the intermediate scales, as e.g.~compatible with the $p$-$f$ compressed scenario discussed in \cref{sec:UNInum}, we can Taylor-expand $m_{M,E}^2$ for $\omega\ll p,f$ and obtain
  	\begin{equation}
  	\begin{aligned}
  	m^2_{T} ~\approx~  p^2 \(\kappa_2^2 + \kappa_4^2\) \,, \qquad 
  	m^2_{M} ~\approx~ p^2 \(\kappa_1^2 + \kappa_3^2\) \,, \qquad
  	m^2_{E} ~\approx~ \frac{\( \kappa_5 \kappa_1 - \kappa_7 \kappa_3 \)^2 + \( \kappa_8 \kappa_1 - \kappa_6 \kappa_3 \)^2}{ \kappa_1^2 + \kappa_3^2} \omega^2\,.   	
  	\end{aligned}\label{vll2}
  	\end{equation}
Even simpler expressions are obtained in the case of $\omega$-$f$ compressed scenario
\begin{equation}\label{vll3}
    	m_{T} ~\approx~  p \kappa_2 \,, \qquad 
  	m_{M} ~\approx~ p \kappa_1 \,, \qquad
  	m_{E} ~\approx~ \omega \sqrt{\kappa_5^2 + \kappa_8^2}\,.   	
\end{equation}
Note, for the analytical approximations in \cref{vll2,vll3} we have assumed that the radiative Yukawa couplings $\kappa_i$ all have comparable sizes. 
We see from these expressions that the two heaviest VLLs are proportional to the $p$ \vev whereas the lightest one -- to the $\omega$ \vev. For example, taking the benchmark scenarios defined in \cref{eq:wf,eq:fp}, where $p~=~600~\ro{TeV}$ and $\omega~=~100~\ro{TeV}$ and assuming $\kappa_i \sim \mathcal{O}\(10^{-2}\)$ we get for both scenarios:
\begin{equation}
    m_{T,M} ~\sim~ \mathcal{O}\(6~\ro{TeV}\), \qquad m_E ~\sim~\mathcal{O}\(1~\ro{TeV}\). 
\end{equation}

\subsection{Neutrinos}\label{sec:neutrino}

Limiting our consideration to the neutral components of the fermionic $L^{i,3}$ bi-triplets only (see \cref{eq:L-tri-triplet}), we briefly discuss the structure of the neutrino sector. Similarly to quarks, one has both tree-level and loop induced contributions to the masses. In the basis
\begin{equation}
    \Psi_N = \(  \nu_\ro{L}^1 ~ \nu_\ro{L}^2 ~ \nu_\ro{L}^3 ~ \mathcal{N}_\ro{L}^1 ~ \mathcal{N}_\ro{L}^2 ~ \mathcal{N}_\ro{L}^3 ~ \mathcal{N}_\ro{R}^1 ~ \mathcal{N}_\ro{R}^2 ~ \mathcal{N}_\ro{R}^3 ~ \phi^1 ~ \phi^2 ~  \phi^3 ~ \nu_\ro{R}^1 ~ \nu_\ro{R}^2 ~ \nu_\ro{R}^3 \)\,,
    \label{eq:vecN}
\end{equation} 
the neutrino mass matrix before EW symmetry breaking can be written in a block-diagonal form as
\begin{align}
\label{eq:Mn}
\mathcal{M}_{N} & ~=~  \(
\begin{array}{cc}
\bm{\overline{M}}_{9 \times 9} \; & \; \bm{0}  \\
\bm{0}  \; &\; \bm{M}_{6 \times 6} \\
\end{array}
\)\,,
\end{align}
with $\bm{M}$ the $6 \times 6$ mass matrix of $\SU{2}{L}$ singlet neutrinos while $\bm{\overline{M}}$ denotes the $9 \times 9$ block of $\SU{2}{L}$ doublet neutral components. 

\begin{figure}[t]
	\centering
	\includegraphics[width=0.6\textwidth]{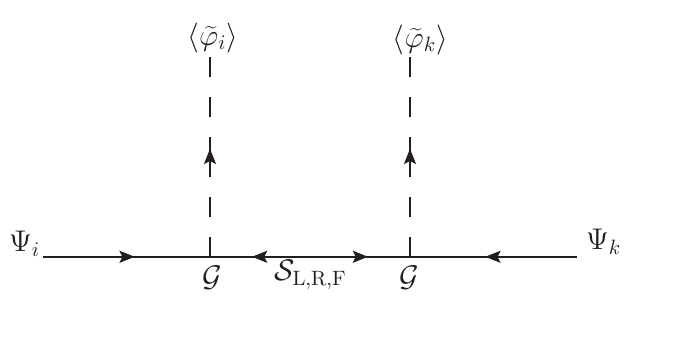}
	\caption{Tree-level diagrams generating the entries of $\bm{M}_{6 \times 6}$ block of the neutrino mass matrix (see \cref{eq:Mn}). The flavour indices $i,k$ denote the last six entries in $\Psi_N$ in \cref{eq:vecN}. The Yukawa coupling $\mathcal{G}$ is equal to $\y{23,24,27,28}^{\Scale[0.5]{\ro{A}}}$ and $\yp{23,24,27,28}^{\Scale[0.5]{\ro{A}}}$ ($\ro{A}=\ro{L,R,F}$). For more details, see \cref{App:Eff-fermion}.}
	\label{fig:tree-N}
\end{figure}
The entries $\bm{M}$ are generated at tree-level
via the topology shown in \cref{fig:tree-N}.
Here we assumed that the gaugino-masses corresponding to
the broken gauge groups at the high scales are significantly larger than the \vevs leading to the breaking.
The corresponding elements are given by
\begin{equation}
    M_{ik} = \dfrac{1}{2} \mathcal{G}^2 \dfrac{\mean{\widetilde{\varphi}_i} \mean{\widetilde{\varphi}_k}}{M_\mathcal{S}}
    \quad
    \textrm{with}
    \quad
    \mean{\widetilde{\varphi}_i} = s_{1,2,3},~\omega,~f,~p\,, 
\end{equation}
where $M_\mathcal{S}$ is the gaugino mass scale, and the Yukawa coupling $\mathcal{G}$ has a D-term origin and is specified in the caption of \cref{fig:tree-N}. Clearly, $\bm{M}$ offers the leading contributions to the total neutrino mass matrix which can be as large as $p^2/M_\mathcal{S}$. In this sector, hierarchies result from possible different sizes among the \vevs $s_{1,2,3},~\omega,~f$ and $p$.

The matrix entries of $\bm{\overline{M}}$ are 
induced at the loop-level in the same way
as for their charged counterparts discussed
in the previous section. In this limit, where
the contributions are dominated by $\kappa_i$,
$f$, $\omega$ and $p$, it is given by 
\begin{align}
\label{eq:Mbar}
\overline{M} & ~=~  \(
\begin{array}{ccccccccc}
0  \; & \; 0 \; & \; 0 \; & \; 0\; & \; 0\; &\; 0 \; & \; 0\; & \; 0\; &\; 0 \\
0  \;      &\; 0\; &\; 0\; &\; 0\; &\; 0\; &\; 0\; &\; 0\; &\; \kappa_7 \omega\; &\; \kappa_5 \omega \\
0 \;      &\; 0\; &\; 0\; &\; 0\; &\; 0\; &\; 0\; &\; 0\; &\; \kappa_6 \omega\; &\; \kappa_8 \omega \\
0 \;      &\; 0\; &\; 0\; &\; 0\; &\; 0\; &\; 0\; &\; 0\; &\; \kappa_1 p\; &\; \kappa_3 f \\
0 \;   &\; 0\; &\; 0\; &\; 0\; &\; 0\; &\; 0\;   &\; \kappa_2 p\; &\; 0\; &\; 0 \\
0       & \; 0\;  &\; 0\; &\; 0\; &\; 0\;  &\; 0\;   & \; \kappa_4 f\;  & \; 0\;  & \; 0\\
0       & \; 0\;  &\; 0\; &\; 0\; &\; \kappa_2 p\;  & \; \kappa_4 f\;  & \; 0\;  &\; 0\;   & \; 0\\
0       & \; \kappa_7 \omega \;  &\; \kappa_6 \omega\; &\; \kappa_1 p\; &\; 0\;  & \; 0\;  & \; 0\;  &\; 0\;   & \; 0
\\
0       & \; \kappa_5 \omega \;  &\; \kappa_8 \omega\; &\; \kappa_3 f\; &\; 0\;  & \; 0\;  & \; 0\;  &\; 0\;   & \; 0
\end{array}
\)\,,
\end{align}
which is a matrix of rank-6. Due to $SU(2)_L$ 
invariance one finds in this limit:
\begin{equation}
    \overline{m}^2_{N_{1,2,3}} ~=~ 0\, \qquad
    \overline{m}^2_{N_{4,5}} ~=~ m^2_E\, \qquad
   \overline{m}^2_{N_{6,7}} ~=~ m^2_M\, \qquad
    \overline{m}^2_{N_{8,9}} ~=~ m^2_T\,.
\end{equation}
At this stage one has 12 massive (6 from $\bm{M}$ and 6 from $\overline{\bm{M}}$)  and three massless neutrinos. In the corresponding mass basis and
denoting these masses as $\mu_i$ ($i=1,\dots,12$) one obtains after electroweak symmetry breaking a seesaw type I
structure for the mass matrix:
\begin{align}
\label{eq:mn-seesaw}
m_\nu & ~=~  \(
\begin{array}{cc}
 \bm{0}_{3 \times 3}  \; &\;
\dfrac{v_{\ro{EW}}}{\sqrt{2}}\(\bm{y}_\nu\)_{12 \times 3} \\
 \dfrac{v_{\ro{EW}}}{\sqrt{2}}\(\bm{y}_\nu^\top\)_{3 \times 12}  \; & \; \(\bm{\mu}_{N}\)_{12 \times 12}  
\end{array}
\)\,,
\end{align}
where $v_\ro{EW}$ represents schematically the electroweak symmetry breaking \vevs. 
 $\bm{y}_\nu$ is a $3 \times 12$ matrix denoting a combination
of various Yukawa couplings, which are radiatively generated via diagrams as in \cref{fig:loop-YukA}.
A more detailed discussion including on how
to fit neutrino data is beyond the scope of this
work and will be presented in a subsequent paper.

\section{Grand Unification} 
\label{Sect:scales}

One of the key features of the considered model is the local nature of the family symmetry implying that the family, strong and electroweak interactions are treated 
on the same footing and are ultimately unified within an $\E{8}$ gauge symmetry. In this section we study the possible hierarchies among soft SUSY, trinification, $\E{6}$ and $\E{8}$ 
breaking scales, denoted by $M_\ro{S}$, $M_3$, $M_6$ and
$M_8$, respectively.
It will also become evident how important are the effects resulting from the five-dimensional terms in 
\cref{eq:5D} which induce threshold corrections at the scale $M_6$. It was shown that without such corrections $M_\ro{S} \ge 10^{11}$ is required \cite{Camargo-Molina:2017kxd}. 
In what follows, we assume for simplicity that 
$p=f=\omega=s_i\equiv M_\ro{S}$. Inspired by the discussion in \cref{sec:CKM} we consider a low-energy EW-scale theory with three light Higgs doublets. We will also include three generations of VLLs and two generations of VLQs with a degenerate mass of order one TeV, 
in agreement with our findings in \cref{sec:numerics}.

For these considerations we use the analytic solutions of the one-loop renormalisation group equations (RGEs)
which are independent of the Yukawa couplings:
\begin{equation}
\alpha_i^{-1}\(\mu{}{}\) = \alf{0} + \frac{b_i}{2 \pi} \log\(\frac{\mu}{\mu_0}\) \,,
\label{eq:RGE}
\end{equation}
where $\alpha_i = g_i^2/(4 \pi)$ and  $b_i$ are 
the one-loop beta-function coefficients. 
The tree-level matching conditions in the gauge sector at every symmetry breaking scale 
and the explicit values of the $b_i$ between the corresponding scales can be found in \cref{app:gauge-match}. 
Note that, between $M_8$ and $M_6$ scales the presence of large representations discussed in \cref{sec:unification} 
results in $b_6 = -1095$ and, thus, a very fast running of the $\E{6}$ gauge coupling. Such a steep running, 
which is governed by
\begin{equation}
\al{6}{M_6} = \alf{8} + \frac{b_6}{2 \pi} \log\(\frac{M_6}{M_8}\)\,,
\label{eq:E6run}
\end{equation}
implies that the $M_8$ and $M_6$ breaking scales are very close to each other but the values of the $\E{8}$ and $\E{6}$ gauge couplings become rather different.
In particular, we find $\alpha_{6}(M_6) < \alpha_{8}$.

We can express the SM gauge couplings at
the $M_Z$ scale in terms of the universal $\mathrm{E}_8$ gauge
coupling and the intermediate symmetry breaking scales:
\begin{equation}
\begin{aligned}
\al{C,L}{M_Z} =& 
\al{6}{M_6} \(1 + \zeta \delta_\mathrm{C,L}\)
+ \frac{b_3}{2\pi}\log\(\frac{M_3}{M_6}\) 
+ \frac{b_\mathrm{C,L}^{(3)}}{2\pi}\log\(\frac{M_\mathrm{S}}{M_3}\)
+ \frac{b_\mathrm{C,L}^{(4)}}{2\pi}\log\(\frac{M_\mathrm{VLF}}{M_\mathrm{S}}\)
\\
&+ \frac{b_\mathrm{C,L}^{(5)}}{2\pi}\log\(\frac{M_{Z}}{M_\mathrm{VLF}}\)\,,
\end{aligned}
\label{eq:aCL}
\end{equation}
\begin{equation}
\begin{aligned}
\al{Y}{M_Z} =& 
\frac{1}{3}\al{6}{M_6} \(5 + \zeta \delta_\mathrm{L} 
+ 4 \zeta \delta_\mathrm{R}\)
+ \frac{5 b_3}{6\pi}\log\(\frac{M_3}{M_6}\)
+ \( \frac{b^{\prime(3)}_\mathrm{R} + b^{\prime(3)}_\mathrm{L}}{6 \pi} + \frac{b^{(3)}_\mathrm{R}}{2 \pi} \) \log\(\frac{M_\mathrm{S}}{M_3}\)
\\
&+ \frac{b^{(4)}_\mathrm{Y}}{2 \pi} \log\(\frac{M_\mathrm{VLF}}{M_\mathrm{S}}\) + \frac{b^{(5)}_\mathrm{Y}}{2 \pi} \log\(\frac{M_{Z}}{M_\mathrm{VLF}}\)\,,
\end{aligned}
\label{eq:aY}
\end{equation}
where $\zeta = M_6/M_8$.
In addition, we take $\alpha_\ro{T}(M_\ro{S})$ as a free parameter
because the corresponding $Z'$ boson with mass
$M_{Z'} = g_\ro{T} M_\ro{S}/2$ will be the lightest of the
additional vector bosons. We find
\begin{equation}
\begin{aligned}
\al{T}{M_\mathrm{S}} =& 
\frac{1}{27}\left[ 16 \alf{8} + \al{6}{M_6} \(29 + \zeta \delta_\mathrm{L} 
+ 28 \zeta \delta_\mathrm{R}\)
+
\frac{29 b_3}{2\pi}\log\(\frac{M_3}{M_6}\) + \frac{2 b^{\prime(1)}_\mathrm{F} + 6 b^{(1)}_\mathrm{F}}{\pi}\log\(\frac{M_3}{M_8}\) 
\] 
\\
&+\(\frac{2 b^{\prime(2)}_\mathrm{F}}{27\pi} +\frac{b^{\prime(3)}_\mathrm{L} + b^{\prime(3)}_\mathrm{R}}{57\pi} + \frac{2 b^{(2)}_\mathrm{F}}{9\pi} + \frac{b^{(3)}_\mathrm{R}}{2\pi} \)\log\( \frac{M_\mathrm{S}}{M_3}\) \,,
\end{aligned}
\label{eq:aT}
\end{equation}
Note that $\delta_\ro{L,R}$ appear in this equation due to the presence of $\U{L,R}$ generators in $\U{T}$, see \cref{tab:gauge-match}, which
themselves originate from $\SU{3}{L,R}$. For the numerical results below we use the following
values for the SM gauge couplings:
\begin{equation}
\al{C}{M_Z} = 8.4\,,\quad \al{L}{M_Z} = 29.6\,,\quad 
\al{Y}{M_Z} = 98.5\,.
\end{equation}

\begin{figure}[t]
		\begin{center}
		\includegraphics[width=.95\textwidth]{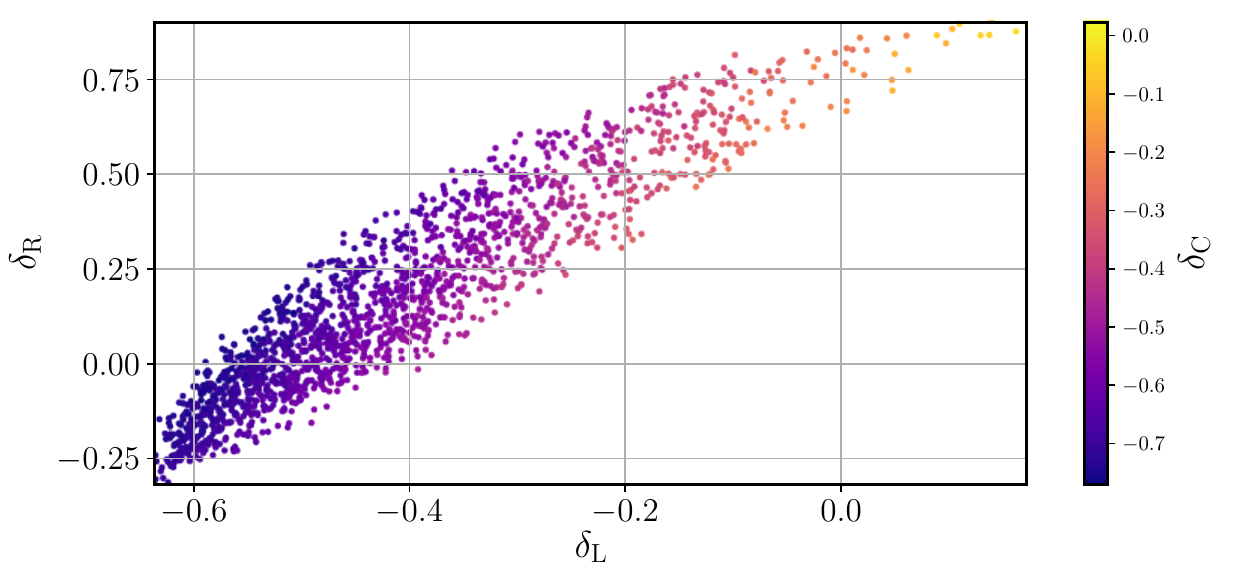}
		\end{center}
		\caption{ Possible values for the non-universality factors $\delta_\ro{L}$, $\delta_\ro{R}$ and $\delta_\ro{C}$.
}\label{fig:deltas}
	\end{figure}

Now, let us address the question on how large the
coefficients $\delta_i,\,~i=\ro{L,R,C}$ have to be to get
a consistent picture while requiring the ranges
for the free parameters as given in \cref{tab:scan}.
For this purpose we numerically invert \cref{eq:E6run,eq:aCL,eq:aY,eq:aT}
in order to determine the $\delta_i$, $\zeta$ and
$M_8$. We find that $0.9\le \zeta <1$ and
$M_8$ is a few times $10^{17}$~GeV, which
is close to the string scale. This clearly demonstrates
the internal consistency with our orbifold assumptions.
\begin{table}[t]
	\begin{center}
		\begin{tabular}{cccc}
			\toprule                     
			$M_\mathrm{S}~\[\mathrm{GeV}\]$ \;&\; $M_3~\[\mathrm{GeV}\]$ \;&\; $\al{T}{M_{Z^\prime}}$ \;&\; $\alf{8}(M_8)$ \\  
			\midrule
			$10^{4} - 10^{6}$  						\;&\; $10^4 - 10^{18}$	\;&\; $10 - 200$ \;&\; $5 - 200$	\\
			\bottomrule
		\end{tabular} 
		\caption{Ranges for parameter scans in the considered compressed scale scenario, $p ~=~ f ~=~ \omega ~=~ s_{1,2,3} ~\equiv ~ M_{\mathrm{S}}$.}
		\label{tab:scan}  
	\end{center}
\end{table}
The results for the $\delta_i$ are presented in  \cref{fig:deltas} and we find that at least one of them 
has to be sizable. 
While in some cases $\delta_\ro{C} \approx \delta_\ro{L}$ leading to a closer universality of the $\SU{3}{C}$ and $\SU{3}{L}$ interactions, the $\SU{3}{R}$ gauge coupling differs always from the other two.
We note here for completeness, that in principle
one should also add the contributions from
integrating out the heavy states corresponding
to the coset of the $E_6$ breaking. However, the masses of the corresponding particles are
of the order of $M_6$ making such contributions
significantly smaller than the required values for the $\delta_i$ and only impact in the regions where
at least one of them is close to zero.
Last but not least we note that
we have not found any solution allowing for a standard unification of the trinification gauge interactions. This is in agreement with ref.~\cite{Camargo-Molina:2017kxd} where it has been shown that this requires $M_\ro{S}\ge 10^{11}$~GeV, well above the values we consider here. 

	\begin{figure}[t]
		\begin{center}
		\includegraphics[width=.95\textwidth]{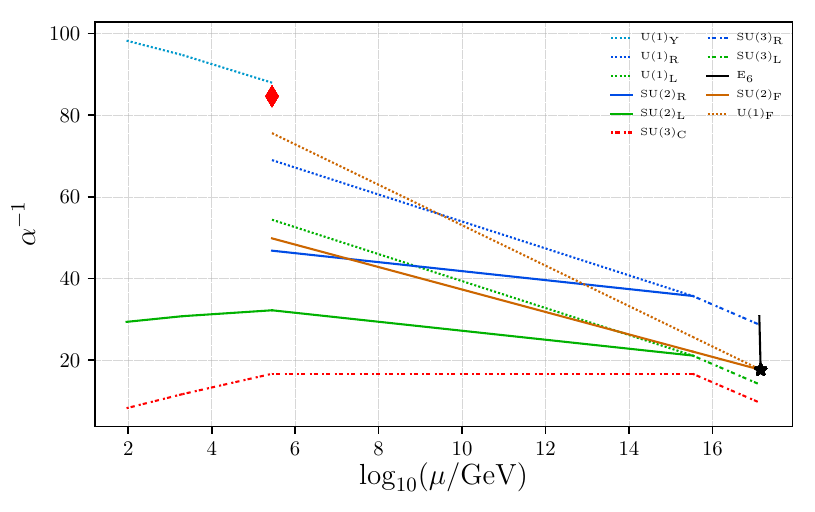}
		\end{center}
		\caption{Running of the gauge couplings in the SHUT model for the first point detailed in \cref{tab:bench}. This example represents scenarios with maximal separation between the $\E{6}$ and trinification breaking scales. While the red diamond indicates the value of the flavour structure constant $\alpha^{-1}_\ro{T}(M_\ro{S})$, the universal $\alpha^{-1}_8(M_8)$ at the unification scale is represented by a black star.
}\label{fig:uni-1}
	\end{figure}	
\begin{figure}[t]
		\begin{center}
		\includegraphics[width=.95\textwidth]{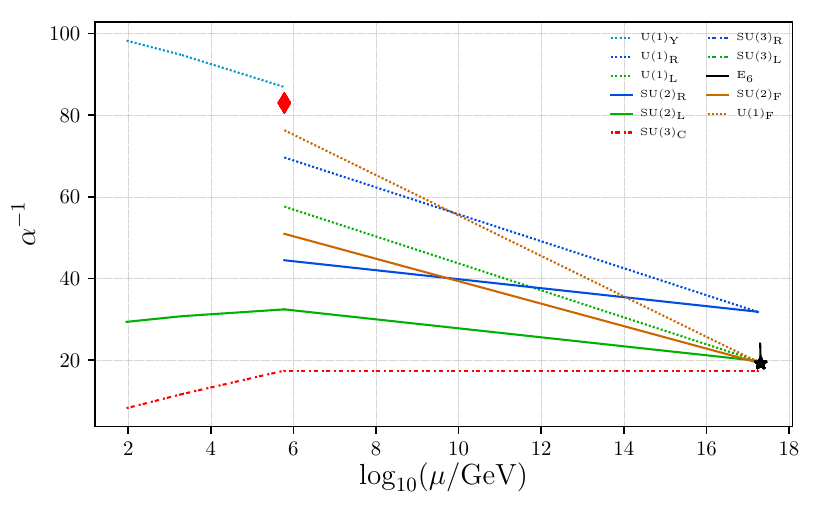}
		\end{center}
		\caption{Running of the gauge couplings in the SHUT model for the second point detailed in \cref{tab:bench}. This example was chosen as a representative solution of the scenarios with gauge couplings as close to universality as possible after $\E{6}$ is broken. In particular we have $\alpha_\ro{L}(M_6) \sim \alpha_\ro{C}(M_6) \sim \alpha_\ro{F}(M_6) \gtrsim \alpha_\ro{R}(M_6)$. The red diamond and the black star have the same meaning as in \cref{fig:uni-1}.
}\label{fig:uni-2}
	\end{figure}	
\begin{table}[t]
	\begin{center}
	\resizebox{\columnwidth}{!}{%
		\begin{tabular}{ccccccccccccc}
			\toprule                     
			$t_8$ & $t_3$ & $t_\ro{S}$ & $\zeta$ & $\alpha_8^{-1}\(M_8\)$ & $\alpha_\ro{T}^{-1}\(M_\ro{S}\)$ & $\delta_\ro{L}$ & $\delta_\ro{R}$ & $\delta_\ro{C}$ & $k_\Psi$ & $k_\Sigma$ & $k_\Sigma^\prime$ & $k_\sigma$ \\  
			\midrule
			$17.16$ & $15.53$ & $5.446$ & $0.928$ & $17.63$ & $84.58$ & $-0.583$ & $-0.0714$ & $-0.740$	& $0.384$ & $-0.418$ & $0.790$ & $0.224$ \\
			$17.31$ & $17.22$ & $5.780$ & $0.973$ & $19.23$ & $82.97$ & $-0.193$ & $0.321$ & $-0.296$	& $0.0953$ & $0.340$ & $-0.420$ & $0.836$ \\
			\bottomrule
		\end{tabular}}
		\caption{Benchmark points used for the
		running of the gauge couplings in \cref{fig:uni-1,fig:uni-2}. The top line corresponds to a parameter space point where $\delta_i$ differ considerably whereas in the bottom
		line their absolute values are of the same order. Here, $t_i = \log_{10}\tfrac{M_i}{\ro{GeV}}$.}
		\label{tab:bench}  
	\end{center}
\end{table}

In \cref{fig:uni-1,fig:uni-2} we show two representative examples of possible 
RG evolution of the gauge couplings
using the parameters of \cref{tab:bench} corresponding
to the cases where (i) $\delta_i$ are quite different
(\cref{fig:uni-1}), and (ii) $\delta_i$ are of similar size albeit with different signs (\cref{fig:uni-2}). These correspond
clearly to different $\E{6}$-breaking \vev configurations at $M_6$
scale which can be seen from \cref{tab:bench}.
A second difference between these scenarios is
the ratio $M_6/M_3$ which in the first case is
about 100 whereas in the second one it is close to unity.
Typical values for $\alpha_8^{-1}(M_8)$ are around $10-30$ as 
represented by a black star. Consequently, the first plot (\cref{fig:uni-1}) represents 
a scenario with a maximised ratio $\alpha_6^{-1}(M_6)/\alpha_8^{-1}(M_8) 
~~\simeq~~ 2$ as found in our numerical scan, whereas in the second one (\cref{fig:uni-2}) this 
ratio is close to unity. We note for completeness that in the considered 
scenarios one typically finds the strength of the gauge-family 
interactions at the soft scale $\alpha_\ro{T}^{-1}(M_\ro{S}) 
\sim \mathcal{O}(100)$ as denoted by a red diamond. This implies that the
corresponding $Z^\prime$ boson can be as light as two TeV or so.
Due to its flavor dependent couplings and a sensitivity 
to the size of $s_{1,2,3}$ \vevs, a detailed investigation will be 
nessesary to obtain bounds from the existing LHC searches.

\section{Conclusions}
\label{Sect:Conclusions}

A consistent first-principle explanation of the measured but seemingly arbitrary features 
of the Standard Model (SM) such as fermion mass spectra and mixings, structure of 
the gauge interactions, proton stability and the properties of the Higgs sector 
in the framework of a single Grand Unified Theory (GUT) remains a challenging long-debated 
programme. 

In this work, as an attempt to address this profound task we have formulated and performed
a first analysis of a novel SUSY $\mathrm{E}_8$-inspired $\mathrm{E}_6 \times \SU{2}{F} \times \U{F}$ GUT framework. 
The underlying guiding principle of our approach is the gauge rank-8 Left-Right-Color-Family (LRCF) 
unification under the $\mathrm{E}_8$ symmetry with a subsequent string-inspired orbifolding mechanism triggering 
the first symmetry reduction step $\mathrm{E}_8 \to \mathrm{E}_6 \times \SU{2}{F} \times \U{F}$. 
The latter is responsible for generating a viable chiral UV complete SUSY theory containing the light 
SM-like fermion and Higgs sectors. One of the emergent properties of the LRCF unification 
is the common origin of the Higgs and matter sectors, both chiral and vector-like, from 
two $\(\bm{27},\bm{2}\)_{(1)}$ and $\(\bm{27},\bm{1}\)_{(-2)}$ superfield representations of 
the $\mathrm{E}_6 \times \SU{2}{F} \times \U{F}$ symmetry, 
offering a SUSY GUT theory with tightly constrained Yukawa interactions. Such a unique feature of the proposed 
framework is in variance with most of the existing GUT formulations in the literature where the fundamental properties 
of the Higgs and matter sectors are typically not or only partially connected with each other. As a by-product
of such formulation, proton decay does not receive any Yukawa-mediated interactions due to
an emergent $B$-parity symmetry, while it can only be mediated by suppressed $\mathrm{E}_6$ gauge 
interactions at the GUT scale.

One of the key distinct properties of the SHUT framework is that neither mass terms nor Yukawa interactions 
for the light chiral $\(\bm{27},\bm{2}\)_{(1)}$ and $\(\bm{27},\bm{1}\)_{(-2)}$ 
superfields emerge in the $\mathrm{E}_6 \times \SU{2}{F} \times \U{F}$ superpotential at a renormalisable level.
However, dimension-4 superpotential terms involving large representations of $\mathrm{E}_6$ generate 
two distinct Yukawa operators in the SUSY $\[\SU{3}{}\]^3\times \SU{2}{F} \times \U{F}$ theory. A hierarchy 
between the $\mathcal{Y}_1$ and $\mathcal{Y}_2$ Yukawa couplings of order $m_t / m_c \sim m_b / m_s 
\sim {\cal O}(100)$ provides the necessary means for reproducing the desired hierarchies in the SM-like quark 
sector readily at the tree-level.
Besides, dimension-5 operators in the gauge sector of the $\mathrm{E}_6 \times \SU{2}{F} \times \U{F}$ 
theory introduce threshold corrections to the $\SU{3}{L,R,C}$ gauge couplings enabling a strong SUSY-protected 
hierarchy between the soft-SUSY breaking, $M_{\rm S}$, and GUT, $M_{\rm 8}\sim 10^{17}$ GeV, scales, 
with $M_\ro{S}$ allowed to be as low as $M_{\rm S}\lesssim 10^3$ TeV.

We have found that if the intermediate scales induced by the soft SUSY breaking sector lie within 
a range of approximately $10^2-10^3~\mathrm{TeV}$, the model contains three families of vector-like 
leptons within the reach of LHC measurements or future High-Energy/High-Luminosity LHC upgrades. 
Our framework features the minimum of three (and maximum of five) light Higgs doublets 
at the electroweak scale providing a Cabibbo mixing consistent with the top-charm and 
bottom-strange mass hierarchies as well as massless first-generation quarks at tree-level.
The inclusion of one-loop corrections with mild hierarchies supply the necessary ingredients 
to potentially generate realistic quark masses and mixing angles consistent with measurements.

Furthermore, we have commented on the possibility for at least one or two light generations 
of VLQs (below $10~\ro{TeV}$ or so) being potentially accessible at the LHC or future colliders. The decoupling 
between light and heavy VLQ generations is dominated by the size of the SHUT superpotential Yukawa coupling 
$\mathcal{Y}_2\ll \mathcal{Y}_1$ -- the same effect that reproduces the top-charm 
and bottom-strange mass hierarchies. This is different from the mass suppression mechanism 
of the light VLLs relative to the $M_{\rm S}$ scale which essentially follows from the quantum 
(loop) effects incorporating the soft SUSY breaking interactions and mass terms.
The size of the soft-SUSY breaking terms and the freedom that they add to the model with 
a total of 35 mass-dimensional parameters provides enough freedom to make the SM Higgs and Yukawa sectors 
consistent with phenomenology and potentially realisable with not too strong fine-tuning.

The SHUT model also offers a rich neutrino sector with the possibility for three sub-eV states and 
twelve heavy ones, with masses within the range of $10^2-10^3~\mathrm{TeV}$. Additional gauge bosons, 
and in particular a $Z^\prime$ with flavour non-universal couplings to different generations, may 
also emerge in the particle spectrum. Such very particular features of the light fermion and gauge 
boson spectrum potentially offer new smoking gun signatures for phenomenological tests of the SHUT 
model at current and future collider experiments.

\section*{Acknowledgments}
The authors would like to thank Ivo de Medeiros Varzielas and João Rosa for discussions on the orbifolding mechanisms. The work of
A.P.M.~has been performed in the framework of COST Action CA16201
``Unraveling new physics at the LHC through the precision frontier'' (PARTICLEFACE).~A.P.M.~is supported by
Funda\c{c}\~ao para a Ci\^encia e a Tecnologia (FCT),
within project UID/MAT/04106/2020 (CIDMA) and by national funds (OE), through FCT, I.P., in the scope
of the framework contract foreseen in the numbers 4, 5 and 6
of the article 23, of the Decree-Law 57/2016, of August 29,
changed by Law 57/2017, of July 19.~A.P.M.~is also supported by the \textit{Enabling Green E-science for the Square Kilometer Array Research Infrastructure} (ENGAGESKA), POCI-01-0145-FEDER-022217, and by the project \textit{From Higgs Phenomenology to the Unification of Fundamental Interactions}, PTDC/FIS-PAR/31000/2017.~R.P.~is partially supported by the Swedish Research Council, contract number 621-2013-428. W.P.\ has been supported by DFG, project nr.\ PO-1337/7-1

\appendix

\section{High-dimensional $\E{6}$ representations}
\label{app:E6-operator}

As we have seen in \cref{Sect:model-defs}, the high-dimensional $\E{6}$ representations are required for generation of threshold effects in the breaking $\E{6}\to [\SU{3}{}]^3$ as well as for a consistent description of the observed light fermion hierarchies already at tree level. For this purpose, as we have seen above, we need at least two different superchiral $\bm{650}$-reps of $\E{6}$. Starting from the product of two $\E{8}$ representations,
\begin{equation}
	\( \bm{248} \otimes \bm{248} \) = \bm{1} \oplus \bm{248} \oplus \bm{3875}\oplus \bm{27000}\oplus \bm{30380}\,,
\end{equation}
one finds in total three different $\bm{650}$-reps by decomposing the corresponding high-dimensional $\E{8}$ reps into representations of $\E{6}$ symmetry such that
\begin{equation}
	\bm{650}_1 \in \bm{3875}\,, \qquad \bm{650}_2 \in\bm{27000}\,, \qquad \bm{650}_3 \in\bm{30380}\,,
	\label{eq:650s}
\end{equation}
where one of them is the same as in the $\E{6}$-product \cref{eq:PhiE6}. Note, there is only one $\bm{2430}$ in the considered $\E{8}$-product,
\begin{equation}
	\bm{2430} \in\bm{27000}\,,
\end{equation}
which should be the same as in \cref{eq:PhiE6}.

Decomposing for example, any two $\bm{650}$ multiplets from \cref{eq:650s} into $\[\SU{3}{}\]^3$ representations yields four independent trinification singlets (two singlets per each $\bm{650}$ multiplet), and an additional one comes from $\bm{2430}$. In essence, there is one $\bm{650}$-plet that gets a \vev along one of the two trinification singlet directions and another $\bm{650}$ from another $\E{8}$ multiplet in \cref{eq:650s} that gets a \vev along the other trinification singlet directions. 
Therefore, a generic breaking of $\E{6}$ down to trinification can follow a linear combination of the following orthogonal directions
\begin{equation}
   \bm{\sigma} \equiv \bm{1}\,, \qquad \bm{\Sigma} \equiv \bm{650}\,, \qquad \bm{\Sigma^\prime} \equiv \bm{650^\prime}\,, \qquad \bm{\Psi} \equiv \bm{2430}\,.
\end{equation}

Note, the considered heavy $\E{6}$ states allow for superpotential interactions of 
the form\footnote{We have used \texttt{LieART} \cite{Feger:2012bs} to determine $\E{6}$ invariant operators.}
\begin{equation}
    \begin{aligned}
    W_{\E{6}} \supset& M_{\Sigma} \mathrm{Tr} \bm{\Sigma}^2 + M_{\Sigma^\prime} \mathrm{Tr} {\bm{\Sigma^\prime}}^{2} + M_{\Psi} \mathrm{Tr} \bm{\Psi}^2 + M_{\Sigma \Sigma^\prime} \mathrm{Tr} \bm{\Sigma} \bm{\Sigma^\prime}
    \\
    &+\lambda_{\Sigma} \mathrm{Tr} \bm{\Sigma}^3 +\lambda_{\Sigma^\prime} \mathrm{Tr} {\bm{\Sigma^\prime}}^{3} + \lambda_{\Psi} \mathrm{Tr} \bm{\Psi}^3
    + \lambda_{\Sigma \Psi} \mathrm{Tr} (\bm{\Sigma}^2 \bm{\Psi}) + \lambda_{\Sigma \Psi}^\prime \mathrm{Tr} (\bm{\Sigma} \bm{\Psi}^2)
    \\
    & + \lambda_{\Sigma^\prime \Psi} \mathrm{Tr} ({\bm{\Sigma^\prime}}^{2} \bm{\Psi}) + \lambda_{\Sigma^\prime \Psi}^\prime \mathrm{Tr} (\bm{\Sigma^\prime} \bm{\Psi}^2)
    + \lambda_{\Sigma^\prime \Sigma} \mathrm{Tr} ({\bm{\Sigma^\prime}}^{2} \bm{\Sigma}) + \lambda_{\Sigma^{\prime} \Sigma}^\prime \mathrm{Tr} (\bm{\Sigma^\prime} \bm{\Sigma}^2) + \dots \,,
    \label{eq:E6super}
    \end{aligned}
\end{equation}
where we have for simplicity omitted the operators containing the $\bm{\sigma}$ superfield as denoted by dots.

\section{Determination of the $b_i$ coefficients, $\U{}$ generators and tree-level matching conditions.}
\label{app:gauge-match}

The one-loop running of the gauge couplings can be generically described by \cref{eq:RGE}. The coefficients $b_i$ depend on the number of states present between two energy scales and on group theoretical factors. For non-abelian groups they can be determined by
\begin{equation}
    b_i = \frac{11}{3} C_2\(G\) - \frac{4}{3} \kappa T\(F\) - \frac{1}{3} T\(S\)\,,
    \label{eq:SUrun}
\end{equation}
where $\kappa = \tfrac{1}{2}$ for Wyel fermions, $C_2\(G\)$ is a group Casimir in the adjoint representation and $T\(F\)$ and $T\(S\)$ are Dynkin indices for fermions and complex scalars respectively. For $\U{}$ groups
\begin{equation}
    b^\prime_i = -\frac{4}{3} \kappa \sum_f \(\frac{Q_f}{2}\)^2 - \frac{1}{3} \sum_s \(\frac{Q_s}{2}\)^2
    \label{eq:U1run}
\end{equation}
with $Q_f$ and $Q_s$ the charges of the fermion and scalars in the theory as shown in appendix. Note that we label the abelian coefficients with a prime, $b^\prime_i$.

\begin{center}
    \textbf{Running between $M_8$ and $M_6$ scales: Region (1)}
\end{center}

The particle content that we consider between the $M_8$ and $M_6$ scales is sumarized in Tabs.~\ref{tab:M8M6vec} and \ref{tab:M8M6chi}.
\begin{table}[htb!]
	\begin{center}
		\begin{tabular}{ccc}
			\toprule                     
			$\E{6}$ \;&\; $\SU{2}{F}$ \;&\; $\U{F}$ \\  
			\midrule
			$\bm{78}$  						\;&\; $\bm{1}$	\;&\; $0$	\\
			\bottomrule
		\end{tabular} 
		\caption{Gauge superfields present between $M_8$ and $M_6$ scales}
		\label{tab:M8M6vec}  
	\end{center}
\end{table}
\begin{table}[htb!]
	\begin{center}
		\begin{tabular}{ccc}
			\toprule                     
			$\E{6}$ \;&\; $\SU{2}{F}$ \;&\; $\U{F}$ \\    
			\midrule
			$\bm{27}$  						\;&\; $\bm{2}$	\;&\; $1$	\\
			$\bm{27}$  						\;&\; $\bm{1}$	\;&\; $-2$	\\
			$\bm{1}$  						\;&\; $\bm{2}$	\;&\; $-1$	\\
			$\bm{78}$  						\;&\; $\bm{1}$	\;&\; $0$	\\
			$\bm{650}\,,\bm{650^\prime}$  	\;&\; $\bm{1}$	\;&\; $0$	\\
			$\bm{2430}$  					\;&\; $\bm{1}$	\;&\; $0$	\\
			\bottomrule
		\end{tabular} 
		\caption{Chiral superfields present between $M_8$ and $M_6$ scales}
		\label{tab:M8M6chi}  
	\end{center}
\end{table}

The Dynkin index of each of such representations, calculated using \texttt{LieART} \cite{Feger:2012bs}, reads
\begin{equation}
 T\(\bm{27}\) = 3\,,\quad T\(\bm{78}\) = 12\,,\quad T\(\bm{650}\) = 150\,,\quad T\(\bm{2430}\) = 810\,,  
\end{equation}
whereas the adjoint Casimir is $C_2\(G\) = 12$. The $\E{6}$ gauge coupling will run with coefficient
\begin{equation}
    b_6 = -1095\,.
\end{equation}
The coefficients of the $\SU{2}{F}$ gauge coupling RG-equation follow from the non-singlet representations of the non-abelian part of the family symmetry whose Dynkin indices are
\begin{equation}
    T\(\bm{2}\) = \frac{1}{2}\, \qquad T\(\bm{3}\) = C_2\(\bm{3}\) = 2\,.
\end{equation}
Replacing it in Eq.~\eqref{eq:SUrun} one obtains
\begin{equation}
    b_\mathrm{F}^{(1)} = -8\,.
\end{equation}
Finally, the coefficient of the $\U{F}$ RG-equation is calculated from the abelian charges in \cref{tab:M8M6chi} which yields
\begin{equation}
    b_\mathrm{F}^{\prime(1)} = -\frac{41}{3}
\end{equation}
where we have used the charge normalization factor $1/\(2 \sqrt{3}\)$.

\begin{center}
    \textbf{Running between $M_6$ and $M_3$ scales: Region (2)}
\end{center}

Once the $\E{6}$ symmetry is broken we are left with the massless bi-triplets $\bm{L}^{(i,3)}$, $\bm{Q}^{(i,3)}_\mathrm{L}$ and $\bm{Q}^{(i,3)}_\mathrm{R}$ as well as with the adjoint $\bm{78}$ components, see Tables~\ref{tab:ChiralSuper} and \ref{tab:ChiralSuperAdj}, respectively. The theory also contains vector supermultiplets transforming under the adjoint representation of $\[\SU{3}{}\]^3 \times \SU{2}{F} \times \U{F}$. The $\SU{3}{}$ adjoint Casimir and Dynkin indices are
\begin{equation}
    T\(\bm{3}\) = \frac{1}{2}\, \qquad T\(\bm{8}\) = C_2\(\bm{8}\) = 3\,,
\end{equation}
which replaced in eq.~\eqref{eq:SUrun} yield the following coefficients
\begin{equation}
    b_3 = -12\,, \qquad b_\mathrm{F}^{(2)} = -8\,, \qquad b_\mathrm{F}^{\prime(2)} = -\frac{41}{3}\,.
\end{equation}

\begin{center}
    \textbf{Running between $M_3$ and $M_\mathrm{S}$ scales: Region (3)}
\end{center}

With the breaking of the trinification symmetry all components of the adjoint superfields become heavy and are integrated out. The only surviving states are those embedded in the trinification bi-triplets as well as the massless gauge supermultiplets according to \cref{tab:ChiralSuper}. Using the $\SU{2}{}$ and $\SU{3}{}$ Dynkin and Casimir indices the coefficients of the RGEs are
\begin{equation}
    \begin{aligned}
    b_\mathrm{C}^{(3)} = 0\,, \qquad & b_\mathrm{L,R}^{(3)} = -3\,, \qquad b_\mathrm{F}^{(3)} = -8 
    \\
    & b^{\prime (3)}_\mathrm{L,R} = -9\,, \qquad b^{\prime (3)}_\mathrm{F} = -\frac{41}{3}
    \,.
    \end{aligned}
\end{equation}

\begin{center}
    \textbf{Running between $M_\mathrm{S}$ and $M_\mathrm{VLF}$ scales: Region (4)}
\end{center}

Below the soft scale the surviving states are three generations of $\SU{2}{L}$ VLL doublets, two generations of $\SU{2}{L}$ VLQ-singlets and three Higgs doublets reproducing the following coefficients
\begin{equation}
   \begin{aligned}
    &b_\mathrm{C}^{(4)} = \frac{19}{3}\,, \qquad b^{(4)}_\mathrm{L} = \frac{11}{6}\,, \qquad b^{(4)}_\mathrm{Y} = -\frac{155}{18}\,.
    \end{aligned} 
\end{equation}

\begin{center}
    \textbf{Running between $M_\mathrm{VLF}$ and $M_{Z}$ scales: Region (5)}
\end{center}

Finally, the running of the gauge couplings after integrating out the vector-like fermions is determined by a SM-like theory with three Higgs doublets. The coefficients of the RGEs are
\begin{equation}
    b_\mathrm{C}^{(5)} = 7\,, \qquad b_\mathrm{L}^{(5)} = \frac{17}{6}\,, \qquad 
b^{(5)}_\mathrm{Y} = -\frac{43}{6}\,.
\end{equation}

If contributions from high-dimensional operators are not relevant, the one-loop running of non-abelian gauge couplings with tree-level matching at each breaking scale is continuous. However, for the case of abelian symmetries, tree-level matching typically introduces discontinuities. Such jumps in the RG-flow are due to a non-trivial combination of generators coming from the original symmetry forming a new set of $\U{}$ generators of the unbroken symmetry. In what follows we provide a summary of the details of the abelian sector of the SHUT model including the tree-level matching conditions of the $\U{}$ gauge couplings. These results, presented in \cref{tab:gauge-match}, were used to calculate the $\U{Y}$ and $\U{T}$ renormalization group equations \eqref{eq:aY} and \eqref{eq:aT} respectively.
\begin{table}[htbp]
\centering
\begin{tabular}{cccc}
\toprule
\vevs & $\U{}$-groups & Generators & Matching condition  \\
\midrule
\multirow{ 2}{*}{$\mean{\widetilde{\phi}^3} = p$} & \multirow{ 2}{*}{$\U{L+R} \times \U{S}$} & $T_\ro{L+R} = T_\ro{L}^8 + T_\ro{R}^8$ & $\alpha_\ro{L+R}^{-1} = {\alpha_\ro{L}^{\prime\,-1}} + {\alpha_\ro{R}^{\prime\,-1}}$   \\
&  & $T_\ro{S} = T_\ro{L}^8 - T_\ro{R}^8 - 2 T_\ro{F}^8$ & $\alpha_\ro{S}^{-1} = {\alpha_\ro{L}^{\prime\,-1}} + {\alpha_\ro{R}^{\prime\,-1}} + 4 {\alpha_\ro{F}^{\prime\,-1}}$ \\ 
\hline
$\mean{\widetilde{\phi}^2} = f$ & $\U{L+R} \times \U{V}$ & $T_\ro{V} = T_\ro{F}^3 - \tfrac{1}{2 \sqrt{3}} T_\ro{S}$ & $\alpha_\ro{V}^{-1} = \alpha_\ro{F}^{-1} + \tfrac{1}{12}{\alpha_\ro{S}^\prime}^{-1}$   \\
\hline
\multirow{ 2}{*}{$\mean{\widetilde{\nu}^1} = \omega$} & \multirow{ 2}{*}{$\U{Y} \times \U{T}$} & $\tfrac12 T_\ro{Y} = T_\ro{R}^3 + \tfrac{1}{\sqrt{3}} T_\ro{L+R}$ & $\alpha_\ro{Y}^{-1} = \tfrac{1}{3}{\alpha_\ro{L+R}^{-1}} + {\alpha_\ro{R}^{-1}}$   \\
&  & $\tfrac16 T_\ro{T} = T_\ro{R}^3 - \tfrac{2}{3} T_\ro{V}$ & $\alpha_\ro{T}^{-1} = \tfrac{4}{9} {\alpha_\ro{V}^{-1}} + {\alpha_\ro{R}^{-1}}$ \\
\hline
$\mean{\widetilde{\mathcal{N}}_\ro{R}^{1,2}} = u_{1,2},\; \mean{\widetilde{\mathcal{N}}_\ro{L}^{2}} = d_{2}$ & $\U{E.M.}$ & $T_\ro{E.M.} = T_\ro{L}^3 - \tfrac12 T_\ro{Y}$ & $\alpha_\ro{E.M.}^{-1} = \alpha_\ro{Y}^{-1} + \alpha_\ro{L}^{-1}$   \\
\bottomrule
\end{tabular}
\caption{Details of the abelian sector of the SHUT model.}
\label{tab:gauge-match}
\end{table}
With the generators on the third column of \cref{tab:gauge-match} it is possible to calculate the $\U{}$ charges of the model eigenstates after each breaking stage. We refer to the appendix of our previous work \cite{Camargo-Molina:2017kxd} for tables containing that information.

\section{Effective Lagrangian below the trinification-breaking scale with tree-level matching} 
\label{app:FullEffectiveLs}

To complete the discussion of \cref{Sect:soft-sector}, we write in this appendix all possible interactions of the gauge $\SU{3}{} \times \[\SU{2}{}\]^3 \times \[\U{}\]^3$ theory, fourth box in \cref{fig:1112abc}, with the corresponding matching conditions at tree-level accuracy.

\subsection{The scalar potential}

While bilinear and trilinear interactions are of soft-SUSY breaking nature and were already discussed in \cref{Sect:soft-sector}, the quartic terms emerge from SUSY F- and D-terms. Due to a large number of possible contractions among gauge indices we separate quartic scalar interactions into five categories. First, we consider the case where all four fields posses \textit{one} common $\SU{N}{}$ index as e.g.
\begin{equation}
\begin{aligned}
V_{\rm sc1} \supset
&\lambda_{k_1}\widetilde{D}_{{\rm L}\,x\,f'}^\ast\widetilde{D}_{\rm L}^{x\,f'} \widetilde{\chi}^{\ast\,r}_{f\,l} \widetilde{\chi}^{f\,l}_{r}
+
\lambda_{k_2}\widetilde{D}_{{\rm L}\,x\,f'}^\ast\widetilde{D}_{\rm L}^{x\,f} \widetilde{\chi}^{\ast\,r}_{f\,l} \widetilde{\chi}^{f'\,l}_r
\\
+&
\overline{\lambda}_{k_1} \widetilde{D}_{{\rm R}\,f'}^{\ast\,x}\widetilde{D}_{{\rm R}\,x}^{f'} \widetilde{\chi}^{\ast\,r}_{f\,l} \widetilde{\chi}^{f\,l}_{r}
+
\overline{\lambda}_{k_2}\widetilde{D}_{{\rm R}\,f'}^{\ast\,x}\widetilde{D}_{{\rm R}\,x}^{f} \widetilde{\chi}^{\ast\,r}_{f\,l} \widetilde{\chi}^{f'\,l}_r
\\
\equiv
&\lam{k_1}{k_2} \widetilde{D}_{{\rm L}\,f'}^\ast\widetilde{D}_{\rm L}^{f'} \widetilde{\chi}^{\ast\,r}_{f\,l} \widetilde{\chi}^{f\,l}_{r} + \(\ro{L} \rightarrow \ro{R}\)\,.
\end{aligned}
\label{sc1}
\end{equation}
In this example, such an index, $f$, belongs to the $\SU{2}{F}$ space. $\SU{3}{C}$, $\SU{2}{L}$ and $\SU{2}{R}$ contractions are denoted by $x$, $l$ and $r$ respectively and only occur once. For ease of notation colour indices are suppressed in the condensed form whereas terms that differ by an interchange of $\SU{2}{L}$ and $\SU{2}{R}$ subscripts are implicitly defined by $\(\ro{L} \rightarrow \ro{R}\)$. Note that in general $\lambda_n \neq \overline{\lambda}_n$.

The second scenario describes interactions among four scalars sharing \textit{two} common gauge indices. In the following example we show a case where the common interactions are $\SU{2}{L}$ and $\SU{2}{F}$ where all possible contractions read
\begin{equation}
\begin{aligned}
V_{\rm sc2} \supset& \lambda_{n_1}\widetilde{\ell}_{{\rm L}\;l'\,f'}^\ast \widetilde{\ell}_{\rm L}^{l'\,f'} \widetilde{q}_{{\rm L}\,x\,f}^{\ast\,l} 
\widetilde{q}_{{\rm L}\,l}^{x\,f}
+
\lambda_{n_2}\widetilde{\ell}_{{\rm L}\;l'\,f'}^\ast \widetilde{\ell}_{\rm L}^{l'\,f}\widetilde{q}_{{\rm L}\,x\,f}^{\ast\,l} 
\widetilde{q}_{{\rm L}\,l}^{x\,f'}
\\
+&
\lambda_{n_3}\widetilde{\ell}_{{\rm L}\;l'\,f'}^\ast \widetilde{\ell}_{\rm L}^{l\,f'}\widetilde{q}_{{\rm L}\,x\,f}^{\ast\,l'} 
\widetilde{q}_{{\rm L}\,l}^{x\,f}
+
\lambda_{n_4}\widetilde{\ell}_{{\rm L}\;l'\,f'}^\ast \widetilde{\ell}_{\rm L}^{l\,f}\widetilde{q}_{{\rm L}\,x\,f}^{\ast\,l'} 
\widetilde{q}_{{\rm L}\,l}^{x\,f'}
\\
+&
\overline{\lambda}_{n_1} \widetilde{\ell}_{{\rm R}\;f'}^{\ast\, r'} \widetilde{\ell}_{{\rm R}\,r'}^{f'} \widetilde{q}_{{\rm R}\,r\,f}^{\ast\,x} 
\widetilde{q}_{{\rm R}\,x}^{r\,f}
+
\overline{\lambda}_{n_2} \widetilde{\ell}_{{\rm R}\;f'}^{\ast\,r'} \widetilde{\ell}_{{\rm R}\,r'}^{f}\widetilde{q}_{{\rm R}\,r\,f}^{\ast\,x} 
\widetilde{q}_{{\rm R}\,x}^{r\,f'}
\\
+&
\overline{\lambda}_{n_3} \widetilde{\ell}_{{\rm R}\;f'}^{\ast\,r'} \widetilde{\ell}_{{\rm R}\,r}^{f'}\widetilde{q}_{{\rm R}\,r'\,f}^{\ast\,x} 
\widetilde{q}_{{\rm R}\,x}^{r\,f}
+
\overline{\lambda}_{n_4} \widetilde{\ell}_{{\rm R}\;f'}^{\ast\,r'} \widetilde{\ell}_{{\rm R}\,r}^{f}\widetilde{q}_{{\rm R}\,r'\,f}^{\ast\,x} 
\widetilde{q}_{{\rm R}\,x}^{r\,f'}
\\
\equiv&
\lam{n_1}{n_4} \widetilde{\ell}_{{\rm L}\;l'\,f}^\ast \widetilde{\ell}_{\rm L}^{l'\,f} \widetilde{q}_{{\rm L}\,f'}^{\ast\,l} 
\widetilde{q}_{{\rm L}\,l}^{3\,f'} 
+
\(\ro{L} \rightarrow \ro{R}\) \,.
\end{aligned}
\label{sc2}
\end{equation}

The third case also considers \textit{two} common indices but four identical fields. Unlike $V_\ro{sc2}$, which has four independent contractions, $V_\ro{sc3}$ only contains two. We illustrate this class of scenarios with quartic self-interactions among $\widetilde{D}_\ro{L,R}$ squarks:
\begin{equation}
\begin{aligned}
V_{\rm sc3} \supset&
\lambda_{j_1}\widetilde{D}_{{\rm L}\;x'\,f'}^\ast \widetilde{D}_{\rm L}^{x'\,f'} \widetilde{D}_{{\rm L}\;x\,f}^\ast \widetilde{D}_{\rm L}^{x\,f}
+
\lambda_{j_2}\widetilde{D}_{{\rm L}\;x'\,f'}^\ast \widetilde{D}_{\rm L}^{x\,f'} \widetilde{D}_{{\rm L}\;x\,f}^\ast \widetilde{D}_{\rm L}^{x'\,f}
\\
+&
\overline{\lambda}_{j_1}\widetilde{D}_{{\rm R}\;f'}^{\ast\;x'} \widetilde{D}_{\rm R\;x'}^{f'} \widetilde{D}_{{\rm R}\;f}^{\ast\;x} \widetilde{D}_{\rm R\;x}^{f}
+
\overline{\lambda}_{j_2}\widetilde{D}_{{\rm R}\;f'}^{\ast\;x'} \widetilde{D}_{\rm R\;x}^{f'} \widetilde{D}_{{\rm R}\;f}^{\ast\;x} \widetilde{D}_{\rm R\;x'}^{f}
\\
\equiv& \lam{j_1}{j_2} \widetilde{D}_{{\rm L}\,f'}^\ast \widetilde{D}_{\rm L}^{f'} \widetilde{D}_{{\rm L}\;f}^\ast \widetilde{D}_{\rm L}^{f}
+
\(\ro{L} \rightarrow \ro{R}\) \,.
\end{aligned}
\label{sc3}
\end{equation}

It is also possible to have four identical fields but sharing three common gauge indices. This is the case of the $\widetilde{\chi}^{\;l\,f}_r$ tri-doublets whose quartic terms can be cast as
\begin{equation}\label{sc4}
\begin{aligned}
V_{\rm sc4} \supset
&\lambda_{m_1}\widetilde{\chi}^{\ast\,r'}_{f'\,l'} \widetilde{\chi}^{f'\,l'}_{r'} \widetilde{\chi}^{\ast\,r}_{f\,l} \widetilde{\chi}^{f\,l}_r
+
\lambda_{m_2}\widetilde{\chi}^{\ast\,r'}_{f'\,l'} \widetilde{\chi}^{f'\,l'}_r \widetilde{\chi}^{\ast\,r}_{f\,l} \widetilde{\chi}^{f\,l}_{r'}
\\
+&
\lambda_{m_3}\widetilde{\chi}^{\ast\,r'}_{f'\,l'} \widetilde{\chi}^{f\,l'}_{r'} \widetilde{\chi}^{\ast\,r}_{f\,l} \widetilde{\chi}^{f'\,l}_{r}
+
\lambda_{m_4}\widetilde{\chi}^{\ast\,r'}_{f'\,l'} \widetilde{\chi}^{f'\,l}_{r'} \widetilde{\chi}^{\ast\,r}_{f\,l} \widetilde{\chi}^{f\,l'}_{r}
\\
\equiv &\lam{m_1}{m_4} \widetilde{\chi}^{\ast\,r'}_{f'\,l'} \widetilde{\chi}^{f'\,l'}_{r'}\widetilde{\chi}^{\ast\,r}_{f\,l} \widetilde{\chi}^{f\,l}_r\,.
\end{aligned}
\end{equation}
Note that in this example there is no $\SU{2}{L} \leftrightarrow \SU{2}{R}$ interchange.

Finally, we refer to the case where there are no common indices among four fields or one reoccurring index with four identical fields. Such scenarios can be illustrated by
\begin{equation}\label{sc5}
V_{\rm sc5} \supset \la{i} \widetilde{\chi}^{\ast\,3\,r}_{l} \widetilde{\chi}^{3\,l}_{r} \widetilde{\phi}^\ast_{f} \widetilde{\phi}^f + \la{j} \widetilde{\phi}^\ast_{f'}\widetilde{\phi}^{f'}    
\widetilde{\phi}^\ast_{f}\widetilde{\phi}^f \,,
\end{equation}
respectively. Note that, for ease of notation, we assume that symmetry factors are implicit in the definition of the various $\la{i}$ and $\lam{i}{j}$. The total scalar quartic potential results from the sum of all five scenarios ~\eqref{sc1}, 
\eqref{sc2}, \eqref{sc3}, \eqref{sc4} and \eqref{sc5} and reads
\begin{equation}
V_4 = V_{\rm sc1} + V_{\rm sc2}+ V_{\rm sc3} + V_{\rm sc4} + V_{\rm sc5}\,.
\end{equation}

In what follows, we use the condensed notation introduced above and write all possible quartic terms. The full $V_\ro{sc1}$ potential interactions read
\begin{equation}\label{eq:Vsc1}
\begin{aligned}
&V_{\rm sc1} =   
\lam{1}{2} \widetilde{q}_{\rm L}^{\ast\,3\,l} \widetilde{q}^3_{{\rm L}\,l} \widetilde{q}_{{\rm R}\,r}^{\ast\,3} \widetilde{q}_{\rm R}^{3\,r}
+ \lam{3}{4} \widetilde{D}_{\rm L}^{\ast\,3} \widetilde{D}_{\rm L}^{3} \widetilde{D}_{\rm R}^{\ast\,3} \widetilde{D}_{\rm R}^3 
+\lam{5}{6} \widetilde{\chi}^{\ast\,r}_{f'\,l} \widetilde{\chi}^{f'\,l}_r \widetilde{\phi}^\ast_{f}\widetilde{\phi}^f  
+\lam{7}{8} \widetilde{\ell}_{{\rm L}\;f'\,l}^\ast \widetilde{\ell}_{\rm L}^{f'\,l} \widetilde{\ell}_{{\rm R}\;f}^{\ast\,r} \widetilde{\ell}_{{\rm R}\,r}^{f} 
\\
&
+\left[ \lam{9}{10} \widetilde{q}_{\rm L}^{\ast\,3\,l} \widetilde{q}^3_{{\rm L}\,l} 
\widetilde{q}_{{\rm R}\,f\,r}^{\ast} \widetilde{q}_{{\rm R}}^{f\,r} 
+\lam{11}{12} \widetilde{D}_{\rm L}^{\ast\,3} \widetilde{D}_{\rm L}^{3} \widetilde{D}_{{\rm R}\;f}^\ast \widetilde{D}_{\rm R}^{f}
+\lam{13}{14}\widetilde{q}_{\rm L}^{\ast\,3\,l} \widetilde{q}^3_{{\rm L}\,l}\widetilde{D}_{{\rm L}\;f}^\ast \widetilde{D}_{\rm L}^{f} 
+\lam{15}{16}\widetilde{q}_{{\rm L}\;f}^{\ast\,l} \widetilde{q}_{{\rm L}\,l}^{f} \widetilde{D}_{\rm L}^{\ast\,3} \widetilde{D}_{\rm L}^{3} 
 \right. \\
&
\left.
+ \lam{17}{18} \widetilde{q}_{\rm L}^{\ast\,3\,l} \widetilde{q}^3_{{\rm L}\,l}\widetilde{D}_{\rm L}^{\ast\,3} \widetilde{D}_{\rm L}^{3}
+\lam{19}{20}\widetilde{q}_{\rm L}^{\ast\,3\,l} \widetilde{q}^3_{{\rm L}\,l} \widetilde{D}_{{\rm R}\;f}^\ast \widetilde{D}_{\rm R}^{f}
+\lam{21}{22}\widetilde{q}_{{\rm L}\;f}^{\ast\,l} \widetilde{q}_{{\rm L}\,l}^{f} \widetilde{D}_{\rm R}^{\ast\,3} \widetilde{D}_{\rm R}^3
+ \lam{23}{24} \widetilde{q}_{{\rm L}\;f}^{\ast\,l'} \widetilde{q}_{{\rm L}\,l'}^{f}  \widetilde{\chi}^{\ast\,3\,r}_{l} \widetilde{\chi}^{3\,l}_{r}
\right. \\
&
\left.
+\lam{25}{26} \widetilde{q}_{\rm L}^{\ast\,3\,l} \widetilde{q}^3_{{\rm L}\,l}\widetilde{D}_{\rm R}^{\ast\,3} \widetilde{D}_{\rm R}^3
+\lam{27}{28} \widetilde{q}_{\rm L}^{\ast\,3\,l'} \widetilde{q}_{{\rm L}\,l'}^3 \widetilde{\chi}^{\ast\,r}_{f\,l} \widetilde{\chi}^{f\,l}_{r}
+ \lam{29}{30} \widetilde{q}_{\rm L}^{\ast\,3\,l'} \widetilde{q}_{{\rm L}\,l'}^3 \widetilde{\chi}^{\ast\,3\,r}_{l} \widetilde{\chi}^{3\,l}_{r}
+ \lam{31}{32} \widetilde{D}_{{\rm L}\,f'}^\ast \widetilde{D}_{\rm L}^{f'} \widetilde{\chi}^{\ast\,r}_{f\,l} \widetilde{\chi}^{f\,l}_{r}
\right. \\
&
\left.
+ \lam{33}{34} \widetilde{q}_{\rm L}^{\ast\,3\,l} \widetilde{q}^3_{{\rm L}\,l} \widetilde{\ell}_{{\rm L}\;f\,l'}^\ast \widetilde{\ell}_{\rm L}^{f\,l'} 
+ \lam{35}{36} \widetilde{q}_{{\rm L}\,f}^{\ast\,l} \widetilde{q}_{{\rm L}\,l}^{f} \widetilde{\ell}_{{\rm L}\;l'}^{\ast\,3} \widetilde{\ell}_{\rm L}^{3\,l'}
+\lam{37}{38} \widetilde{q}_{\rm L}^{\ast\,3\,l} \widetilde{q}^3_{{\rm L}\,l}  \widetilde{\ell}_{{\rm L}\;l'}^{\ast\,3} \widetilde{\ell}_{\rm L}^{3\,l'}
+ \lam{39}{40} \widetilde{q}_{{\rm R}\;f'}^{\ast\;r} \widetilde{q}_{{\rm R}\;\,r}^{f'}  \widetilde{\ell}_{{\rm L}\;f\,l}^\ast \widetilde{\ell}_{\rm L}^{f\,l}   
\right. \\
&
\left. 
+ \lam{41}{42 }\widetilde{D}_{{\rm L}\,f'}^\ast \widetilde{D}_{\rm L}^{f'} \widetilde{\ell}_{{\rm L}\;f\,l}^\ast \widetilde{\ell}_{\rm L}^{f\,l} 
+\lam{43}{44} \widetilde{D}_{{\rm R}\;f'}^\ast \widetilde{D}_{\rm R}^{f'}  \widetilde{\ell}_{{\rm L}\;f\,l}^\ast \widetilde{\ell}_{\rm L}^{f\,l}
+ \lam{45}{46} \widetilde{q}_{{\rm L}\,f'}^{\ast\,l} \widetilde{q}_{{\rm L}\,l}^{f'} \widetilde{\phi}^\ast_{f}\widetilde{\phi}^f 
+\lam{47}{48} \widetilde{D}_{{\rm L}\,f'}^\ast \widetilde{D}_{\rm L}^{f'}\widetilde{\phi}^\ast_{f}\widetilde{\phi}^f 
\right. \\
&
\left. 
+ \lam{49}{50} \widetilde{\chi}^{\ast\,3\,r}_{l'} \widetilde{\chi}^{3\,l'}_{r}  \widetilde{\ell}_{{\rm L}\;f\,l}^\ast \widetilde{\ell}_{\rm L}^{f\,l}
+ \lam{51}{52}  \widetilde{D}_{{\rm L}\;f}^\ast \widetilde{D}_{\rm L}^{f} \widetilde{\ell}_{{\rm L}\;f\,l}^\ast \widetilde{\ell}_{\rm L}^{f\,l} 
+\lam{53}{54}  \widetilde{\chi}^{\ast\,r}_{f\,l'} \widetilde{\chi}^{f\,l'}_{r}  \widetilde{\ell}_{{\rm L}\;l}^{\ast\,3} \widetilde{\ell}_{\rm L}^{3\,l} 
+\lam{55}{56} \widetilde{\chi}^{\ast\,3\,r}_{l'} \widetilde{\chi}^{3\,l'}_{r} \widetilde{\ell}_{{\rm L}\;l}^{\ast\,3} \widetilde{\ell}_{\rm L}^{3\,l}
\right. \\
&
\left.
+ \lam{57}{58} \widetilde{D}_{\rm L}^{\ast\,3} \widetilde{D}_{\rm L}^{3}  \widetilde{D}_{{\rm L}\;f}^\ast \widetilde{D}_{\rm L}^{f} 
+\lam{59}{60} \widetilde{\ell}_{{\rm L}\;f'\,l}^\ast \widetilde{\ell}_{\rm L}^{f'\,l} \widetilde{\phi}^\ast_{f}\widetilde{\phi}^f 
+ \left( \lam{61}{62} \widetilde{\phi}^f \widetilde{\chi}^{f'\,l}_r  \widetilde{\ell}_{{\rm L}\;f'\,l}^\ast \widetilde{\ell}_{{\rm R}\;f}^{\ast\,r} 
+ \lam{63}{64} \widetilde{\chi}^{\ast\,3\,r}_{l'} \widetilde{\chi}^{f\,l'}_{r}  \widetilde{\ell}_{{\rm L}\;f\,l}^\ast \widetilde{\ell}_{\rm L}^{3\,l}
\right. \right. \\
&
\left. \left.
+ \lam{65}{66}  \widetilde{\chi}^{\ast\,3\,r}_{l'} \widetilde{\chi}^{f\,l'}_{r} \widetilde{q}_{{\rm L}\,f}^{\ast\,l} \widetilde{q}^3_{{\rm L}\,l} 
+ \lam{67}{68} \widetilde{\ell}_{{\rm L}\;f\,l'}^\ast \widetilde{\ell}_{\rm L}^{3\,l'} \widetilde{q}_{\rm L}^{\ast\,3\,l} \widetilde{q}_{{\rm L}\,l}^{f}
+ \lam{69}{70} \widetilde{D}_{{\rm L}\,f'}^\ast \widetilde{q}_{{\rm L}\,l}^{f'} \widetilde{\ell}_{{\rm R}\;f}^{\ast\,r} \widetilde{\chi}^{f\,l}_{r}
+ \lam{71}{72} \widetilde{D}_{{\rm L}\,f'}^\ast \widetilde{q}_{{\rm L}\,l}^{f'} \widetilde{\ell}_{\rm L}^{f\,l}  \widetilde{\phi}^\ast_{f}
\right. \right. \\
&
\left. \left.
+ \lam{73}{74} 
\widetilde{D}_{{\rm L}\;f}^\ast
\widetilde{D}_{\rm L}^{3}
\widetilde{q}_{{\rm R}\,r}^{\ast\,3}
\widetilde{q}_{{\rm R}\;\,r}^{f}  
+ \lam{75}{76} \widetilde{\ell}_{{\rm L}\;l'}^{\ast\,3} \widetilde{\ell}_{\rm L}^{3\,l'}  \widetilde{\ell}_{{\rm L}\;f\,l}^\ast \widetilde{\ell}_{\rm L}^{f\,l}
+ {\rm c.c.} \right) 
+
\(\ro{L} \rightarrow \ro{R}\)  \right] \,.
\end{aligned}
\end{equation}
For the second scenario we have
\begin{equation}\label{eq:Vsc2}
\begin{aligned}
&V_{\rm sc2} =  
\lam{77}{80} \widetilde{q}_{{\rm L}\,f'}^{\ast\,l} \widetilde{q}_{{\rm L}\,l}^{f'} \widetilde{q}_{{\rm R}\,f\,r}^{\ast} 
\widetilde{q}_{{\rm R}}^{f\,r}
+\lam{81}{84} \widetilde{D}_{{\rm L}\,f'}^\ast \widetilde{D}_{\rm L}^{f'} \widetilde{D}_{{\rm R}\;f}^\ast \widetilde{D}_{\rm R}^{f} 
+\lam{85}{88} \widetilde{\chi}^{\ast\,3\,r'}_{l'} \widetilde{\chi}^{3\,l'}_{r'} \widetilde{\chi}^{\ast\,r}_{f\,l} \widetilde{\chi}^{f\,l}_{r}
\\
&
+\left[ \lam{89}{92}  \widetilde{q}_{{\rm L}\;f'}^{\ast\,l} \widetilde{q}_{{\rm L}\,l}^{f'} 
\widetilde{D}_{{\rm L}\;f}^\ast \widetilde{D}_{\rm L}^{f}
+ \lam{93}{96}\widetilde{q}_{{\rm L}\;f'}^{\ast\,l} \widetilde{q}_{{\rm L}\,l}^{f'} \widetilde{D}_{{\rm R}\;f}^\ast \widetilde{D}_{\rm R}^{f}
+ \lam{97}{100} \widetilde{q}_{\rm L}^{\ast\,3\,l'} \widetilde{q}_{{\rm L}\,l'}^3 \widetilde{q}_{{\rm L}\,f}^{\ast\,l} \widetilde{q}_{{\rm L}\,l}^{f}
\right. \\
&
\left. 
+\lam{101}{104} \widetilde{q}_{{\rm L}\;f'}^{\ast\,l'} \widetilde{q}_{{\rm L}\,l'}^{f'} \widetilde{\chi}^{\ast\,r}_{f\,l} \widetilde{\chi}^{f\,l}_{r}
+ \lam{105}{108}\widetilde{q}_{{\rm L}\,f'}^{\ast\,l} \widetilde{q}_{{\rm L}\,l}^{f'} \widetilde{\ell}_{{\rm L}\;f\,l'}^\ast 
\widetilde{\ell}_{\rm L}^{f\,l'} 
+\lam{109}{112}  \widetilde{\chi}^{\ast\,r}_{f'\,l'} \widetilde{\chi}^{f'\,l'}_{r} \widetilde{\ell}_{{\rm L}\;f\,l}^\ast \widetilde{\ell}_{\rm L}^{f\,l}
+
\(\ro{L} \rightarrow \ro{R}\) \right]\,.
\end{aligned}
\end{equation}
For the case of four identical fields the potential reads
\begin{equation}\label{eq:Vsc3}
\begin{aligned}
&V_{\rm sc3} =   
\lam{113}{114} \widetilde{\chi}^{\ast\,3\,r'}_{l'} \widetilde{\chi}^{3\,l'}_{r'} \widetilde{\chi}^{\ast\,3\,r}_{l} \widetilde{\chi}^{3,l}_{r} 
+
\left[ \lam{115}{116}\widetilde{q}_{\rm L}^{\ast\,3\,l'} 
\widetilde{q}_{{\rm L}\,l'}^3 \widetilde{q}_{\rm L}^{\ast\,3\,l} \widetilde{q}^3_{{\rm L}\,l} \right. \\
&
\left. 
+ \lam{117}{118} \widetilde{D}_{{\rm L}\,f'}^{\ast} \widetilde{D}_{\rm L}^{f'} \widetilde{D}_{{\rm L}\;f}^\ast \widetilde{D}_{\rm L}^{f}
+
\lam{119}{120}\widetilde{\ell}_{{\rm L}\;f'\,l'}^\ast \widetilde{\ell}_{\rm L}^{f'\,l'} \widetilde{\ell}_{{\rm L}\;f\,l}^\ast \widetilde{\ell}_{\rm L}^{f\,l} 
+
\(\ro{L} \rightarrow \ro{R}\)  \right] \,,
\end{aligned}
\end{equation}
while for the scenario with three reoccurring indices it looks like
\begin{equation}\label{eq:Vsc4}
\begin{aligned}
&V_{\rm sc4} =   
\lam{121}{124} \widetilde{\chi}^{\ast\,r'}_{f'\,l'} \widetilde{\chi}^{f'\,l'}_{r'} \widetilde{\chi}^{\ast\,r}_{f\,l} \widetilde{\chi}^{f\,l}_{r}
+
\left[ \lam{125}{128} 
\widetilde{q}_{{\rm L}\;f'}^{\ast\,l'} \widetilde{q}_{{\rm L}\,l'}^{f'} \widetilde{q}_{{\rm L}\,f}^{\ast\,l} \widetilde{q}_{{\rm L}\,l}^{f} 
+
\(\ro{L} \rightarrow \ro{R}\)\right]\,.
\end{aligned}
\end{equation}
Finally, quartic interactions with one single contraction of group indices read
\begin{equation}\label{eq:Vsc5}
\begin{aligned}
& V_{\rm sc5} =   
\la{129} \widetilde{\chi}^{\ast\,3\,r}_{l} \widetilde{\chi}^{3\,l}_{r} \widetilde{\phi}^\ast_{f}\widetilde{\phi}^f 
+ \la{130} \widetilde{\chi}^{\ast\,r}_{f\,l} \widetilde{\chi}^{f\,l}_{r} \widetilde{\phi}^{\ast\,3} \widetilde{\phi}^3 
+ \la{131} \widetilde{\chi}^{\ast\,3\,r}_{l} \widetilde{\chi}^{3\,l}_{r}  \widetilde{\phi}^{\ast\,3} \widetilde{\phi}^3
+ \la{132}  \widetilde{\ell}_{{\rm L}\;l}^{\ast\,3} \widetilde{\ell}_{\rm L}^{3\,l} 
\widetilde{\ell}_{\rm R}^{\ast\,3\,r} \widetilde{\ell}_{{\rm R}\,r}^3 
\\
&
+
\la{133} \widetilde{\phi}^\ast_{f'}\widetilde{\phi}^{f'}    \widetilde{\phi}^\ast_{f}\widetilde{\phi}^f 
+
\la{134}  \widetilde{\phi}^\ast_{f}\widetilde{\phi}^f \widetilde{\phi}^{\ast\,3} \widetilde{\phi}^3  
+ 
\la{135} \widetilde{\phi}^{\ast\,3} \widetilde{\phi}^3\,\widetilde{\phi}^{\ast\,3} \widetilde{\phi}^3 
+
\(\la{136} \widetilde{\chi}^{\ast\,3\,r}_{l} \widetilde{\chi}^{f\,l}_{r} \widetilde{\phi}^\ast_{f} \widetilde{\phi}^3
\right. \\
&
\left.
+\la{137} \widetilde{\ell}_{{\rm L}\;f\,l}^\ast \widetilde{\ell}_{\rm L}^{3\,l} \widetilde{\ell}_{\rm R}^{\ast\,3\,r} \widetilde{\ell}_{{\rm R}\,r}^{f} 
+
\la{161} \widetilde{\phi}^3 \widetilde{\chi}^{3\,l}_{r} \widetilde{\ell}_{{\rm L}\;l}^{\ast\,3} \widetilde{\ell}_{\rm R}^{\ast\,3\,r}
+ \ro{c.c.} \)
+ \left[  
\la{138}  \widetilde{D}_{\rm L}^{\ast\,3} \widetilde{D}_{\rm L}^3 \widetilde{D}_{\rm L}^{\ast\,3} \widetilde{D}_{\rm L}^3
+ \la{139} \widetilde{D}_{\rm L}^{\ast\,3} \widetilde{D}_{\rm L}^3  \widetilde{\chi}^{\ast\,r}_{f\,l} \widetilde{\chi}^{f\,l}_{r} 
\right. \\
&
\left.
+ \la{140} \widetilde{D}_{{\rm L}\;f}^\ast \widetilde{D}_{\rm L}^{f} \widetilde{\chi}^{\ast\,3\,r}_{l} \widetilde{\chi}^{3\,l}_{r} 
+ \la{141}  \widetilde{D}_{\rm L}^{\ast\,3} \widetilde{D}_{\rm L}^3 \widetilde{\chi}^{\ast\,3\,r}_{l} \widetilde{\chi}^{3\,l}_{r}
+ \la{142} \widetilde{q}_{{\rm R}\,r}^{\ast\,3} \widetilde{q}_{\rm R}^{3\,r}  \widetilde{\ell}_{{\rm L}\;f\,l}^\ast \widetilde{\ell}_{\rm L}^{f\,l}
+ \la{143}  \widetilde{q}_{{\rm R}\,f\,r}^{\ast} \widetilde{q}_{{\rm R}}^{f\,r}  \widetilde{\ell}_{{\rm L}\;l}^{\ast\,3} \widetilde{\ell}_{\rm L}^{3\,l} 
\right. \\
&
\left. 
+ \la{144} \widetilde{q}_{{\rm R}\,r}^{\ast\,3} \widetilde{q}_{\rm R}^{3\,r} \widetilde{\ell}_{{\rm L}\;l}^{\ast\,3} \widetilde{\ell}_{\rm L}^{3\,l} 
+ \la{145}  \widetilde{D}_{{\rm L}\;f}^\ast \widetilde{D}_{\rm L}^{f} \widetilde{\ell}_{{\rm L}\;l}^{\ast\,3} \widetilde{\ell}_{\rm L}^{3\,l} 
+ \la{146} \widetilde{D}_{\rm L}^{\ast\,3} \widetilde{D}_{\rm L}^3  \widetilde{\ell}_{{\rm L}\;l}^{\ast\,3} \widetilde{\ell}_{\rm L}^{3\,l}
\right. \\
&
\left.
+ \la{147} \widetilde{D}_{\rm R}^{\ast\,3} \widetilde{D}^3_{\rm R}  \widetilde{\ell}_{{\rm L}\;f\,l}^\ast \widetilde{\ell}_{\rm L}^{f\,l}
+ \la{148}  \widetilde{D}_{{\rm R}\;f}^\ast \widetilde{D}_{\rm R}^{f} \widetilde{\ell}_{{\rm L}\;l}^{\ast\,3} \widetilde{\ell}_{\rm L}^{3\,l} 
+ \la{149}  \widetilde{D}_{\rm R}^{\ast\,3} \widetilde{D}^3_{\rm R}  \widetilde{\ell}_{{\rm L}\;l}^{\ast\,3} \widetilde{\ell}_{\rm L}^{3\,l}
+ \la{150} \widetilde{q}_{\rm L}^{\ast\,3\,l} \widetilde{q}^3_{{\rm L}\,l} \widetilde{\phi}^\ast_{f}\widetilde{\phi}^f  
\right. \\
&
\left.
+ \la{151} \widetilde{q}_{{\rm L}\,f}^{\ast\,l} \widetilde{q}_{{\rm L}\,l}^{f} \widetilde{\phi}^{\ast\,3} \widetilde{\phi}^3 
+ \la{152} \widetilde{q}_{\rm L}^{\ast\,3\,l} \widetilde{q}^3_{{\rm L}\,l}  \widetilde{\phi}^{\ast\,3} \widetilde{\phi}^3
+ \la{153}\widetilde{D}_{\rm L}^{\ast\,3} \widetilde{D}_{\rm L}^3 \widetilde{\phi}^\ast_{f}\widetilde{\phi}^f 
+ \la{154} \widetilde{D}_{{\rm L}\;f}^\ast \widetilde{D}_{\rm L}^{f} \widetilde{\phi}^{\ast\,3} \widetilde{\phi}^3 
\right. \\
&
\left.
+ \la{155}\widetilde{D}_{\rm L}^{\ast\,3} \widetilde{D}_{\rm L}^3  \widetilde{\phi}^{\ast\,3} \widetilde{\phi}^3 
+ \la{156} \widetilde{\ell}_{{\rm L}\;l'}^{\ast\,3} \widetilde{\ell}_{\rm L}^{3\,l'} \widetilde{\ell}_{{\rm L}\;l}^{\ast\,3} \widetilde{\ell}_{\rm L}^{3\,l}
+ \la{157} \widetilde{\ell}_{{\rm L}\;l}^{\ast\,3} \widetilde{\ell}_{\rm L}^{3\,l} \widetilde{\ell}_{{\rm R}\;f}^{\ast\,r} \widetilde{\ell}_{{\rm R}\,r}^{f}
+\la{158} \widetilde{\ell}_{{\rm L}\;l}^{\ast\,3} \widetilde{\ell}_{\rm L}^{3\,l} \widetilde{\phi}^\ast_{f}\widetilde{\phi}^f 
\right. \\
&
\left.
+\la{159} \widetilde{\ell}_{{\rm L}\;f\,l}^\ast \widetilde{\ell}_{\rm L}^{f\,l} \widetilde{\phi}^{\ast\,3} \widetilde{\phi}^3 
+\la{160} \widetilde{\ell}_{{\rm L}\;l}^{\ast\,3} \widetilde{\ell}_{\rm L}^{3\,l} \widetilde{\phi}^{\ast\,3} \widetilde{\phi}^3
+ \la{162} \widetilde{D}_{{\rm L}\;f}^\ast \widetilde{D}_{\rm L}^3 \widetilde{D}_{\rm R}^{\ast\,3} \widetilde{D}_{\rm R}^{f}
\right.  \\
&
\left. 
+ \la{163} \widetilde{q}_{{\rm L}\,f}^{\ast\,l} \widetilde{q}^3_{{\rm L}\,l} \widetilde{q}_{{\rm R}\,r}^{\ast\,3} \widetilde{q}_{{\rm R}}^{f\,r}
+\left(
\la{164} \widetilde{\ell}_{{\rm L}\;f\,l}^\ast \widetilde{\ell}_{\rm L}^{3\,l} \widetilde{\phi}^{\ast\,3} \widetilde{\phi}^f
+ \la{165}  \widetilde{\ell}_{{\rm L}\;f\,l}^\ast \widetilde{\phi}^f \widetilde{\chi}^{3\,l}_{r} \widetilde{\ell}_{\rm R}^{\ast\,3\,r}
+ \la{166} \widetilde{\chi}^{f\,l}_{r} \widetilde{\ell}_{{\rm R}\;f}^{\ast\,r} \widetilde{\ell}_{{\rm L}\;l}^{\ast\,3} \widetilde{\phi}^3
\right. \right. \\
&
\left. \left.
+ \la{167} \widetilde{\ell}_{{\rm L}\;f\,l}^\ast \widetilde{\ell}_{\rm L}^{3\,l} \widetilde{D}_{\rm L}^{\ast\,3} \widetilde{D}_{\rm L}^{f}
+ \la{168} \widetilde{\phi}^{\ast\,3} \widetilde{\phi}^f \widetilde{D}_{{\rm L}\;f}^\ast \widetilde{D}_{\rm L}^3 
+ \la{169} \widetilde{D}_{{\rm L}\;f}^\ast \widetilde{D}_{\rm L}^3 \widetilde{q}_{\rm L}^{\ast\,3\,l} \widetilde{q}_{{\rm L}\,l}^{f} 
+ \la{170} \widetilde{D}_{\rm L}^{\ast\,3} \widetilde{q}^3_{{\rm L}\,l} \widetilde{\ell}_{{\rm R}\;f}^{\ast\,r} \widetilde{\chi}^{f\,l}_{r}
\right. \right. \\
&
\left. \left. 
+ \la{171} \widetilde{D}_{\rm L}^{\ast\,3} \widetilde{q}_{{\rm L}\,l}^{f} \widetilde{\ell}_{{\rm R}\;f}^{\ast\,r} \widetilde{\chi}^{3\,l}_{r}
+ \la{172} \widetilde{D}_{\rm L}^{\ast\,3} \widetilde{q}^3_{{\rm L}\,l} \widetilde{\ell}_{\rm R}^{\ast\,3\,r} \widetilde{\chi}^{3\,l}_{r}
+ \la{173} \widetilde{D}_{\rm L}^{\ast\,3} \widetilde{q}^3_{{\rm L}\,l} \widetilde{\ell}_{\rm L}^{f\,l} \widetilde{\phi}^\ast_{f}
+ \la{174} \widetilde{D}_{\rm L}^{\ast\,3} \widetilde{q}_{{\rm L}\,l}^{f} \widetilde{\ell}_{\rm L}^{3\,l} \widetilde{\phi}^\ast_{f} 
\right. \right.  \\
&
\left. \left. 
+ \la{175} \widetilde{D}_{\rm L}^{\ast\,3} \widetilde{q}^3_{{\rm L}\,l} \widetilde{\ell}_{\rm L}^{3\,l} \widetilde{\phi}^{\ast\,3} 
+ \la{176} \widetilde{D}_{\rm L}^{\ast\,3} \widetilde{D}_{\rm L}^{f} \widetilde{\ell}_{{\rm R}\;f}^{\ast\,r} \widetilde{\ell}_{{\rm R}\,r}^3
+ \la{177} \widetilde{D}_{{\rm L}\, f}^{\ast} \widetilde{q}^3_{{\rm L}\,l} \widetilde{\ell}_{\rm R}^{\ast\,3\,r} \widetilde{\chi}^{f\,l}_{r}
+ \la{178} \widetilde{D}_{{\rm L}\, f}^{\ast} \widetilde{q}_{{\rm L}\,l}^{f} \widetilde{\ell}_{\rm R}^{\ast\,3\,r} \widetilde{\chi}^{3\,l}_{r}
\right. \right.  \\
&
\left. \left. 
+ \la{179} \widetilde{D}_{{\rm L}\, f}^{\ast} \widetilde{q}^3_{{\rm L}\,l} \widetilde{\ell}_{\rm L}^{f\,l} \widetilde{\phi}^{\ast\,3}
+\la{180} \widetilde{D}_{{\rm L}\, f}^{\ast} \widetilde{q}_{{\rm L}\,l}^{f} \widetilde{\ell}_{\rm L}^{3\,l} \widetilde{\phi}^{\ast\,3}
+\la{181} \widetilde{\chi}^{\ast\,3\,r}_{l} \widetilde{\chi}^{f\,l}_{r} \widetilde{D}_{{\rm L}\;f}^\ast \widetilde{D}^3_{\rm L} 
+ \la{182} \widetilde{\phi}^{\ast\,3} \widetilde{\phi}^f \widetilde{q}_{{\rm L}\,f}^{\ast\,l} \widetilde{q}^3_{{\rm L}\,l}
\right. \right. \\
&
\left. \left.
+ \la{183} \widetilde{D}_{\rm L}^{\ast\,3} \widetilde{D}_{\rm L}^{f} \widetilde{\ell}_{{\rm L}\,f\,l}^\ast \widetilde{\ell}_{\rm L}^{3\,l}
+ \la{184} \widetilde{\ell}_{{\rm L}\;f\,l}^\ast \widetilde{\ell}_{\rm L}^{l} \widetilde{q}_{{\rm R}\,r}^{\ast\,3} \widetilde{q}_{{\rm R}}^{f\,r}
+ {\rm c.c.} \right)
+\(\ro{L} \rightarrow \ro{R}\) \right]  
\,.
\end{aligned}
\end{equation}
\begin{table}[htb!]
	\begin{center}
		\Scale[0.95]{
		\begin{tabular}{cc}
			\toprule                     
			Matching value \;&\; Quartic coupling\\  
			\midrule
    		$-\tfrac{1}{24} \(3 g_\ro{L}^2 + 3 g_\ro{R}^2 + 3 g_\ro{F}^2 - {g^\prime_\ro{L}}^2 - {g_\ro{R}^\prime}^2 - {g_\ro{F}^\prime}^2\)$  			\;&\; $\la{121}$
    		\\
			\hline
    		$-\tfrac{1}{24} \(2 g_\ro{C}^2 + 3 g_\ro{L}^2 + 3 g_\ro{F}^2 - {g^\prime_\ro{L}}^2  - {g_\ro{F}^\prime}^2\)$  			\;&\; $\la{125}$
    		\\
			\hline
    		$-\tfrac{1}{24} \(2 g_\ro{C}^2 + 3 g_\ro{R}^2 + 3 g_\ro{F}^2 - {g^\prime_\ro{R}}^2  - {g_\ro{F}^\prime}^2\)$  			\;&\; $\lap{125}$
    		\\
			\hline
    		$-\tfrac{1}{24} \(3 g_\ro{L}^2 + 3 g_\ro{R}^2 - {g^\prime_\ro{L}}^2 - {g_\ro{R}^\prime}^2 + 2 {g_\ro{F}^\prime}^2\)$  			\;&\; $\la{85}$
    		\\
    		\hline
    		$-\tfrac{1}{24} \(3 g_\ro{L}^2 + 3 g_\ro{R}^2 - {g^\prime_\ro{L}}^2 - {g_\ro{R}^\prime}^2 - 4 {g_\ro{F}^\prime}^2\)$  			\;&\; $\la{113}$
    		\\
   			\bottomrule
		\end{tabular} }
		\caption{Scalar quartic couplings matching conditions with five or six D-term interactions. The $\lap{i}$ couplings refer to the $\(\ro{L} \rightarrow \ro{R}\)$ part of $V_4$.}
		\label{tab:match}  
	\end{center}
\end{table}

Tree-level matching conditions for quartic couplings are obtained by solving the condition
\begin{equation}\label{eq:Lambda-Match}
    V_4 = V_\ro{SUSY}
\end{equation}
where $V_\ro{SUSY} = V_\ro{F} + V_\ro{D}$. While $V_\ro{F}$ refers to the F-term potential determined by the superpotential \eqref{super3}, $V_\ro{D}$ describes the scalar D-term interactions. For example, if we take
\begin{equation}
    \begin{aligned}
    \(-F^\ast_{q_\ro{L}^3}\)^{l}_x = \frac{\del W}{\del \widetilde{q}_\ro{L}^3}
    &= \mathcal{Y}_1 \varepsilon_{ij} \( \widetilde{\chi}^{i\,l}_r \widetilde{q}_{\ro{R}\;x}^{j\;r}
    +
    \widetilde{\ell}_\ro{L}^{i\;l} \widetilde{D}_{\ro{R}\;x}^j \)
    \end{aligned}
\end{equation}
the part of the scalar potential corresponding to these F-terms read
\begin{equation}
    \begin{aligned}
    V_\ro{SUSY} \supset \abs{F_{q_\ro{L}^3}}^2 &= \abs{\mathcal{Y}_1}^2 \varepsilon_{ij} \varepsilon^{km} \[
    \widetilde{\chi}_{k\,l}^{\ast\;r}
    \widetilde{\chi}^{i\,l}_r  \widetilde{q}_{\ro{R}\;m\;r}^{\ast\;x}
    \widetilde{q}_{\ro{R}\;x}^{j\;r}
    +
    \widetilde{\ell}_{\ro{L}\;k\;l}^\ast
    \widetilde{\ell}_\ro{L}^{i\;l}
    \widetilde{D}_{\ro{R}\;m}^{\ast\;x}
    \widetilde{D}_{\ro{R}\;x}^j
    +\(
    \widetilde{\chi}^{i\,l}_r \widetilde{q}_{\ro{R}\;x}^{j\;r}
    \widetilde{\ell}_{\ro{L}\;k\;l}^\ast
    \widetilde{D}_{\ro{R}\;m}^{\ast\;x}
    + \ro{c.c.}
    \)
    \]
    \\
    &=
    \abs{\mathcal{Y}_1}^2 
    \[
    \widetilde{\chi}_{i\,l}^{\ast\;r}
    \widetilde{\chi}^{i\,l}_r  \widetilde{q}_{\ro{R}\;j\;r}^{\ast\;x}
    \widetilde{q}_{\ro{R}\;x}^{j\;r}
    -
    \widetilde{\chi}_{j\,l}^{\ast\;r}
    \widetilde{\chi}^{i\,l}_r  \widetilde{q}_{\ro{R}\;i\;r}^{\ast\;x}
    \widetilde{q}_{\ro{R}\;x}^{j\;r} 
    +
    \widetilde{\ell}_{\ro{L}\;i\;l}^\ast
    \widetilde{\ell}_\ro{L}^{i\;l}
    \widetilde{D}_{\ro{R}\;j}^{\ast\;x}
    \widetilde{D}_{\ro{R}\;x}^j
    -
    \widetilde{\ell}_{\ro{L}\;j\;l}^\ast
    \widetilde{\ell}_\ro{L}^{i\;l}
    \widetilde{D}_{\ro{R}\;i}^{\ast\;x}
    \widetilde{D}_{\ro{R}\;x}^j
    \right. \\
    &
    \left.
    +
    \(
    \widetilde{\chi}^{i\,l}_r \widetilde{q}_{\ro{R}\;x}^{j\;r}
    \widetilde{\ell}_{\ro{L}\;i\;l}^\ast
    \widetilde{D}_{\ro{R}\;j}^{\ast\;x}
    -
    \widetilde{\chi}^{i\,l}_r \widetilde{q}_{\ro{R}\;x}^{j\;r}
    \widetilde{\ell}_{\ro{L}\;j\;l}^\ast
    \widetilde{D}_{\ro{R}\;i}^{\ast\;x}
    + \ro{c.c.}
    \)
    \]
    \end{aligned}
\end{equation}
from where we see that $\la{43}$, $\overline{\la{69}}$ and $\overline{\la{103}}$ have a common F-term contribution equal to $\abs{\mathcal{Y}_1}^2$ while for $\la{44}$, $\overline{\la{70}}$ and $\overline{\la{104}}$ it is $-\abs{\mathcal{Y}_1}^2$. On the other hand, $D$-term contributions can be generically determined based on $\SU{N}{}$ generators properties as well as on the $\U{}$ factors of each representation. For example, a generic D-term expansion for two fundamental $\SU{N}{A}$ scalar representations $A^i$ and $B^i$ reads
\begin{equation}\label{eq:D-expansion}
    \begin{aligned}
    V_\ro{D} \supset&\frac{g^2_\ro{A}}{2} \(T^a\)^i_j \(T^a\)^k_l A_i^\ast A^j\;B_k^\ast B^l =  \frac{g^2_\ro{A}}{4} \( \delta^i_l \delta^k_j - \frac{1}{N} \delta^i_j \delta^k_l \) A_i^\ast A^j\;B_k^\ast B^l 
    \\
    =& \frac{g^2_\ro{A}}{4} \(A_i^\ast B^i\) \(A^j\;B_j^\ast\) - \frac{g^2_\ro{A}}{4N} \(A_i^\ast A^i\)\;\(B_k^\ast B^k\)\,.
    \end{aligned}
\end{equation}
The same results applies if $A$ and $B$ are both anti-fundamental. However, for the case where either $A$ is fundamental and $B$ anti-fundamental, or vice-versa, a global $-1$ factor steaming from the anti-fundamental generators must multiply \eqref{eq:D-expansion}. Finally, for $\U{A}$ D-terms, we recall that abelian charges are determined from the branching of $\SU{3}{A}$ triplets and anti-triplets down to its $\SU{2}{A} \times \U{A}$ subgroup as
\begin{equation}
    \begin{aligned}
    \bm{3} \to \bm{2}_{1} \oplus \bm{1}_{-2} \qquad
    \textrm{and}
    \qquad
    \bm{\overline{3}} \to \bm{\overline{2}}_{-1} \oplus \bm{1}_{2}\,.
    \end{aligned}
\end{equation}
With a charge-normalization factor of $\tfrac{1}{2\sqrt{3}}$ the abelian D-terms read
\begin{equation}
    V_\ro{D} \supset\frac{{g^\prime}^2_\ro{A}}{24} \(A_i^\ast A^i\)\;\(B_k^\ast B^k\)
\end{equation}
if $A$ and $B$ are both either fundamental or anti-fundamental doublets;
\begin{equation}
    V_\ro{D} \supset -\frac{{g^\prime}^2_\ro{A}}{24} \(A_i^\ast A^i\)\;\(B_k^\ast B^k\)
\end{equation}
if $A$ is fundamental and $B$ anti-fundamental or vice-versa;
\begin{equation}
    V_\ro{D} \supset \frac{{g^\prime}^2_\ro{A}}{6} \(A_i^\ast A^i\)\;\(B_k^\ast B^k\)
\end{equation}
if $A$ and $B$ are singlets embedded both in either triplets or anti-triplets of $\SU{3}{A}$;
\begin{equation}
    V_\ro{D} \supset -\frac{{g^\prime}^2_\ro{A}}{6} \(A_i^\ast A^i\)\;\(B_k^\ast B^k\)
\end{equation}
if $A$ and $B$ are both singlets but one belongs to a triplet whereas the other to an anti-triplet;
\begin{equation}
    V_\ro{D} \supset \frac{{g^\prime}^2_\ro{A}}{12} \(A_i^\ast A^i\)\;\(B_k^\ast B^k\)
\end{equation}
if $A$ is a doublet and $B$ a singlet, or vice-versa, with one belonging to a $\SU{3}{A}$ triplet whereas the other to an anti-triplet;
\begin{equation}
    V_\ro{D} \supset -\frac{{g^\prime}^2_\ro{A}}{12} \(A_i^\ast A^i\)\;\(B_k^\ast B^k\)
\end{equation}
if $A$ is a doublet and $B$ a singlet, or vice-versa, with both embedded in either a triplet or an anti-triplet.

Using the method described above we have determined the tree-level matching conditions for $V_4$ showing the results in
\crefrange{tab:match}{tab:match-4}.
\begin{table}[htb!]
	\begin{center}
		\Scale[0.95]{
		\begin{tabular}{cc}
			\toprule                     
			Matching value \;&\; Quartic coupling\\  
			\midrule
			$\tfrac{1}{24} \(3 g_\ro{F}^2 + 4 {g^\prime_\ro{L}}^2 + 4 {g_\ro{R}^\prime}^2 + {g_\ro{F}^\prime}^2\)$  		\;&\; $\la{133}$
    		\\
			\hline
    		$-\tfrac{1}{24} \(2 g_\ro{C}^2 + 3 g_\ro{L}^2 -  {g^\prime_\ro{L}}^2 + 2 {g_\ro{F}^\prime}^2\)$  		\;&\; $\la{97}$
    		\\
			\hline			
			$-\tfrac{1}{24} \(2 g_\ro{C}^2 + 3 g_\ro{L}^2 -  {g^\prime_\ro{L}}^2 - 4 {g_\ro{F}^\prime}^2\)$  			\;&\; $\la{115}$
			\\
			\hline			
			$-\tfrac{1}{24} \(2 g_\ro{C}^2 + 3 g_\ro{R}^2 -  {g^\prime_\ro{R}}^2 - 4 {g_\ro{F}^\prime}^2\)$  			\;&\; $\lap{115}$
			\\
			\hline			
			$-\tfrac{1}{24} \(2 g_\ro{C}^2 + 3 g_\ro{F}^2 - 4 {g^\prime_\ro{L}}^2 -  {g_\ro{F}^\prime}^2\)$  			\;&\; $\la{117}$
			\\
			\hline			
			$-\tfrac{1}{24} \(2 g_\ro{C}^2 + 3 g_\ro{F}^2 - 4 {g^\prime_\ro{R}}^2 -  {g_\ro{F}^\prime}^2\)$  			\;&\; $\lap{117}$
			\\
			\hline			
			$-\tfrac{1}{24} \(2 g_\ro{C}^2 + 3 g_\ro{R}^2 -  {g^\prime_\ro{R}}^2 + 2 {g_\ro{F}^\prime}^2\)$  			\;&\; $\lap{97}$
			\\
			\hline
    		$-\tfrac{1}{24} \(2 g_\ro{C}^2 + 3 g_\ro{F}^2 + 2 {g^\prime_\ro{L}}^2 - {g_\ro{F}^\prime}^2\)$  		\;&\; $\la{89}$
    		\\
			\hline			
			$-\tfrac{1}{24} \(2 g_\ro{C}^2 + 3 g_\ro{F}^2 + 2 {g^\prime_\ro{R}}^2 - {g_\ro{F}^\prime}^2\)$  			\;&\; $\lap{89}$
			\\
			\hline			
			$\tfrac{1}{24} \(3 g_\ro{L}^2 - 3 g_\ro{F}^2 -  {g^\prime_\ro{L}}^2 + {g_\ro{F}^\prime}^2\)$  			\;&\; $\la{101}$, $\la{105}$
			\\
			\hline			
			$\tfrac{1}{24} \(3 g_\ro{R}^2 - 3 g_\ro{F}^2 - {g^\prime_\ro{R}}^2 + {g_\ro{F}^\prime}^2\)$  			\;&\; $\lap{101}$, $\lap{105}$
			\\
			\hline			
			$-\tfrac{1}{24} \(3 g_\ro{L}^2 + 3 g_\ro{F}^2 -  {g^\prime_\ro{L}}^2 - {g_\ro{F}^\prime}^2\)$  			\;&\; $\la{109}$, $\la{119}$
			\\
			\hline			
			$-\tfrac{1}{24} \(3 g_\ro{R}^2 + 3 g_\ro{F}^2 - {g^\prime_\ro{R}}^2 - {g_\ro{F}^\prime}^2\)$  			\;&\; $\lap{109}$, $\lap{119}$
			\\
			\hline
    		$-\tfrac{1}{24} \(3 g_\ro{L}^2 - {g^\prime_\ro{L}}^2 + 2 {g_\ro{R}^\prime}^2 + 2 {g_\ro{F}^\prime}^2\)$  				\;&\; $\la{49}$, $\la{53}$
    		\\
			\hline			
			$-\tfrac{1}{24} \(3 g_\ro{R}^2 - {g^\prime_\ro{R}}^2 + 2 {g_\ro{L}^\prime}^2 + 2 {g_\ro{F}^\prime}^2\)$  			\;&\; $\lap{49}$, $\lap{53}$
			\\
			\hline
    		$-\tfrac{1}{24} \(3 g_\ro{L}^2 - {g^\prime_\ro{L}}^2 + 2 {g_\ro{R}^\prime}^2 - 4 {g_\ro{F}^\prime}^2\)$  			\;&\; $\la{55}$
    		\\
			\hline			
			$-\tfrac{1}{24} \(3 g_\ro{R}^2 - {g^\prime_\ro{R}}^2 + 2 {g_\ro{L}^\prime}^2 - 4  {g_\ro{F}^\prime}^2\)$  			\;&\; $\lap{55}$
			\\
			\hline			
			$-\tfrac{1}{24} \(3 g_\ro{F}^2 -2 {g^\prime_\ro{F}}^2 + 2 {g_\ro{L}^\prime}^2 + 2 {g_\ro{R}^\prime}^2\)$  				\;&\; $\la{5}$, $\la{7}$
			\\
			\hline			
			$-\tfrac{1}{24} \(3 g_\ro{F}^2 - {g^\prime_\ro{F}}^2 + 2 {g_\ro{L}^\prime}^2 - 4 {g_\ro{R}^\prime}^2\)$  			\;&\; $\la{59}$
			\\
			\hline			
			$-\tfrac{1}{24} \(3 g_\ro{F}^2 - {g^\prime_\ro{F}}^2 - 4 {g_\ro{L}^\prime}^2 + 2 {g_\ro{R}^\prime}^2\)$				\;&\; $\lap{59}$
			\\
			\hline			
			$-\tfrac{1}{24} \(3 g_\ro{L}^2 - {g^\prime_\ro{L}}^2 - 4 {g_\ro{R}^\prime}^2 + 2 {g_\ro{F}^\prime}^2\)$				\;&\; $\la{75}$
			\\
			\hline			
			$-\tfrac{1}{24} \(3 g_\ro{R}^2 - {g^\prime_\ro{R}}^2 - 4 {g_\ro{L}^\prime}^2 + 2 {g_\ro{F}^\prime}^2\)$				\;&\; $\lap{75}$
			\\
			\hline			
			$\tfrac{1}{24} \(3 g_\ro{L}^2 + {g^\prime_\ro{L}}^2 + 4 {g_\ro{R}^\prime}^2 + 4 {g_\ro{F}^\prime}^2\)$				\;&\; $\la{156}$
			\\
			\hline			
			$\tfrac{1}{24} \(3 g_\ro{R}^2 + {g^\prime_\ro{R}}^2 + 4 {g_\ro{L}^\prime}^2 + 4 {g_\ro{F}^\prime}^2\)$				\;&\; $\lap{156}$
			\\
   			\bottomrule
		\end{tabular} }
		\caption{Scalar quartic couplings matching conditions with four D-term interactions. The $\lap{i}$ couplings refer to the $\(\ro{L} \rightarrow \ro{R}\)$ part of $V_4$.}
		\label{tab:match-1}  
	\end{center}
\end{table}
\begin{table}[htb!]
	\begin{center}
	\Scale[0.95]{
		\begin{tabular}{cc}
			\toprule                     
			Matching value \;&\; Quartic coupling\\  
			\midrule
			$\tfrac{1}{24} \( 2 g_\ro{C}^2 - 3 g_\ro{F}^2 +  {g_\ro{F}^\prime}^2\)$  			\;&\; $\la{77}$, $\la{81}$, $\la{93}$, $\lap{93}$	\\
			\hline
			$-\tfrac{1}{12} \( g_\ro{C}^2 +  {g_\ro{L}^\prime}^2 +  {g_\ro{F}^\prime}^2\)$  						\;&\; $\la{13}$, $\la{15}$, $-\tfrac{1}{2} \la{138}$	\\
			\hline
			$-\tfrac{1}{12} \( g_\ro{C}^2 +  {g_\ro{R}^\prime}^2 +  {g_\ro{F}^\prime}^2\)$  			\;&\; $\lap{13}$, $\lap{15}$, $-\tfrac{1}{2} \lap{138}$	\\
			\hline
			$-\tfrac{1}{12} \( g_\ro{C}^2 -2  {g_\ro{L}^\prime}^2 +  {g_\ro{F}^\prime}^2\)$				\;&\; $\la{57}$
			\\
			\hline
			$-\tfrac{1}{12} \( g_\ro{C}^2 -2  {g_\ro{R}^\prime}^2 +  {g_\ro{F}^\prime}^2\)$  			\;&\; $\lap{57}$
			\\
			\hline			
			$-\tfrac{1}{12} \( g_\ro{C}^2 +  {g_\ro{L}^\prime}^2 -2  {g_\ro{F}^\prime}^2\)$  				\;&\; $\la{17}$
			\\
			\hline
			$-\tfrac{1}{12} \( g_\ro{C}^2 +  {g_\ro{R}^\prime}^2 -2  {g_\ro{F}^\prime}^2\)$  				\;&\; $\lap{17}$
			\\
			\hline			
			$\tfrac{1}{24} \(3 g_\ro{L}^2 -  {g_\ro{L}^\prime}^2 -2  {g_\ro{F}^\prime}^2\)$  				\;&\; $\la{23}$, $\la{27}$, $\la{33}$, $\la{35}$
			\\
			\hline
			$\tfrac{1}{24} \(3 g_\ro{R}^2 -  {g_\ro{R}^\prime}^2 +2  {g_\ro{F}^\prime}^2\)$  				\;&\; $\lap{23}$, $\lap{27}$, $\lap{33}$, $\lap{35}$
			\\
			\hline			
			$-\tfrac{1}{24} \(3 g_\ro{F}^2 -  {g_\ro{F}^\prime}^2 -2  {g_\ro{L}^\prime}^2\)$  				\;&\; $\la{31}$, $\lap{39}$, $\la{41}$, $\la{45}$, $\la{51}$
			\\
			\hline
			$-\tfrac{1}{24} \(3 g_\ro{F}^2 -  {g_\ro{F}^\prime}^2 -2  {g_\ro{R}^\prime}^2\)$  				\;&\; $\lap{31}$, $\la{39}$, $\lap{41}$, $\lap{45}$, $\lap{51}$
			\\
			\hline			
			$\abs{\mathcal{Y}_2}^2
			-\tfrac{1}{24} \(3 g_\ro{F}^2 -  {g_\ro{F}^\prime}^2 +4  {g_\ro{L}^\prime}^2\)$  				\;&\; $\lap{43}$, $\la{47}$
			\\
			\hline
			$\abs{\mathcal{Y}_1}^2
			-\tfrac{1}{24} \(3 g_\ro{F}^2 -  {g_\ro{F}^\prime}^2 +4  {g_\ro{R}^\prime}^2\)$  				\;&\; $\la{43}$, $\lap{47}$
			\\
			\hline			
			$\tfrac{1}{24} \(3 g_\ro{L}^2 -  {g_\ro{L}^\prime}^2 +4  {g_\ro{F}^\prime}^2\)$  				\;&\; $\la{29}$, $\la{37}$
			\\
			\hline
			$\tfrac{1}{24} \(3 g_\ro{R}^2 -  {g_\ro{R}^\prime}^2 +4  {g_\ro{F}^\prime}^2\)$  				\;&\; $\lap{29}$, $\lap{37}$
			\\
			\hline
			$-\tfrac{1}{12} \(  {g_\ro{L}^\prime}^2 + {g_\ro{R}^\prime}^2 +
			{g_\ro{F}^\prime}^2\)$  		\;&\; $\la{129}$, $\la{130}$, $-\tfrac{1}{2} \la{135}$, $\la{157}$, $\lap{157}$
			\\
			\hline
			$-\tfrac{1}{12} \(  {g_\ro{L}^\prime}^2 + {g_\ro{R}^\prime}^2 - 2
			{g_\ro{F}^\prime}^2\)$  		\;&\; $\la{131}$, $\la{132}$
			\\
			\hline
			$\tfrac{1}{12} \(  2 {g_\ro{L}^\prime}^2 + 2 {g_\ro{R}^\prime}^2 -
			{g_\ro{F}^\prime}^2\)$  		\;&\; $\la{134}$
			\\
			\hline
			$\tfrac{1}{12} \(  2 {g_\ro{R}^\prime}^2 -  {g_\ro{L}^\prime}^2 -
			{g_\ro{F}^\prime}^2\)$  		\;&\; $\la{158}$, $\la{159}$
			\\
			\hline
			$\tfrac{1}{12} \(  2 {g_\ro{L}^\prime}^2 -  {g_\ro{R}^\prime}^2 -
			{g_\ro{F}^\prime}^2\)$  		\;&\; $\lap{158}$, $\lap{159}$
			\\
			\hline
			$\tfrac{1}{12} \(  2 {g_\ro{R}^\prime}^2 -  {g_\ro{L}^\prime}^2 - 2
			{g_\ro{F}^\prime}^2\)$  		\;&\; $\la{160}$
			\\
			\hline
			$\tfrac{1}{12} \(  2 {g_\ro{L}^\prime}^2 -  {g_\ro{R}^\prime}^2 - 2
			{g_\ro{F}^\prime}^2\)$  		\;&\; $\lap{160}$
			\\
   			\bottomrule
		\end{tabular} }
		\caption{Scalar quartic couplings matching conditions with three D-term interactions. The $\lap{i}$ couplings refer to the $\(\ro{L} \rightarrow \ro{R}\)$ part of $V_4$.}
		\label{tab:match-2}  
	\end{center}
\end{table}
\begin{table}[htb!]
	\begin{center}
		\Scale[0.95]{
		\begin{tabular}{cc}
			\toprule                     
			Matching value \;&\; Quartic coupling\\  
			\midrule
			$\tfrac{1}{4} \(g_\ro{L}^2 + g_\ro{R}^2\)$  					\;&\; $\la{114}$	\\
			\hline
			$\tfrac{1}{4} \(g_\ro{C}^2 + g_\ro{L}^2\)$  					\;&\; $\la{116}$	\\
			\hline
			$\tfrac{1}{4} \(g_\ro{C}^2 + g_\ro{R}^2\)$  					\;&\; $\lap{116}$	\\
			\hline
			$\tfrac{1}{4} \(g_\ro{C}^2 + g_\ro{F}^2\)$  					\;&\; $\la{118}$, $\lap{118}$	\\
			\hline
			$\tfrac{1}{4} \(g_\ro{L}^2 + g_\ro{F}^2\)$  					\;&\; $\la{120}$
			\\
			\hline
			$\tfrac{1}{4} \(g_\ro{R}^2 + g_\ro{F}^2\)$  					\;&\; $\lap{120}$
			\\
			\hline
			$\tfrac{1}{12} \(g_\ro{C}^2 + 2 {g^\prime_\ro{F}}^2\)$  						\;&\; $\la{1}$, $\la{3}$, $\la{25}$, $\lap{25}$	\\
			\hline
			$\tfrac{1}{12} \(g_\ro{C}^2 - {g^\prime_\ro{F}}^2 \)$  						\;&\; $\la{9}$, $\lap{9}$, $\la{11}$, $\lap{11}$, $\la{19}$, $\lap{19}$, $\la{21}$, $\lap{21}$	\\
			\hline
			$\tfrac{1}{12} \( 2 {g^\prime_\ro{L}}^2 - {g^\prime_\ro{F}}^2 \)$  			\;&\; $\la{139}$, $2\la{140}$
			\\
			\hline
			$\tfrac{1}{12} \( 2 {g^\prime_\ro{R}}^2 - {g^\prime_\ro{F}}^2 \)$  			\;&\; $\lap{139}$, $2\lap{140}$
			\\
			\hline
			$\tfrac{1}{12} \( {g^\prime_\ro{L}}^2 + 2 {g^\prime_\ro{F}}^2 \)$  			\;&\; $\la{140}$, $\lap{144}$, $\la{146}$, $\la{152}$
			\\
			\hline
			$\tfrac{1}{12} \( {g^\prime_\ro{R}}^2 + 2 {g^\prime_\ro{F}}^2 \)$  			\;&\; $\lap{140}$, $\la{144}$, $\lap{146}$, $\lap{152}$
			\\
			\hline
			$\tfrac{1}{12} \( {g^\prime_\ro{L}}^2 - {g^\prime_\ro{F}}^2 \)$  			\;&\; $\lap{142}$, $\lap{143}$, $\la{145}$, $-\tfrac{1}{2}\lap{149}$, $\la{150}$, $\la{151}$, $-\tfrac{1}{2}\la{155}$
			\\
			\hline
			$\tfrac{1}{12} \( {g^\prime_\ro{R}}^2 - {g^\prime_\ro{F}}^2 \)$  			\;&\; $\la{142}$, $\la{143}$, $\lap{145}$, $-\tfrac{1}{2}\la{149}$, $\lap{150}$, $\lap{151}$, $-\tfrac{1}{2}\lap{155}$
			\\
			\hline
			$\abs{\mathcal{Y}_2}^2 - \tfrac{1}{12} \( 2 {g^\prime_\ro{R}}^2 + {g^\prime_\ro{F}}^2 \)$  			\;&\; $\la{147}$, $\la{148}$, $\lap{153}$, $\la{154}$
			\\
			\hline
			$\abs{\mathcal{Y}_2}^2 - \tfrac{1}{12} \( 2 {g^\prime_\ro{L}}^2 + {g^\prime_\ro{F}}^2 \)$  			\;&\; $\lap{148}$, $\lap{154}$
			\\
			\hline
			$\abs{\mathcal{Y}_1}^2 - \tfrac{1}{12} \( 2 {g^\prime_\ro{L}}^2 + {g^\prime_\ro{F}}^2 \)$  			\;&\; $\lap{147}$, $\la{153}$
			\\
   			\bottomrule
		\end{tabular} }
		\caption{Scalar quartic couplings matching conditions with two D-term interactions. The $\lap{i}$ couplings refer to the $\(\ro{L} \rightarrow \ro{R}\)$ part of $V_4$.}
		\label{tab:match-3}  
	\end{center}
\end{table}
\begin{table}[htb!]
	\begin{center}
		\Scale[0.95]{
		\begin{tabular}{cc}
			\toprule                     
			Matching value \;&\; Quartic coupling\\  
			\midrule
			$\abs{\mathcal{Y}_1}^2$  		\;&\; $\lap{69}$, $-\lap{70}$, $-\lap{71}$, $\lap{72}$, $-\lap{104}$, $-\lap{108}$, $\la{170}$, $\la{173}$
			\\
			\hline
            \multirow{3}{*}{$\abs{\mathcal{Y}_2}^2$}\;&\; $-\lap{66}$, $-\lap{68}$, $\la{69}$, $-\la{70}$, $-\la{71}$, $\la{72}$, $-\la{80}$, $-\la{84}$,\\
            \;&\;$-\la{96}$, $-\lap{96}$, $-\la{104}$, $-\la{108}$, $-\lap{168}$, $\lap{170}$, $-\lap{171}$, $\lap{173}$\\
            \;&\;$-\lap{174}$, $-\lap{176}$, $-\lap{177}$, $\la{178}$, $\lap{178}$, $-\lap{179}$, $\la{180}$, $\lap{180}$\\
			\hline
            \multirow{2}{*}{$\mathcal{Y}_1 \mathcal{Y}_2^\ast$}\;&\; $\la{66}$, $\la{68}^\ast$, $-\la{74}$, $-\lap{74}^\ast$, $-\la{162}$, $-\la{163}$,\\
            \;&\;$\la{168}$, $\la{171}^\ast$, $\la{174}^\ast$, $\la{176}^\ast$, $\la{177}$, $\la{179}$\\
			\hline
			$\abs{\mathcal{Y}_1}^2 - \tfrac{1}{4} g_\ro{C}^2 $  						\;&\; $\la{10}$, $\la{12}$, $\la{20}$, $\lap{22}$	\\      
			\hline
			$\abs{\mathcal{Y}_2}^2 - \tfrac{1}{4} g_\ro{C}^2 $  						\;&\; $\lap{10}$, $\lap{12}$, $\lap{20}$, $\la{22}$, $\la{79}$, $\la{83}$, $\la{95}$, $\lap{95}$	
			\\
			\hline
			$\abs{\mathcal{Y}_2}^2 - \tfrac{1}{4} g_\ro{L}^2 $  						\;&\; $\la{24}$, $\la{28}$, $\la{34}$, $\la{36}$, $\la{103}$, $\la{107}$	\\
			\hline
			$\abs{\mathcal{Y}_1}^2 - \tfrac{1}{4} g_\ro{R}^2 $  						\;&\; $\lap{103}$, $\lap{107}$
			\\
			\hline
			$\abs{\mathcal{Y}_2}^2 - \tfrac{1}{4} g_\ro{R}^2 $  						\;&\; $\lap{24}$, $\lap{28}$, $\lap{34}$, $\lap{36}$	
			\\
			\hline
			$\abs{\mathcal{Y}_1}^2 - \tfrac{1}{4} g_\ro{F}^2 $  						\;&\; $-\la{44}$, $-\lap{48}$
			\\      
			\hline
			$\abs{\mathcal{Y}_2}^2 - \tfrac{1}{4} g_\ro{F}^2 $  						\;&\; $-\lap{44}$, $-\la{48}$	
			\\
			\hline
            \multirow{2}{*}{$\tfrac{1}{4} g_\ro{C}^2$}\;&\; $-\la{2}$, $-\la{4}$, $\la{14}$, $\lap{14}$, $\la{16}$, $\lap{16}$, $-\la{18}$, $-\lap{18}$,\\
            \;&\;$-\la{26}$, $-\lap{26}$, $\la{58}$, $\lap{58}$, $\la{91}$, $\lap{91}$, $\la{98}$, $\lap{98}$, $\la{126}$, $\lap{126}$\\
			\hline
            \multirow{2}{*}{$\tfrac{1}{4} g_\ro{L}^2$}\;&\; $-\la{30}$, $-\la{38}$, $\la{50}$, $\lap{54}$, $\la{56}$, $\la{76}$, $\la{86}$, $\la{99}$,\\
            \;&\;$\la{110}$, $\la{122}$, $\la{127}$\\
			\hline
            \multirow{2}{*}{$\tfrac{1}{4} g_\ro{R}^2$}\;&\; $-\lap{30}$, $-\lap{38}$, $\lap{50}$, $\lap{54}$, $\lap{56}$, $\lap{76}$, $\la{87}$, $\lap{99}$,\\
            \;&\;$\lap{110}$, $\la{123}$, $\lap{127}$\\
			\hline
            \multirow{4}{*}{$\tfrac{1}{4} g_\ro{F}^2$}\;&\; $\la{6}$, $\la{8}$, $\la{32}$, $\lap{32}$, $\la{40}$, $\lap{40}$, $\la{42}$, $\lap{42}$,\\
            \;&\; $\la{46}$, $\lap{46}$, $\la{52}$, $\lap{52}$, $\la{60}$, $\lap{60}$, $\la{78}$, $\la{82}$,\\
            \;&\; $\la{90}$, $\lap{90}$, $\la{94},$ $\lap{94}$, $\la{102}$, $\lap{102}$, $\la{106}$, $\lap{106}$, $\la{111}$, $\lap{111}$,\\
            \;&\; $\la{124}$, $\la{128}$, $\lap{128}$\\
			\hline
            \multirow{6}{*}{$0$}\;&\; $\la{61}$, $\la{62}$, $\la{63}$, $\lap{63}$, $\la{64}$, $\lap{64}$, $\la{65}$, $\lap{65}$,\\
            \;&\; $\la{67}$, $\lap{67}$, $\la{73}$, $\lap{73}$, $\la{88}$, $\la{92}$, $\lap{92}$,\\
            \;&\; $\la{100}$, $\lap{100}$, $\la{112},$ $\lap{112}$, $\la{136}$, $\la{137}$, $\la{161}$, $\la{164}$, $\lap{164}$,\\
            \;&\; $\la{165}$, $\lap{165}$, $\la{166}$, $\lap{166}$, $\la{167}$, $\lap{167}$, $\la{169}$, $\lap{169}$,\\
            \;&\; $\la{172}$, $\lap{172}$, $\la{175}$, $\lap{175}$, $\la{181}$, $\lap{181}$, $\la{182}$, $\lap{182}$,\\
            \;&\; $\la{183}$, $\lap{183}$, $\la{184}$, $\lap{184}$\\
   			\bottomrule
		\end{tabular} }
		\caption{Scalar quartic couplings matching conditions with one or zero D-term interactions. The $\lap{i}$ couplings refer to the $\(\ro{L} \rightarrow \ro{R}\)$ part of $V_4$.}
		\label{tab:match-4}  
	\end{center}
\end{table}
%

\subsection{The effective Yukawa Lagrangian}
\label{App:Eff-fermion}

In addition to fundamental-chiral fermions, which were broadly discussed in \cref{Sect:fermion-sector}, the SHUT model also contains fermionic states coming from the chiral-adjoint and gaugino sectors. We refer to our previous work \cite{Camargo-Molina:2017kxd} for thorough details. In this appendix we preserve the original notation which we recall in what follows:
Soft-scale weak-singlet fermions embedded in $\SU{3}{L}$, $\SU{3}{R}$ and $\SU{2}{F}\times \U{F}$ are denoted as $\mathcal{S}_\ro{L,R,F}$ respectively. All doublets acquire D-term masses receiving T-GUT values thus not included below the soft scale. Triplet fermions are denoted as $\mathcal{T}_\ro{L,R,F}$ and finally, $\SU{3}{C}$ octets, which are mostly gluino-like, are identified as $g^a$. The effective Lagrangian contains both quadratic and Yukawa interactions which we cast as
\begin{align}
\mathcal{L}_{\rm fermi} = \mathcal{L}_{\rm M} + \mathcal{L}_{\rm Yuk}\,.
\label{eq:Lfermi-LR}
\end{align}  
For the mass terms we have
\begin{align}
\begin{aligned}
\label{LM}
\mathcal{L}_{\rm M}  = \sum_{\ro{A}=\ro{L,R,F}}
\[\tfrac{1}{2} m_{S_{\rm A}} S_{\rm A} S_{\rm A} 
+ \tfrac{1}{2} m_{\mathcal{T}_{\rm A}} \mathcal{T}^i_{\rm A} \mathcal{T}^i_{\rm A} \]
+
\tfrac{1}{2} m_{g} g^a g^a
+
\sum_{\ro{A} \neq \ro{A}^\prime} m_{\ro{A} \ro{A}^\prime} S_\ro{A} S_{\ro{A}^\prime}
+
\ro{c.c.}\,,
\end{aligned}
\end{align}
with $\ro{A}, \ro{A}^\prime = \ro{L}, \ro{R}, \ro{F}$, while for the Yukawa ones we write for convenience,
\begin{align}
\mathcal{L}_{\rm Yuk} = \mathcal{L}_{\rm 3c} + \mathcal{L}_{\rm 2c} + \mathcal{L}_{\rm 1c} + \mathcal{L}_{\mathcal{S}} + 
\mathcal{L}_{\mathcal{T}} + \mathcal{L}_{\rm \widetilde{g}}\,,
\end{align}
where the first three terms, which involve only the fields from the fundamental representations of the trinification group, denote three, 
two and one $\SU{2}{}$ contractions, respectively, whereas the last ones describe the Yukawa interactions of the singlet $\mathcal{S}$, 
triplet $\mathcal{T}$ and octet $g$ fermions. Similarly to the the scalar potential, whenever we have $(\ro{L} \to \ro{R})$ the Yukawa couplings should be identified as $\y{i} \to \yp{i}$.
The terms with three $\SU{2}{}$ contractions are given by
\begin{align}
\begin{split}
\label{L3c}
\mathcal{L}_{\rm 3c}  = 
\varepsilon_{f f'} \left( \y{1} {q}_{\rm R}^{f\,r} \widetilde{\chi}^{3\,l}_{r} q_{{\rm L}\,l}^{f'}
+\[
\y2 \widetilde{q}_{\rm R}^{3\,r}  \chi^{f\,l}_{r} q_{{\rm L}\,l}^{f'}
+ \y3 \widetilde{q}_{\rm R}^{f\,r}\, \chi^{3\,l}_{r}  q_{{\rm L}\,l}^{f'} 
+ \y4 q_{\rm R}^{3\,r} \widetilde{\chi}^{f\,l}_{r} q_{{\rm L}\,l}^{f'}
+ \(\ro{L} \rightarrow \ro{R}\)
\]+ {\rm c.c.} \right) \,,
\end{split}
\end{align}
those with two $\SU{2}{}$ contractions are written as
\begin{align}
\begin{split}
\label{L2c}
\mathcal{L}_{\rm 2c}  &= 
\varepsilon_{f f'} \left[\y{5} \widetilde{D}_{\rm R}^3 q_{{\rm L}\,l}^{f} \ell_{\rm L}^{f'\,l}
+\y{6} \widetilde{D}_{\rm R}^{f}  q_{{\rm L}\,l}^{f'} \ell_{\rm L}^{3\,l}
+\y{7} \widetilde{D}_{\rm R}^{f} q^3_{{\rm L}\,l} \ell_{\rm L}^{f'\,l}
+\y{8} D_{\rm R}^3 \widetilde{q}_{{\rm L}\,l}^{f} \ell_{\rm L}^{f'\,l}
+\y{9} D_{\rm R}^{f}  \widetilde{q}_{{\rm L}\,l}^{f'} \ell_{\rm L}^{3\,l}
 \right.  \\
&
\left.  
+\y{10} D_{\rm R}^{f} \widetilde{q}^3_{{\rm L}\,l} \ell_{\rm L}^{f'\,l}
+\y{11} D^3_{\rm R}  q_{{\rm L}\,l}^{f} \widetilde{\ell}_{\rm L}^{f'\,l} 
+\y{12} D_{\rm R}^{f} q_{{\rm L}\,l}^{f'} \widetilde{\ell}_{\rm L}^{3\,l}
+\y{13} D_{\rm R}^{f} q^3_{{\rm L}\,l} \widetilde{\ell}_{\rm L}^{f'\,l} 
+\(\ro{L} \rightarrow \ro{R}\)  \right]
+ {\rm c.c.}\,,
\end{split}
\end{align}
and for those with one $\SU{2}{}$ contraction we have
\begin{align}
\begin{split}
\label{L1c}
\mathcal{L}_{\rm 1c}  &= 
\varepsilon_{f f'}  \left( 
\y{14} D_{\rm R}^f \widetilde{\phi}^3 D_{\rm L}^{f'}
+ \left[
\y{15} \widetilde{D}^3_{\rm R} \phi^f  D_{\rm L}^{f'}
+\y{16} \widetilde{D}_{\rm R}^{f} \phi^{f'} D^3_{\rm L} 
+\y{17}  \widetilde{D}_{\rm R}^{f} \phi^3  D_{\rm L}^{f'}
\right. \right. \\
&
\left. \left. 
+\y{18} D^3_{\rm R}  \widetilde{\phi}^f D_{\rm L}^{f'} + \(\ro{L} \rightarrow \ro{R}\) \right] \right) + {\rm c.c.} \,.
\end{split}
\end{align}
The part of the Lagrangian involving the singlets $\mathcal{S}_{\rm L,R,F}$ reads
\begin{align}
\begin{split}
\label{LS}
\mathcal{L}_{\mathcal{S}}  &= \sum_{\ro{A} = \ro{L,R,F}} \(
 \left[  
\y{19}^{\Scale[0.5]{\ro{A}}} \widetilde{q}^{\ast\,l}_{{\rm L}\,f}\mathcal{S}_{\rm A} {q}_{{\rm L}\,l}^{f}
+\y{20}^{\Scale[0.5]{\ro{A}}} \widetilde{q}^{\ast\,3\,l}_{\rm L}\mathcal{S}_{\rm A} {q}^3_{{\rm L}\,l}
+\y{21}^{\Scale[0.5]{\ro{A}}} \widetilde{D}_{{\rm L}\,f}^\ast \mathcal{S}_{\rm A} {D}_{\rm L}^{f}
+\y{22}^{\Scale[0.5]{\ro{A}}} \widetilde{D}_{\rm L}^{\ast\;3} \mathcal{S}_{\rm A} {D}^3_{\rm L}
+\y{23}^{\Scale[0.5]{\ro{A}}} \widetilde{\ell}_{{\rm L}\,f\,l}^\ast \mathcal{S}_{\rm A} {\ell}_{\rm L}^{f\,l}
\right. \right. \\
&
\left. \left.
+\y{24}^{\Scale[0.5]{\ro{A}}}  \widetilde{\ell}_{{\rm L}\,l}^{\ast \,3} \mathcal{S}_{\rm A}{\ell}_{\rm L}^{\,3\,l}
+ \(\ro{L} \rightarrow \ro{R}\)
\right]
+\y{25}^{\Scale[0.5]{\ro{A}}} \widetilde{\chi}^{\ast \,r}_{f\,l} \mathcal{S}_{\rm A} \chi^{f\,l}_{r}
+\y{26}^{\Scale[0.5]{\ro{A}}} \widetilde{\chi}^{\ast \,3 \,r}_{l}\mathcal{S}_{\rm A} \chi^{3\,l}_{r}
+\y{27}^{\Scale[0.5]{\ro{A}}} {\widetilde{\phi}^\ast_f} \mathcal{S}_{\rm A} {\phi}^{f}
+\y{28}^{\Scale[0.5]{\ro{A}}} {\widetilde{\phi}^{\ast\;3}} \mathcal{S}_{\rm A} {\phi^3} \)
\\
&+
{\rm c.c.}
\,,
\end{split}
\end{align}
where $\(\ro{L} \rightarrow \ro{R} \)$ is only acting on the components of the fundamental tri-triplet superfields. For interactions involving the triplets $\mathcal{T}^i_{\rm L,R,F}$ we have
\begin{align}
\begin{split}
\label{LT}
\mathcal{L}_{\mathcal{T}}  &=  \left({\sigma^i_{\Scale[0.5]{ \ro{L} }}}\right)^{l}_{l'}   \left(
\y{29}  \widetilde{q}^{\ast\,l'}_{{\rm L}\,f}\mathcal{T}^i_\ro{L} {q}_{{\rm L}\,l}^{f}
+\y{30} \widetilde{q}^{\ast\,3 \,l'}_{\rm L} \mathcal{T}^i_\ro{L}  {q}^3_{{\rm L}\,l}
+\y{31} \widetilde{\chi}^{\ast \,r}_{f\,l} \mathcal{T}^i_\ro{L} \chi^{f\,l'}_{r}
+\y{32} \widetilde{\chi}^{\ast\,3 \,r}_{l} \mathcal{T}^i_\ro{L} \chi^{3\,l'}_{r}
+\y{33} \widetilde{\ell}_{{\rm L}\,f\,l}^\ast \mathcal{T}^i_\ro{L} {\ell}_{\rm L}^{f\,l'}
\right. \\
&
\left.
+\y{34} \widetilde{\ell}_{{\rm L}\,l}^{\ast \,3} \mathcal{T}^i_\ro{L} {\ell}_{\rm L}^{3\,l'}
\right)
\\
&
+
\left({\sigma^i_{\Scale[0.5]{ \ro{R}  }}}\right)^{r'}_{r}   \left(
\y{35}  \widetilde{q}^{\ast}_{{\rm R}\,f\,r'}\mathcal{T}^i_\ro{R} {q}_{{\rm R}}^{f\,r}
+\y{36} \widetilde{q}^{\ast\,3}_{{\rm R}\,r'} \mathcal{T}^i_\ro{R}  {q}^{3\,r}_{{\rm R}}
+\y{37} \widetilde{\chi}^{\ast \,r}_{f\,l} \mathcal{T}^i_\ro{R} \chi^{f\,l}_{r'}
+\y{38} \widetilde{\chi}^{\ast\,3 \,r}_{l} \mathcal{T}^i_\ro{R} \chi^{3\,l}_{r'}
+\y{39} \widetilde{\ell}_{{\rm R}\,f}^{\ast\,r} \mathcal{T}^i_\ro{R} {\ell}_{{\rm R}\,r'}^{f}
\right. \\
&
\left.
+\y{40} \widetilde{\ell}_{{\rm R}}^{\ast \,3\, r} \mathcal{T}^i_\ro{R} {\ell}_{{\rm R}\,r'}^{3}
\right)
\\
&
+
\left({\sigma^i_{\Scale[0.5]{ \ro{F}  }}}\right)^{f}_{f'}   \left(
\y{41}  \widetilde{q}^{\ast\,l}_{{\rm L}\,f}\mathcal{T}^i_\ro{F} {q}_{{\rm L}\,l}^{f'}
+\y{42}  \widetilde{q}^{\ast}_{{\rm R}\,f\,r}\mathcal{T}^i_\ro{F} {q}_{{\rm R}}^{f'\,r}
+\y{43} \widetilde{\chi}^{\ast \,r}_{f\,l} \mathcal{T}^i_\ro{F} \chi^{f'\,l}_{r}
+\y{44} \widetilde{\ell}_{{\rm L}\,f\,l}^\ast \mathcal{T}^i_\ro{F} {\ell}_{\rm L}^{f'\,l}
+\y{45} \widetilde{\ell}_{{\rm R}\,f}^{\ast\,r} \mathcal{T}^i_\ro{F} {\ell}_{{\rm R}\,r}^{f'}
\right. \\
&
\left.
+\y{46}  \widetilde{D}^{\ast}_{{\rm L}\,f}\mathcal{T}^i_\ro{F} {D}_{\rm L}^{f'}
+\y{47}  \widetilde{D}^{\ast}_{{\rm R}\,f}\mathcal{T}^i_\ro{F} {D}_{{\rm R}}^{f'}
+\y{48} \widetilde{\phi}_{f}^{\ast} \mathcal{T}^i_\ro{F} {\phi}^{f'}
\right)
+ {\rm c.c.}\,,
\end{split}
\end{align}
with ${\sigma^i_{\Scale[0.5]{ \ro{L,R,F} }}}$ the generators of the $\SU{2}{L,R,F}$ interactions and where summation over the adjoint index $i$ is implicit. Finally, the Yukawa interactions involving gluinos are given by
\begin{align}
\begin{split}
\label{Lglu}
&\mathcal{L}_{\widetilde{g}}  = 
\y{49} \widetilde{q}^{\ast \,l}_{{\rm L}\,f}{\bm{T}^a} g^a q_{{\rm L}\,l}^{f}
+\y{50} \widetilde{q}^{\ast \,3 \,l}_{\rm L} {\bm{T}^a} g^a{q}^3_{{\rm L}\,l}
+\y{15} \widetilde{D}_{{\rm L}\,f}^\ast {\bm{T}^a} g^a {D}_{\rm L}^{f}
+\y{52} \widetilde{D}_{\rm L}^{\ast\,3}  {\bm{T}^a} g^a{D}^3_{\rm L}
+ \( \ro{L} \rightarrow \ro{R} \)
+ {\rm c.c.}
\,.
\end{split}
\end{align}
The tree-level matching conditions for the fermion sector are sumarized in \cref{tab:match-Y}.
\begin{table}[htb!]
	\begin{center}
		\Scale[0.95]{
		\begin{tabular}{cc}
			\toprule                     
			Matching value \;&\; Yukawa coupling\\  
			\midrule
			$\mathcal{Y}_1$  		\;&\; $\yp{2}$, $\yp{5}$, $\y{7}$, $\yp{8}$, $\y{10}$, $\yp{11}$, $\y{13}$, $\y{16}$ 
			\\
			\hline            \multirow{2}{*}{$\mathcal{Y}_2$}\;&\; $\y{1}$, $-\y{2}$, $\y{3}$, $-\y{4}$, $-\y{5}$, $\y{6}$, $\yp{6}$, $-\yp{7}$, $-\yp{8}$, $\y{9}$, $\yp{9}$, $-\yp{10}$,\\
			\;&\; $-\y{11}$, $\y{12}$, $\yp{12}$, $\yp{13}$, $\y{14}$, $-\y{15}$, $\y{17}$, $-\y{18}$ \\
			\hline
			$\sqrt{2} g_\ro{L}^\prime$  		\;&\; $\y{19}^{\Scale[0.5]{\ro{L}}}$, $\y{20}^{\Scale[0.5]{\ro{L}}}$, $-\tfrac{1}{2}\y{21}^{\Scale[0.5]{\ro{L}}}$, $-\tfrac{1}{2}\y{22}^{\Scale[0.5]{\ro{L}}}$, $-\y{23}^{\Scale[0.5]{\ro{L}}}$, $\tfrac{1}{2}\yp{23}^{\Scale[0.5]{\ro{L}}}$, $-\y{24}^{\Scale[0.5]{\ro{L}}}$, $\tfrac{1}{2}\yp{24}^{\Scale[0.5]{\ro{L}}}$, $-\y{25}^{\Scale[0.5]{\ro{L}}}$,
			$-\y{26}^{\Scale[0.5]{\ro{L}}}$, $\tfrac{1}{2}\y{27}^{\Scale[0.5]{\ro{L}}}$, $\tfrac{1}{2}\y{28}^{\Scale[0.5]{\ro{L}}}$
			\\
			\hline
			$\sqrt{2} g_\ro{R}^\prime$  		\;&\; $-\yp{19}^{\Scale[0.5]{\ro{R}}}$, $-\yp{20}^{\Scale[0.5]{\ro{R}}}$, $\tfrac{1}{2}\yp{21}^{\Scale[0.5]{\ro{R}}}$, $\tfrac{1}{2}\yp{22}^{\Scale[0.5]{\ro{R}}}$, $\yp{23}^{\Scale[0.5]{\ro{R}}}$, $-\tfrac{1}{2}\y{23}^{\Scale[0.5]{\ro{R}}}$, $\yp{24}^{\Scale[0.5]{\ro{R}}}$, $-\tfrac{1}{2}\y{24}^{\Scale[0.5]{\ro{R}}}$, $\y{25}^{\Scale[0.5]{\ro{R}}}$,
			$\y{26}^{\Scale[0.5]{\ro{R}}}$, $-\tfrac{1}{2}\y{27}^{\Scale[0.5]{\ro{R}}}$, $-\tfrac{1}{2}\y{28}^{\Scale[0.5]{\ro{R}}}$
			\\
			\hline
			$\sqrt{2} g_\ro{F}^\prime$  		\;&\; $-\y{19}^{\Scale[0.5]{\ro{F}}}$, $\tfrac{1}{2}\y{20}^{\Scale[0.5]{\ro{F}}}$, $-\y{21}^{\Scale[0.5]{\ro{F}}}$, $\tfrac{1}{2}\y{22}^{\Scale[0.5]{\ro{F}}}$, $-\y{23}^{\Scale[0.5]{\ro{F}}}$, $\tfrac{1}{2}\y{24}^{\Scale[0.5]{\ro{F}}}$, $-\y{25}^{\Scale[0.5]{\ro{F}}}$,
			$\tfrac{1}{2}\y{26}^{\Scale[0.5]{\ro{F}}}$, $-\y{27}^{\Scale[0.5]{\ro{F}}}$, $\tfrac{1}{2}\y{28}^{\Scale[0.5]{\ro{F}}}$
			\\
			\hline
			$-\sqrt{2} g_\ro{L}$  		\;&\; $\y{29}$ to $\y{34}$, 
			\\
			\hline
			$-\sqrt{2} g_\ro{R}$  		\;&\; $\y{35}$ to $\y{40}$, 
			\\
			\hline
			$-\sqrt{2} g_\ro{F}$  		\;&\; $\y{41}$ to $\y{48}$, 
			\\
			\hline
			$-\sqrt{2} g_\ro{C}$  		\;&\; $\y{49}$ to $\y{52}$ and $\yp{49}$ to $\yp{52}$ 
			\\
   			\bottomrule
		\end{tabular} }
		\caption{Yukawa couplings matching conditions. The $\yp{i}$ couplings refer to the $\(\ro{L} \rightarrow \ro{R}\)$ part of $\mathcal{L}_\ro{fermi}$.}
		\label{tab:match-Y}  
	\end{center}
\end{table}

\FloatBarrier

\bibliographystyle{JHEP}
\bibliography{bib}

\end{document}